\documentclass[twocolumn,trackchanges]{aastex631}
\usepackage{lipsum}
\usepackage{amssymb}
\usepackage{hyperref}
\usepackage{listings}
\usepackage{xcolor}
\usepackage{amsmath}

\lstdefinelanguage{SQL}{
    morekeywords={SELECT, FROM, WHERE, AND, AS, LOG10},
    sensitive=false,
    morecomment=[l]{--},
    morestring=[b]',
}

\begin{document}
\makeatletter
\let\frontmatter@title@above=\relax
\makeatother

\newcommand\lsim{\mathrel{\rlap{\lower4pt\hbox{\hskip1pt$\sim$}}
\raise1pt\hbox{$<$}}}
\newcommand\gsim{\mathrel{\rlap{\lower4pt\hbox{\hskip1pt$\sim$}}
\raise1pt\hbox{$>$}}}

\newcommand{\CS}[1]{{\color{red} CS: #1}}
\newcommand{\KEB}[1]{{\color{cyan} KEB: #1}}
\newcommand{\SN}[1]{{\color{blue} SN: #1}}

\title{\Large 10,000 Resolved Triples from Gaia: Empirical Constraints on Triple Star Populations}

\shorttitle{10,000 Resolved Triples from Gaia}
\shortauthors{Shariat et al.}

\author[0000-0003-1247-9349]{Cheyanne Shariat}
\affiliation{Department of Astronomy, California Institute of Technology, 1200 East California Boulevard, Pasadena, CA 91125, USA}

\author[0000-0002-6871-1752]{Kareem El-Badry}
\affiliation{Department of Astronomy, California Institute of Technology, 1200 East California Boulevard, Pasadena, CA 91125, USA}

\author[0000-0002-9802-9279]{Smadar Naoz}
\affiliation{Department of Physics and Astronomy, University of California, Los Angeles, Los Angeles, CA 90095, USA}
\affiliation{Mani L. Bhaumik Institute for Theoretical Physics, University of California, Los Angeles, Los Angeles, CA 90095, USA }


\correspondingauthor{Cheyanne Shariat}
\email{cshariat@caltech.edu}

\begin{abstract}
We present a catalog of $\sim 10,000$ resolved triple star systems within 500 pc of the Sun, constructed using Gaia data. The triples include main-sequence, red giant, and white dwarf components spanning separations of 10 to 50,000 au. A well-characterized selection function allows us to constrain intrinsic demographics of the triple star population. We find that 
(a) all systems are compatible with being hierarchical and dynamically stable; 
(b) mutual orbital inclinations are isotropic for wide triples but show modest alignment as the systems become more compact; 
(c) primary masses follow a Kroupa initial mass function weighted by the triple fraction; 
(d) inner binary orbital periods, eccentricities, and mass ratios mirror those of isolated binaries, including a pronounced twin excess (mass ratios greater than 0.95) out to separations of 1000+ au, suggesting a common formation pathway; 
(e) tertiary mass ratios follow a power-law distribution with slope -1.4; 
(f) tertiary orbits are consistent with a log-normal period distribution and thermal eccentricities, subject to dynamical stability. Informed by these observations, we develop a publicly available prescription for generating mock triple star populations. Finally, we estimate the catalog's completeness and infer the intrinsic triple fraction, which rises steadily with primary mass: from $5\%$ at $\lesssim 0.5\,{\rm M_\odot}$ to $35\%$ at $2\,{\rm M_\odot}$. The public catalog provides a robust testbed for models of triple star formation and evolution.
\end{abstract}

\section{Introduction}\label{sec:introduction}
Most stars both form and live out their main-sequence lives with at least one stellar companion. The companion fraction of stars increases with their mass, where $\sim50\%$ of solar-type stars and $\sim100\%$ of massive stars reside in binaries or higher-order multiples \citep{DM91, Raghavan2010, Moe17}. However, multiple star systems are born inside dusty star-forming regions, making direct observations challenging \citep[][]{Offner23}. Their formation involves several complex physical processes -- such as turbulent fragmentation, disk fragmentation, migration, and dynamical interactions -- each operating across a range of spatial and temporal scales \citep[e.g.,][]{Offner10, Bate12, Kratter16, Lee19, Offner23}. These early formation processes leave lasting imprints on the architectures of multiple star systems, shaping their mass distributions, orbital configurations, and multiplicity \citep[see][for recent reviews]{Tokovinin21, Offner23}. As such, studying the present-day population of stellar multiples offers a valuable and accessible window into their formation and subsequent evolution. 

A major challenge in studying multiple star populations lies in assembling large, statistically robust samples with characterized selection biases. Previous efforts have made significant progress by combining high-resolution imaging, radial velocity monitoring, and astrometry to create surveys of hierarchical multiples \citep[e.g.,][]{Tokovinin1997, DM91, Tokovinin06, Tokovinin08, Raghavan2010, Duchene13, Tokovinin14a, Tokovinin14b, Borkovits16_keplertriples, Moe17, Winters19,Tokovinin22_resolvedtriples,Tokovinin23_gaia}. While combining multiple detection techniques improves sensitivity to a wider range of periods and masses, it also produces heterogeneous selection functions that are difficult to quantify.
Some studies addressed this by constructing volume-limited samples with relatively high completeness \citep[e.g.,][]{Raghavan2010, Tokovinin14a, Moe17, Winters19, Tokovinin22_resolvedtriples}, but the samples remain modest in size, especially for triples and higher-order multiples.

{\it Gaia} has opened a new era for studying stellar multiplicity. Its precise astrometry, uniform all-sky coverage, and well-characterized selection function enable systematic surveys of multiple star systems at an unprecedented scale \citep{Gaia_Collab}.
Early studies used {\it Gaia} DR2 and DR3 astrometry to produce large catalogs of wide binaries in the solar neighborhood \citep[e.g.,][see latter for a review on {\it Gaia} binaries]{Andrews17, EB18, Tian20, Hartman20, EB21_widebin, EB_review}. Immediately, these statistical catalogs revealed new discoveries about binary orbital periods \citep[e.g.,][]{EB18}, mass ratios \citep[e.g.,][]{EB_19_twins}, metallicities \citep[e.g.,][]{EB19_metal, Hwang21_metal, Niu22_WBmetals}, and eccentricities \citep[e.g.,][]{Hwang22_ecc, Hamilton22_widebinecc}.
However, in focusing on binaries, most of these studies explicitly removed triples from their samples to ensure clean selection. As a result, triple and higher-order multiples remain comparatively underexplored with {\it Gaia}, despite their critical role in stellar dynamics and evolution.
Recent efforts have begun to fill this gap. Specifically, \citet{Tokovinin22_resolvedtriples} presents a sample of resolved triples within $100$ pc, and \citet{Tokovinin23_gaia} used {\it Gaia} diagnostics to infer unresolved subsystems in binaries. 
Nevertheless, these samples remain modest in size. Furthermore, a major limitation in these studies is the difficulty of distinguishing true gravitationally bound systems from chance alignments and moving groups, particularly for wide pairs where contamination becomes significant. In this work, we use {\it Gaia} data to create a large catalog of resolved triples with a focus on high purity.

Beyond probing star formation, triple systems offer a laboratory for studying secular dynamics coupled to stellar evolution.
One key dynamical phenomenon among hierarchical triples is the eccentric Kozai-Lidov (EKL) mechanism, whereby the tertiary star torques the more compact inner binary, driving periodic oscillations in its eccentricity and inclination \citep[][see latter for a review]{vonZeipel1910,Kozai1962,Lidov1962,Naoz2016}. Through the interplay of EKL dynamics and stellar evolution, triple stellar evolution serves as a key formation channel for a broad range of astrophysical phenomena in the Galaxy \citep[][]{Perets25}, including short-period contact binaries \citep[e.g.,][]{Tokovinin06,Laos20,Fabrycky07,Naoz2014}, cataclysmic variables \citep[][]{Knigge22,Shariat25CV}, stellar mergers with a white dwarf component \citep[][]{Perets12,Naoz2014,Hamers22,Toonen20,Toonen2022,Heintz22,Shariat23,Kummer23, Heintz24, Shariat25Merge}, blue stragglers \citep[][]{Perets09,Naoz2014,Gao23,Shariat25Merge,Leiner25}, and X-ray binaries \citep[][]{NaozLMXB,Shariat24LMXB}. Large, statistically robust samples of triple star systems are essential for testing and refining these dynamical formation pathways.

In this paper, we construct a catalog of $\sim$10,000 resolved triple star systems from {\it Gaia}, with a homogeneous selection function. We use this catalog as an observational anchor to calibrate models of triple star evolution and to understand the role of triples in Galactic stellar populations. The remainder of the paper is structured as follows. In Section \ref{sec:creating_catalog}, we describe our method of constructing the catalog from {\it Gaia} astrometry. Section \ref{sec:properties} outlines basic properties of the triple population. 
Section \ref{sec:results} presents our main analysis, where we examine the origin of triple masses (Section \ref{subsec:triple_masses}), orbital separations (Section \ref{subsec:triple_separations}), and eccentricities (Section \ref{subsec:triple_ecc}).
In Section \ref{subsec:how_to_sample}, we build on these results to develop an observationally motivated prescription for sampling triple parameters. In Section \ref{sec:discussion}, we place our results in the context of previous triple catalogs and highlight notable systems, including exoplanet-hosting triples.
Finally, we conclude in Section \ref{sec:conclusions} with a summary of our main conclusions. Additional details about the data and modeling are discussed in Appendices \ref{app:chance_alignment}, \ref{app:unmatched_pairs}, 
\ref{app:ang_res_criteria},
\ref{app:WB_control},
\ref{app:projected_separations},
\ref{app:missing_binaries},
\ref{app:exoplanets}, and extra catalogs of higher-order multiples (4 or more stars) are provided in Appendix
\ref{app:resolved_quads} and \ref{app:higher_order_mults}.

\section{Constructing the Catalog}\label{sec:creating_catalog}

Using {\it Gaia} data, we construct a catalog of resolved triple star systems by extending the approach developed for wide binaries in \citet{EB21_widebin} to identify triples. Their original wide binary catalog filters out higher-order multiples by removing sources that matched with two different companions. Here, we relax this constraint to include systems with three or more stars. We summarize our selection process below.

First, we query {\it Gaia} for all sources within $500$~pc (${\tt parallax}> 2$ mas) that have well-measured parallaxes, proper motion, and G magnitude measurements. Specifically, we require that ${\tt parallax\_over\_error}>5$, ${\tt parallax\_error}<2$ mas, and a {\tt phot\_g\_mean\_mag} is measured for the source. These requirements are identical to the initial query from \citet{EB21_widebin}, except for our distance cut, which is smaller than their cut of $1000$~pc. We adopt a distance cut of $500$~pc because beyond this distance, {\it Gaia} is only sensitive to inner binaries wider than $\sim500$~au, which is a vanishingly small subset of the triple population. Moreover, wider inner binaries are accompanied by even wider tertiaries, increasing the fraction of chance alignments.
The ADQL query for our initial {\it Gaia} selection is:

\begin{lstlisting}[language=SQL]
SELECT *
FROM 
    gaiadr3.gaia_source
WHERE 
    parallax > 2 AND
    parallax_over_error > 5 AND
    parallax_error < 2 AND
    phot_g_mean_mag is not null
\end{lstlisting}

This {\it Gaia} search yields 23,525,994 sources, which entails $N(N-1)(N-2)/6\approx2\times10^{21}$ possible triple combinations and $N(N-1)/2\approx2\times10^{14}$ binary combinations. Using a nearest neighbors search, we identify nearby stars within $1$~pc of one another:

\begin{equation}
    \frac{\theta}{{\rm arcsec}} \leq 206.265 \times \frac{\varpi}{{\rm mas}}.
\end{equation}

Here, $\theta$ is the angular separation between two sources and $\varpi$ is the measured parallax. Beyond $1$~pc, most binaries would have already been disrupted due to Galactic tides and/or flybys \citep[e.g.,][]{Binney08,Kaib2014}, so most pairs would be spurious. We further filter these initial companion candidates, keeping only those that share a parallax and proper motion within uncertainties. The parallax cut is 
\begin{equation}
    |\varpi_1 - \varpi_2| < n \sqrt{\sigma_{\rm \varpi,1}^2 + \sigma_{\rm \varpi,2}^2},
\end{equation}
where $\sigma_{\rm \varpi,i}$ is the uncertainty in parallax of the $i$-th component ($i$ = 1 or 2) in the pair. Following \citet{EB21_widebin} we choose $n=3$ for pairs with $\theta>4''$ and $n=6$ for pairs with $\theta<4''$. Choosing a looser cut at $\theta<4''$ reflects the fact that the chance alignment probability there is low and the true error in parallax will be systematically underestimated at closer angular separations due to blending.

We further require that the proper motion ($\mu_i$) of any two sources be consistent with a bound Keplerian orbit within $2$ sigma. Thereby, for a given proper motion difference ($\Delta \mu$) and its corresponding uncertainty ($\sigma_{{\rm \Delta\mu}}$) we require that
\begin{equation}\label{eq:proper_motion_cut}
    \Delta\mu - \Delta\mu_{\rm orbit} \leq 2\sigma_{{\rm \Delta\mu}} , 
\end{equation}\label{eq:deltamu_criteria}
where $\Delta\mu_{\rm orbit}$ is the maximum proper motion difference expected due to purely orbital motion.
We calculate $\Delta \mu$ and $\sigma_{\Delta\mu}$ from the reported {\it Gaia} astrometry with
\begin{equation}\label{eq:deltamu}
    \Delta\mu = [(\mu^*_{\rm \alpha,1} - \mu^*_{\rm \alpha,2})^2 + (\mu_{\rm \delta,1} - \mu_{\rm \delta,2})^2]^{1/2} \ ,
\end{equation}
and
\begin{equation}
\sigma_{\Delta \mu} = \frac{1}{\Delta \mu} 
\sqrt{ 
\left( \sigma_{\mu^*_{\alpha,1}}^2 + \sigma_{\mu^*_{\alpha,2}}^2 \right) \Delta \mu_{\alpha}^2 
+ 
\left( \sigma_{\mu_{\delta,1}}^2 + \sigma_{\mu_{\delta,2}}^2 \right) \Delta \mu_{\delta}^2 
} \ ,
\end{equation}
where 
$ \Delta \mu_{\alpha}^2 = (\mu^*_{\alpha,1} - \mu^*_{\alpha,2})^2 $
and 
$ \Delta \mu_{\delta}^2 = (\mu_{\delta,1} - \mu_{\delta,2})^2$, and $\mu^*_{\alpha,i}=\mu_{\alpha,i}\cos{\delta_i}$ and $\mu_{\alpha,i}$ ($\mu_{\delta,i}$) is the component of the proper motion in the right ascension (declination) direction. The maximum proper motion difference expected for a circular orbit of total mass $5$~M$_\odot$ is \citep{EB18}
\begin{equation}
    \Delta\mu_{\rm orbit} = 0.44~{\rm mas~yr^{-1}} \times \left(\frac{\varpi}{{\rm mas}}\right)^{3/2}\left(\frac{\theta}{{\rm arcsec}}\right)^{-1/2} \ ,
\end{equation}\label{eq:mu_orbit}
which we adopt as the conservative upper limit for $\Delta\mu$. Since most systems in our sample have total masses below $5~{\rm M_\odot}$, this cut is safely inclusive. Moreover, it allows more massive systems to enter the sample because it assumes the limit where maximal orbital motion is fully projected into the plane of the sky and provides a $2\sigma$ buffer.
This filter also removes a majority of unresolved pairs, which generally have larger orbital velocities and therefore larger proper motion differences. Moreover, this cut can also exclude some resolved triples, particularly when the inner binary is close and its high orbital motion skews the proper motions of both components. In such cases, the tertiary may appear to have a large proper motion difference from both inner stars, causing it to fail the consistency check. This introduces a bias against compact inner binaries with wide tertiaries, and should be considered when using this catalog. We discuss this further in the following section.

We exclude sources in crowded regions and clusters following the method of \citet{EB21_widebin}, which we summarize here. For each source, we count the number of neighboring stars within $5$pc that (i) are brighter than $G=18$, (ii) have proper motions within $5{\rm km~s^{-1}}$ at the $2\sigma$ level, and (iii) have parallaxes consistent within $2\sigma$. Sources with more than 30 such neighbors are removed, effectively eliminating those in dense environments like clusters and moving groups.

To further exclude stars in moving groups and open clusters, we count the number of nearby astrometric companions per source. We reject candidates with more than $4$ such companions, allowing for multiple star systems (i.e., two or three resolved companions) but removing most dense co-moving groups/clusters.

After the above cuts, our sample of candidate pairs reduces to $938,127$ pairs, of which $103,598$ ($11\%$) contain duplicate sources (i.e., one source is matched to more than one companion). At this stage, \citet{EB21_widebin} filter out duplicates to generate their sample of wide binaries. Here, we exclusively focus on these duplicate matches since they are initial candidates for triple and higher-order multiplicity systems. Before identifying the multiple systems among our large sample of paired stars, we first filter out chance alignments between the pairs.

\subsection{Chance Alignment Probability}\label{subsec:chance_alignments}
Sources are chance alignments if they share a similar proper motion and parallax but are not gravitationally bound. In principle, identifying false matches is challenging, but methods of reliably quantifying chance alignment probabilities have been developed and tested. We follow the formulation of \citet{EB21_widebin}, where we appoint the ${\tt R\_chance_\_align}$ ($\mathcal{R}$) parameter as an estimate of the chance alignment probability for a given pair. 

For each candidate pair among our initial sample, we calculate $\mathcal{R}$ by applying the Kernel Density Estimate (KDE) described in the Appendix of \citet{EB21_widebin}. This KDE uses the (1) angular separation, (2) distance, (3) parallax difference uncertainty, (4) local sky density, (5) tangential velocity, (6) parallax difference over error, and (7) proper motion difference over error to derive $\mathcal{R}$. The local sky density is estimated by $\Sigma_{\rm 18}$, which represents the number of sources per square degree that both (1) pass the cuts of our initial query and (2) are brighter than $G = 18$. 
The KDE acts on a {\it shifted} catalog, which is the same as the initial catalog with all sources shifted in declination. Any pairs found in the shifted catalog are, by construction, chance alignments. To decrease Poisson noise, we produce $15$ different realizations of the shifted chance alignment catalog, with each one shifting the declination of stars by a random amount $\mathcal{U}(-0.5,0.5)$ degrees. In $500$~pc sample, the source density does vary significantly on $\lsim0.5^{\circ}$ scales, so the shifted catalogs preserve chance alignment statistics while ensuring that all matched pairs are purely spurious.  
These realizations are combined during the KDE calculation, and the final density is divided by $15$ to reflect the number of samples. In this KDE space, a given 7-dimensional vector $\vec{x}$ will have a KDE-estimated density of chance alignments given by $\mathcal{N}_{\rm chance~alignments}(\vec{x})$ and a total number of potential binary candidates given by $\mathcal{N}_{\rm candidates}(\vec{x})$. Thereby, $\mathcal{R}$ is defined by 
\begin{equation}\label{eq:r_chance}
    \mathcal{R} = \frac{\mathcal{N}_{\rm chance~alignments}(\vec{x})}{\mathcal{N}_{\rm candidates}(\vec{x})} \ .
\end{equation}
Note that the above definition allows for $\mathcal{R}>1$, meaning that it is not strictly a probability. However, as \citet{EB21_widebin} showed, $\mathcal{R}$ tracks the true chance alignment probability closely (see also Appendix \ref{app:chance_alignment}).

Now that each of the $938,127$ pairs has a designated $\mathcal{R}$ value, we search for triples among them. Qualitatively, we identify a triple as two or more pairs that share a common star. For example, take three stars A, B, and C. If we find three pairs, A-B, A-C, and B-C, then ABC makes a triple. This is the case for most triples. In some cases, we find A-B and A-C, from which we infer the B-C pair, for example. The missing pairs have consistent proper motions and parallaxes with their companions but are missed in our search because their observed proper motion difference exceeds the $\Delta\mu_{\rm orbit}$ threshold. This excess is caused by additional orbital motion from the third star in the system, which biases the pair's apparent relative motion (see Appendix \ref{app:unmatched_pairs}).

To perform the triple search efficiently among the $\sim1$ million pairs, we use a graph method. Pairs are represented as an undirected graph, where each node corresponds to a unique {\it Gaia} source, and each edge represents a binary connection between two stars. Edges store metadata of the binary, such as physical separation and chance alignment probability, and the nodes store the {\it Gaia} parameters for a given source. We then extract all connected components of size exactly $N=3$, each of which represents a candidate resolved triple system. The search yields $37494$ candidate triple star systems, without any duplicates (i.e., one source in more than one triple). The same process can be used to identify resolved hierarchies with higher multiplicity. For example, an $N=4$ search reveals $3175$ yields quadruple candidates.

Among the triple candidates, there exist at least two pairs, each of which has a $\mathcal{R}$ statistic associated with it. We take a conservative approach and choose the largest $\mathcal{R}$ between the pairs in the triple to represent the single $\mathcal{R}_{\rm triple}$ of the entire triple. Namely, $\mathcal{R}_{\rm triple} = max\{\mathcal{R}_{ij}
\}$, where $i,j\in(1,2,3)$ are sub-pairs in the triple. Then, we only consider real triples as those with $\mathcal{R}_{\rm triple}<0.1$: i.e., less than a $10\%$ probability of being a chance alignment. In Appendix \ref{app:chance_alignment} we test whether $\mathcal{R}_{\rm triple}$ can be interpreted as a chance alignment probability for the entire triple system. By performing a triple search among the shifted source catalog, we create a sample of purely spurious triples. As we show in Figure \ref{fig:R_chance_plot}, our definition of $\mathcal{R}_{\rm triple}$ indeed closely matches the intrinsic chance alignment probability.

After applying a cut of $\mathcal{R}_{\rm triple} < 0.1$, we retain 9,767 triples, with a median chance alignment probability of 0.001. Based on the $\mathcal{R}_{\rm triple}$ distribution, we estimate that $\sim99\%$ of the remaining triples are truly gravitationally bound, indicating a high purity in our sample.
There are also $12142$ triples with $\mathcal{R}_{\rm triple}<0.5$.  
Our search also yields $314$ resolved quadruples with ${\tt R\_chance_\_align} < 0.1$. These systems are not explored further here.

For the purposes of this study, we focus on the high-purity triple sample, consisting of $9767$ confidently bound systems within the $500$~pc search volume.
Our method of filtering out chance alignments is strict and meant to create a sample with low contamination. However, our strict cuts exclude a fraction of physically bound triples as well. Weaker cuts on $\mathcal{R}_{\rm triple}$ would yield a larger number of real triples but also introduce more chance alignments (e.g., Figure \ref{fig:R_chance_plot}). For completeness, we publish the full catalog of $37494$ triple candidates with reported $\mathcal{R}_{\rm triple}$ for each system,  but we advise users of the catalog to think carefully about the effects of unrecognized chance alignments, and to use $\mathcal{R}_{\rm triple} < 0.1$ by default if they have not done so.

\begin{figure*}
    \centering
    \includegraphics[width=0.90\textwidth]{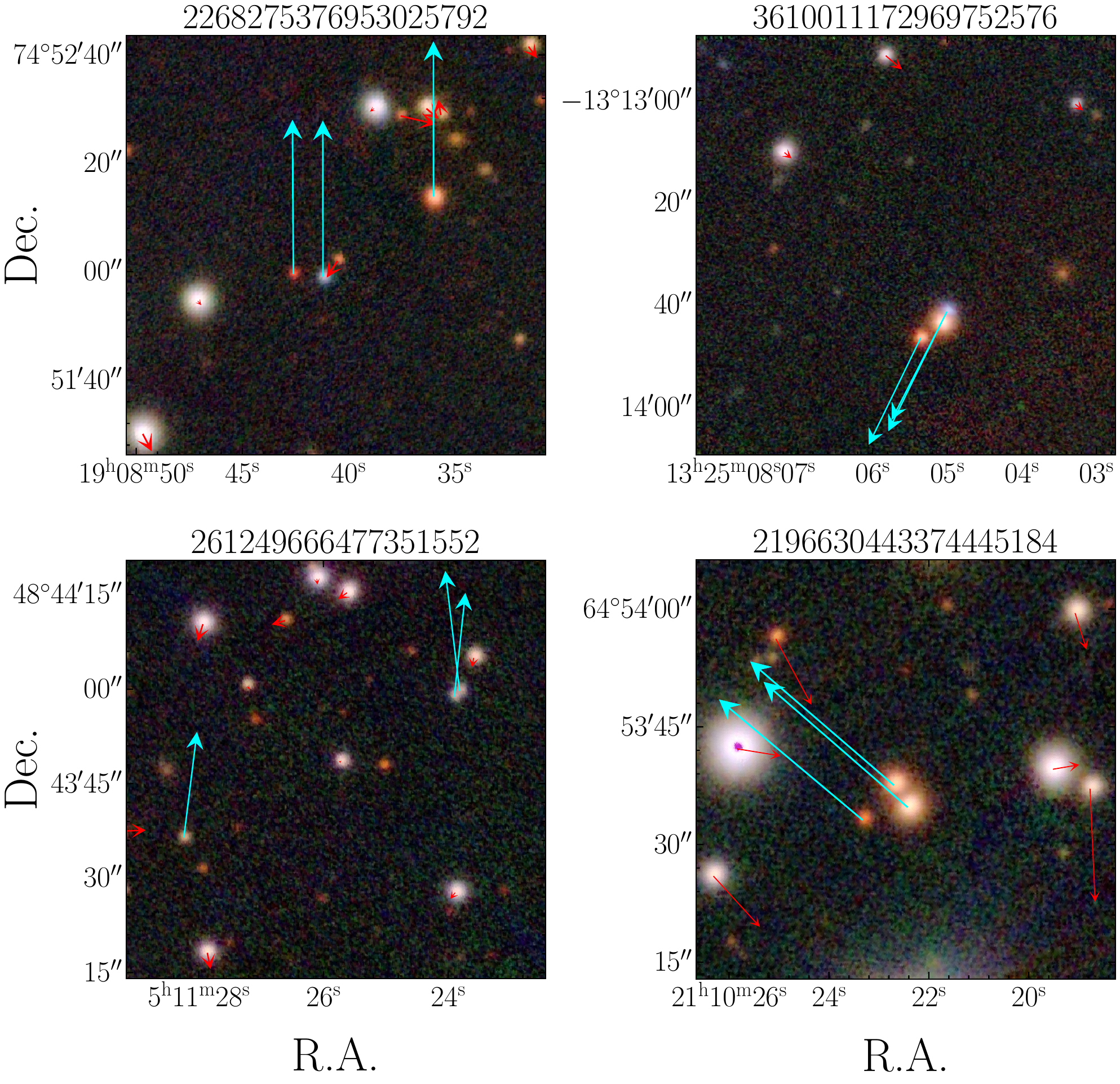}
    \caption{Images of resolved triples with proper motion vectors. The images show proper motion arrows atop PanSTARRs {\it gri} images for all {\it Gaia} sources in the image. The triple has blue proper motion vectors, while the stars unassociated with the triple have red proper motion arrows. The {\it Gaia} DR3 source id for the primary star is displayed above each image. {\bf Left:} Visually hierarchical triples where the bottom system contains three white dwarfs and the top system contains one white dwarf component in the inner binary.  {\bf Right:} Visually non-hierarchical triples, where the ratio between their projected separations is less than 2. The top system contains a white dwarf, whereas the bottom system has three main-sequence components. Such non-hierarchical systems are rare in our sample and likely result from projection effects, but we show them here for illustration purposes.}
    \label{fig:triple_images}
\end{figure*}

In Figure \ref{fig:triple_images} we show PanSTARRS {\it gri} images of example resolved triples in our sample. We show the proper motion vectors for all sources (red) and for the three stars in the triple (blue). This Figure contains examples of visually hierarchical triples in the left column. The top left contains one white dwarf in the inner binary, while the bottom left is a newly identified triple white dwarf system (described further in Section \ref{subsec:demographics}). In the right two panels, we show examples of visually non-hierarchical triples; the top right triple also contains one WD component.

\section{Properties of Resolved Triples}\label{sec:properties}
In this section, we outline basic properties of our resolved triples sample. We reference the three stars in a given triple through the following scheme. The two closer stars form the `inner binary' of the triple, and among them, the `Primary' is the brighter component (in the {\it Gaia} G band) and `Secondary' is the fainter one. Then, the third, more distant star is denoted as the `Tertiary Star.' Note that, due to projection effects, the star with the largest apparent separation is not always the true tertiary. However, such misidentifications will be exceedingly rare in a statistical sample.

\begin{figure*}
    \centering
    \includegraphics[width=0.85\textwidth]{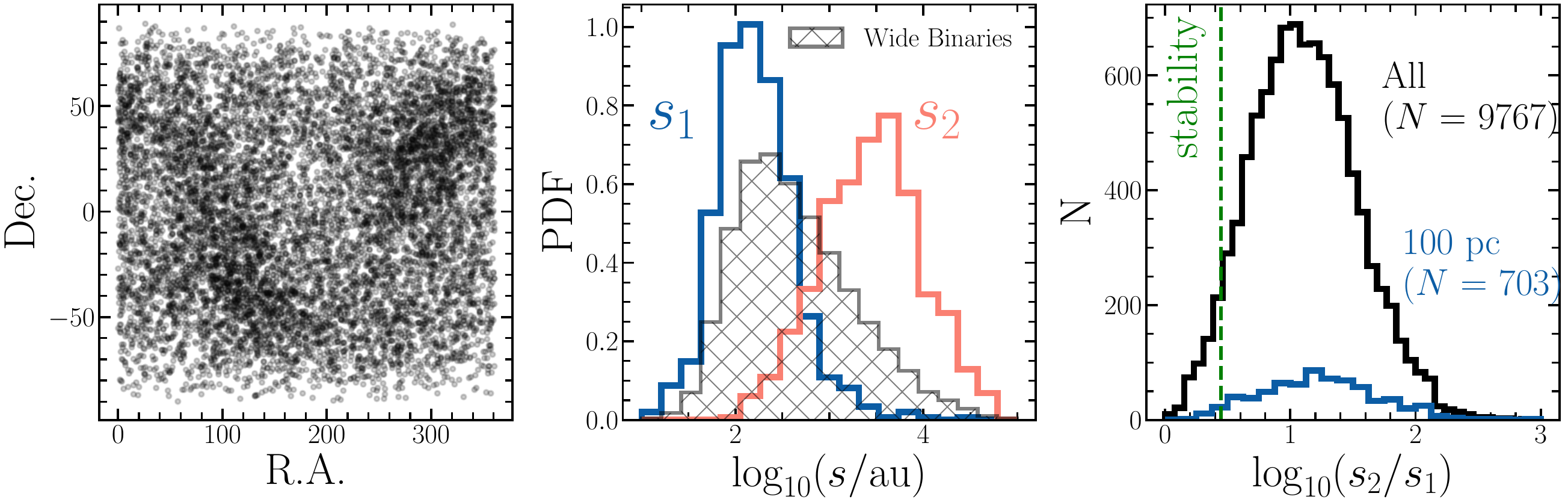}\\[1em]
    \includegraphics[width=0.87\textwidth]{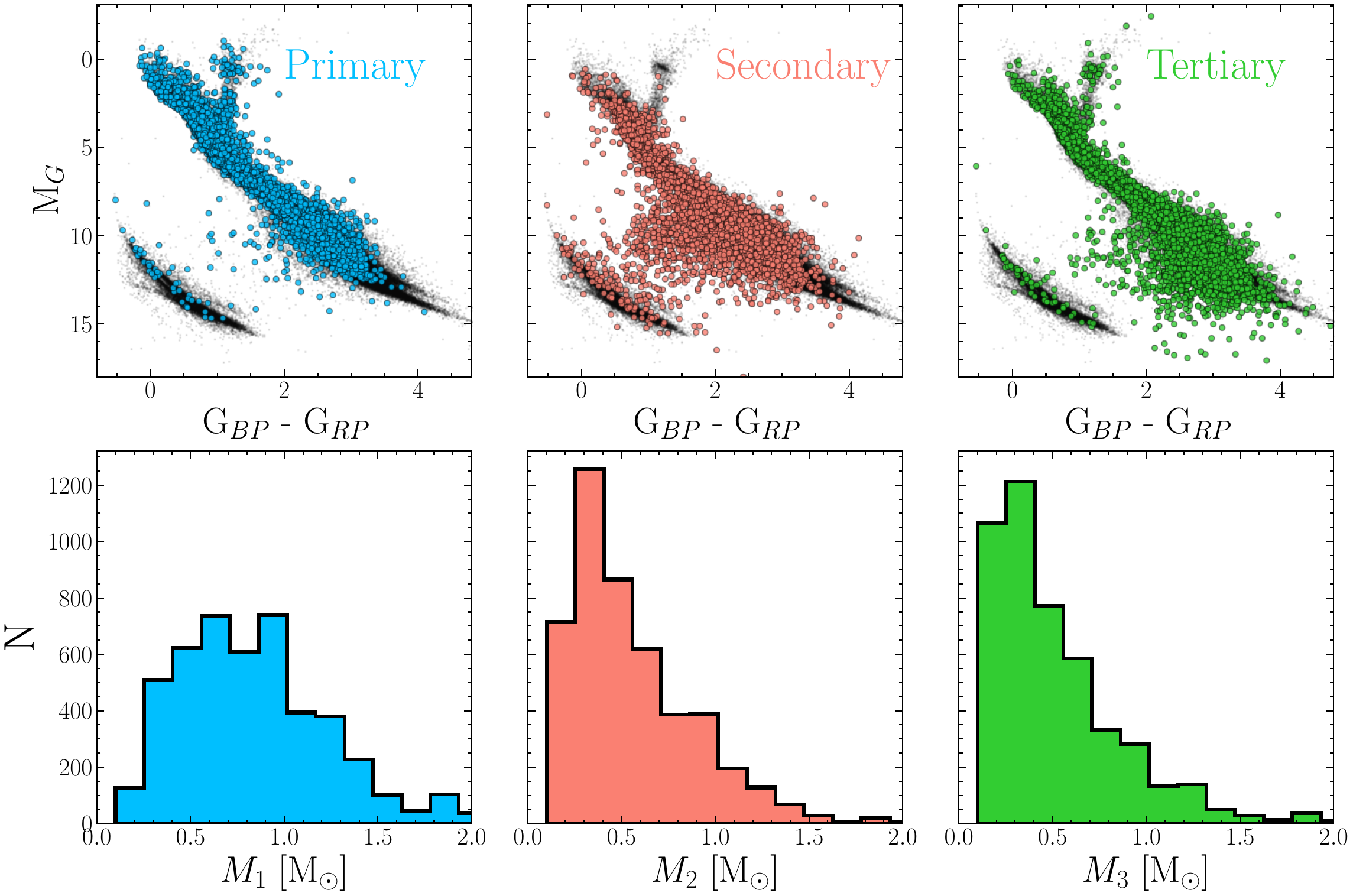}
    \caption{Basic properties of the resolved triples sample. {\bf Top:} From left to right, we show the on-sky positions of all triples, the distribution of separations for the $100$~pc sample {\bf (inner $s_1$ in blue, outer $s_2$ in red)}, and the histogram of separation ratios between the outer and inner components. In the middle column of this row, we also plot the $100$~pc wide binary separations for comparison. In the right column, we include the $100$~pc distribution in blue and show the threshold for dynamical stability ($s_2/s_1 = 2.8$) in green. Triples with $s_2/s_1 < 2.8$ are unstable in projection, though most, if not all, are due to projection effects. {\bf Middle:} HR diagrams for all three stars in the triple. Most systems between the white dwarf and main-sequence tracks are due to BP/RP color blending.
    {\bf Bottom:} Mass distributions for all main-sequence stars in triples. $M_1$ ($M_2$) is the mass of the brighter (fainter) star in the inner binary, and $M_3$ is the tertiary's mass.}
    \label{fig:summary_plots}
\end{figure*}
\subsection{Demographics of the Sample}\label{subsec:demographics}

Figure \ref{fig:summary_plots} presents various basic properties of the resolved triple sample. In the top row, from left to right, we show the on-sky positions of all triples, the distribution of separations for the $100$~pc sample, and the distribution of separation ratios between the outer and inner components.
The leftmost plot illustrates that the local triples are found all over the sky, with overdensities corresponding to the Galactic plane. The dearth of stars near (RA, Dec) = ($250^\circ$, $-25^\circ$) is an artifact of {\it Gaia}'s scanning law. The middle plot displays the projected separations of the inner and outer components for the triples within $100$~pc. In hierarchical triples, the two relevant orbital separations are the inner binary separation ($s_1$) and outer binary separation ($s_2$); the latter is between the center of mass of the inner binary and the tertiary. The blue (red) curve here plots the $s_1$ ($s_2$). For comparison, we also plot the separation distribution of $100$~pc wide binaries \citep[][]{EB21_widebin}, shown by the gray histogram. Choosing the $100$~pc subsample minimizes selection biases while retaining a sizable sample ($\sim$700 triples), such that the displayed distribution roughly reflects the intrinsic triple distribution for ($s\gsim100$~au).

The inner separation distribution has a mean of $\overline{\log{s_1}} = 2.4$ ($234$~au) and standard deviation $\sigma_{\log{s_1}} = 0.4$ dex. The outer separation distribution has $\overline{\log{s_2}} = 3.6$ ($4,000$~au) and standard deviation $\sigma_{\log{s_2}} = 0.5$ dex. Note that, at least for the inner binary, the mean separation and sigma are mostly set by the {\it Gaia} angular resolution, not the intrinsic distributions. 
For a detailed discussion on the origin of triple separations, see Section \ref{subsec:triple_separations}.

The top right plot in Figure \ref{fig:summary_plots} shows the ratio of outer to inner projected separations ($s_2/s_1$) in triples. The black curve represents all triples in our $500$ pc sample, and the blue curve shows the $100$~pc sample only. In the latter volume-limited sample, the separation ratio has a mean of $\overline{\log(s_2/s_1)} = 1.2$, corresponding to $s_2/s_1 \approx 15$. Here we also draw a line marking $s_2/s_1 = 2.8$ to demonstrate that some {\it visually} unstable triples exist. As we will later show in Section \ref{subsec:triple_separations}, all of these systems are consistent with being intrinsically stable ($a_2/a_1 > 2.8$) but have small $s_2/s_1$ when projected in the plane of the sky.

In the middle row of Figure \ref{fig:summary_plots}, we show the {\it Gaia} color-magnitude diagrams for all triple components. The primary (secondary) star is defined as the brighter (fainter) source in the inner binary, and the tertiary is the more distant third object. Most stars lie on the main sequence, with a few hundred falling on the white dwarf (WD) or red giant (RG) tracks. Some sources fall between the white dwarf and main sequence tracks, often because their photometry is blended with that of their close companions. However, a subset of the sources between the WD and MS tracks have well-measured photometry and are separated from their companion ($\theta>5''$), making them potential candidates for unresolved WDMS or higher order multiples. Note that sources without a measured ${\tt bp\_rp}$ color are not included in these plots. 

In the bottom row of Figure \ref{fig:summary_plots}, we plot the mass distributions for stars on the main-sequence (MS). The masses correspond to the primary ($M_1$), secondary ($M_2$), and tertiary ($M_3$) stars as defined above. Masses are derived using the absolute G magnitude, $M_G$. From the empirical table of \citet{Pecaut13}\footnote{\url{https://www.pas.rochester.edu/~emamajek/EEM_dwarf_UBVIJHK_colors_Teff.txt}}, we create a linearly-interpolated grid of masses and $M_G$, and for each of the MS stars in our triple sample, we map their $M_G$ to a stellar mass. While this method is imperfect, it generally provides a reasonable estimate within $0.1$~M$_\odot$. Note that for unresolved pairs, this calculated mass will be overestimated. 

Primary masses are weighted towards larger masses than a simple Kroupa IMF since the triple fraction increases with primary mass \citep[e.g.,][]{Offner23}. As we demonstrate in later sections, this distribution can be reproduced by convolving the Kroupa IMF with the triple fraction (Section \ref{subsec:triple_masses}). The largest $M_1$ in our triple sample is $5.1~{\rm M_\odot}$, and among all triples, the median $M_1$ is $0.8~{\rm M_\odot}$. 

All triples in our sample have measured G magnitudes by the initial {\it Gaia} query, but $39$, $4320$, and $101$ of the primaries, secondaries, and tertiaries, respectively, do not have a {\tt bp\_rp} color. For these systems, we still provide a mass estimate in the public catalog using their $M_G$ measurements, assuming that most would lie on the main-sequence (MS). While the majority will indeed be MS stars, we caution that these estimates will be unreliable for white dwarfs and giants since {\it Gaia} photometry alone cannot confirm their MS status. Other photometric surveys (e.g., PanSTARRs) could help disentangle these MS status of these sources. All mass-related analyses in this paper consider only MSMS-MS triples, where mass measurements are reliable for all three stars. Requiring {\tt bp\_rp} colors for all components excludes nearly half the sample. However, this has minimal impact on our conclusions because our choice of the {\it Gaia} resolvability criteria accounts for our requirement of a reported {\tt bp\_rp} \citep[][]{EB_review}. Including all sources in our sample and using the looser resolvability criteria \citep[`no cuts' from][]{EB_review} does not change our results.

Following \citet{EB18} and \citet{EB21_widebin}, we categorize our triples into broad stellar types based on their CMD position. WDs are defined as objects with $M_{\rm G} > 3.25(G_{\rm BP}-G_{\rm RP}) + 9.625$ while anything above this line on the CMD is classified as `MS'. Note that the `MS' cut is only meant to separate WDs with non-WDs, so those classified as 'MS' include unresolved binaries/multiples, red giants, and brown dwarfs. Red giants can be selected easily using additional CMD cuts, but we do not apply them here to avoid having too many categories of triples. Moreover, some sources near the boundary of this cutoff could be WDs or unresolved WDWD binaries. For sources without a measured {\tt bp\_rp} color, we denote the stellar type by `??'. We define the triple type as the combination of the three stars' types, where, for example, a `WDMS-MS' triple has a WDMS inner binary with an MS tertiary. Note that the type of the inner binary does not necessarily list the brighter component first. In the published catalog, we include the individual star types along with the triple types. We show the various categories of triple stellar types in Table \ref{tab:triple_types}.

From Table \ref{tab:triple_types} we find that our sample contains an array of different triples, with both `WD' and `MS' components. About half of the triples have three main-sequence components, and $216$ ($2.5\%$) have at least one WD component. Nearly half of the triples have one component without a measured ${\tt bp\_rp}$ color in {\it Gaia}, mostly due to nearby companions that obstruct precise ${\tt bp\_rp}$ photometry measurements. However, the stellar types for most of these sources can be distinguished using other photometric surveys, such as Pan-STARRs, which we do not pursue in this work. 

\subsubsection{Triple White Dwarfs}

White dwarf triples offer an opportunity to study the outcomes of long-term triple dynamics with mass loss and stellar evolution, some of which may even be progenitors of double degenerate Type Ia supernovae \citep[e.g.,][]{Katz12, Toonen18, Shariat23}.
Two of the high-confidence ($\mathcal{R_{\rm triple}}<0.1$) sources are triple white dwarfs. The primary stars in these three systems are Gaia DR3 261249666477351552 and Gaia DR3 4190499986125543168.

The first system is a newly identified resolved WD triple where all three stars lie on the bottom of the WD branch on the {\it Gaia} CMD. The inner binary projected separation is $170$~au, and the tertiary is $6500$~au away from the inner binary. The components are Gaia DR3 261249666477351552,	261249670772776832, and	261249636413169920\footnote{\citet{Heintz22} previously identified this system as a candidate triple but excluded it from their final sample due to one component being classified as spurious. In contrast, our analysis confirms that all three components are physically associated, with a chance alignment probability of $4.7\times10^{-4}$.}. A Pan-STARRS image of this system with displayed proper motion vectors is presented in the bottom left of Figure \ref{fig:triple_images}. 

The other resolved triple WD in our sample (with primary component 4190499986125543168) was previously identified with {\it Gaia} DR2 \citep{Perpinya19_resolvedtripWD}. This system contains three WDs, each with mass $\approx0.6~{\rm M_\odot}$ and inner (outer) separation of $303$ ($6400$) au \citep{Perpinya19_resolvedtripWD}. The inner (outer) separation we derive from {\it Gaia} DR3 is $301$ au ($6102$ au), which is not too different from the DR2 measurements. 

Other WD triples with smaller angular separations likely exist in our catalog but lack ${\tt bp\_rp}$ colors due to nearby companions. These systems could potentially be identified by cross-matching with other photometric surveys. Unresolved double white dwarfs also likely exist among the resolved triples, as hinted by systems directly above the WD track (Figure \ref{fig:summary_plots}).

\begin{deluxetable}{lccc}
\tablecaption{Classification of Stellar Triples\label{tab:triple_types}}
\tablehead{
\colhead{Classification} &
\colhead{N$_{\rm \mathcal{R}<0.1}$} &
\colhead{N$_{\rm \mathcal{R}<0.5}$} &
\colhead{N$_{\rm All~Candidates}$}
}
\startdata
MSMS-MS               & 4704                       & 6065                       & 24279                     \\
MS??-MS               & 4696                       & 5617                       & 10778                     \\
WDMS-MS               & 137                        & 181                        & 793                       \\
MS??-??               & 79                         & 84                         & 246                       \\
MSMS-??               & 44                         & 60                         & 601                       \\
MSMS-WD               & 30                         & 44                         & 451                       \\
MS??-WD               & 22                         & 28                         & 125                       \\
????-MS               & 19                         & 20                         & 60                        \\
WD??-MS               & 15                         & 17                         & 38                        \\
WDWD-MS               & 15                         & 19                         & 52                        \\
WDMS-WD               & 3                          & 3                          & 33                        \\
WDWD-WD               & 2                          & 2                          & 11                        \\
????-??               & 1                          & 2                          & 4                         \\
\tableline
Total                 & 9767                       & 12142                      & 37471                    
\enddata
\vspace{1mm}
{\bf Note:} Each star in the triple is classified as `MS' or `WD' depending on whether they fall above or below the cutoff of $M_{\rm G} = 3.25(G_{\rm BP}-G_{\rm RP}) + 9.625$. `??' indicates that the source has no measured color in {\it Gaia} DR3. The first set of classification denotes the stellar type of the inner binary, and the text after `-' is the tertiary stellar type. $\mathcal{R}$ represents the ${\tt R\_chance_\_align}$ parameter.
\end{deluxetable}

\begin{figure*}
    \centering
    \includegraphics[width=0.98
    \textwidth]{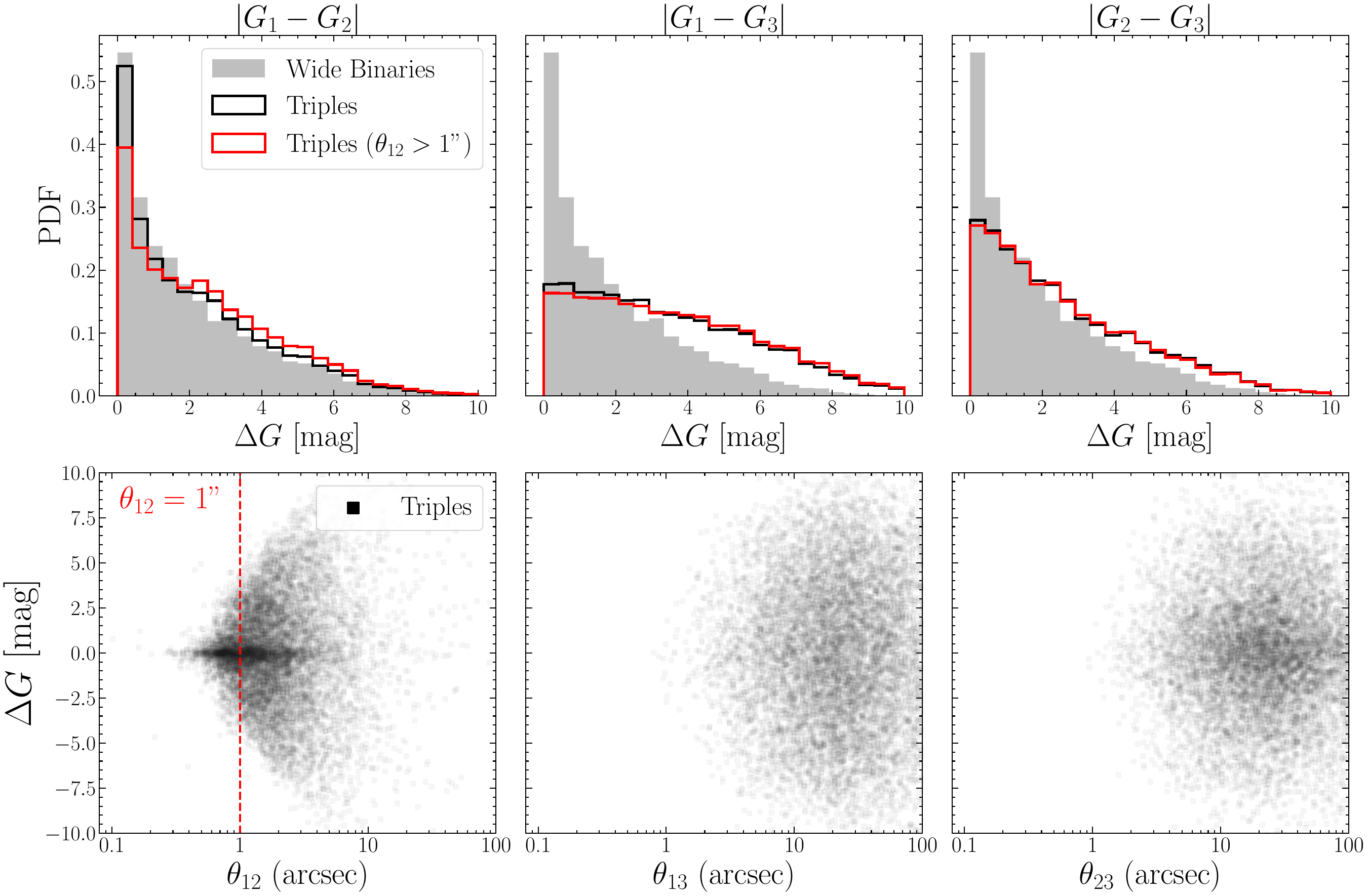}
    \caption{Apparent twin excess in triple inner binaries. We show the difference in apparent G magnitudes between components in resolved triples. From left to right, the columns show the $\Delta G$ between the inner two stars, inner primary (1) and the tertiary (3), and inner secondary (2) and the tertiary. {\bf Top:} The distribution of $\Delta G$ between triple components for all resolved triples (black) and those where the angular separation of the inner binary is greater than one arcsecond (red). The gray shaded curve shows the distribution for all wide binaries, where a twin excess has been established. {\bf Bottom:} The difference in G magnitudes as a function of the angular separation for the components in the triples. The bottom left plot filters only for triples where the angular separation is greater than one arcsecond, while the other two plots have no restriction on $\theta$. Triple inner binaries show a striking and statistically significant excess of equal-mass twins.
     } \label{fig:deltaG_all} 
\end{figure*}

\subsection{Twin Excess in Wide Triples}\label{subsec:twins}
An intriguing aspect of binary mass ratios is the observed excess of near-equal-mass ($0.95<q<1$) `twin' binaries. Early studies suggest that the fraction of twins is strong in populations of close binaries \citep[$a\lsim0.2$~au;][]{Lucy97,Tokovinin00,Kounkel19} and tends to decrease with binary separation \citep[e.g.,][]{Tokovinin14b,Moe17}. \citet{EB_19_twins} show that a statistically robust twin excess is observed at separations of $\sim1000$+~au.

The physical origin of this twin excess remains debated. 
For binaries that form closer than $s\sim200$~au, accretion from a shared gas reservoir among protostars tends to cause an excess of mass ratios near unity \citep[e.g.,][]{Kroupa95a,Kroupa95b,Bate97,Bate00,Tokovinin00,White01,Marks11,Ochi05,Young15}. This `competitive' accretion model applies to binaries that co-evolve in a shared gas reservoir during their early stages, such as a circumbinary disk with typical length scales less than $\sim100$~au. The observed presence of a twin excess at $s\gsim1000$~au \citep[][]{EB_19_twins}, therefore, might suggest that these systems formed at close or intermediate separations and have since dynamically widened to their present-day separations \citep[e.g.,][]{Kouwenhoven10,EB_19_twins,Hwang22_twinecc}. Turbulent fragmentation in molecular clouds may also preferentially produce equal-mass binaries \citep{Tobin16, Guszejnov17}, which can generally operate on $>1000$~au scales. 

As previous studies have shown, the presence of a twin excess offers a valuable probe of early star formation processes. Compared to binaries, the formation of triple systems is less well studied and involves additional complexity due to the increased multiplicity and dynamical evolution. Leveraging our large sample of resolved triples, we investigate whether a twin excess is present among any pair of components in our triple sample, how it depends on separation, and what it reveals about the formation mechanisms of multiple star systems.

Since all stars in a given triple system have roughly the same distance and interstellar dust absorption, near-equal mass main-sequence stars (i.e., twins) will exhibit nearly equal apparent $G$ magnitudes. Therefore, a twin pair can be identified by systems with $|G_1 - G_2| \approx 0$.
In Figure \ref{fig:deltaG_all}, we plot the difference in the apparent G magnitudes ($\Delta G$) between all of the pairs in the triple. This includes the apparent brightness difference between the two stars in the inner binary ($G_1 - G_2$, left column), the inner primary and the tertiary ($G_1 - G_3$, middle column), and the inner secondary with the tertiary ($G_2 - G_3$, right column). Here `1' and `2' refer to the brighter and fainter star in the inner binary, while `3' refers to the tertiary star.

The top panel of Figure \ref{fig:deltaG_all} shows the distribution of $\Delta G$ in triples (black curve) and compares them to wide binaries (gray distribution). The wide binaries plotted here are a subset of the $500$~pc wide binary sample that have the same separations and distances as the triple inner binaries in our sample (Appendix \ref{app:WB_control}). With shared separations and distances, both the binary control sample and the triple inner binaries are subject to similar systematic biases. The red histogram here shows the triple inner binaries with angular separations ($\theta_{12}$) greater than $1''$. At separations below $1''$, the ability to resolve a wide binary is extremely sensitive to the flux ratio and tends to bias towards equal-flux pairs \citep[e.g.,][]{EB18}. Above $1''$, this bias is significantly weaker \citep[but still non-negligible;][]{EB_review}, so any twin excess is more likely intrinsic to the population instead of a systematic bias. 

Among the three components in the triples, there is a strong twin excess among inner binaries. The leftmost column of Figure \ref{fig:deltaG_all} shows that twin excess is present out to $\theta \sim 5''$, where {\it Gaia} is sensitive almost all pairs with $\Delta G <8''$ \citep[][]{EB_review}. Such a twin excess is not observed between the tertiary and any of the inner binary components. In the $100$ pc resolved triple sample, a similar twin excess is observed \citep{Tokovinin22_resolvedtriples}, as is the case in wide binaries at similar separations \citep[][]{EB_19_twins, EB21_widebin}.

While the twin phenomena is already suggested by Figure \ref{fig:deltaG_all} out to large angular separations, the degree to which it extends is unclear. We therefore explore how the twin fraction evolves as a function of both the angular and physical separation of inner binaries. 

\begin{figure}
    \centering
    \includegraphics[width=0.99
    \columnwidth]{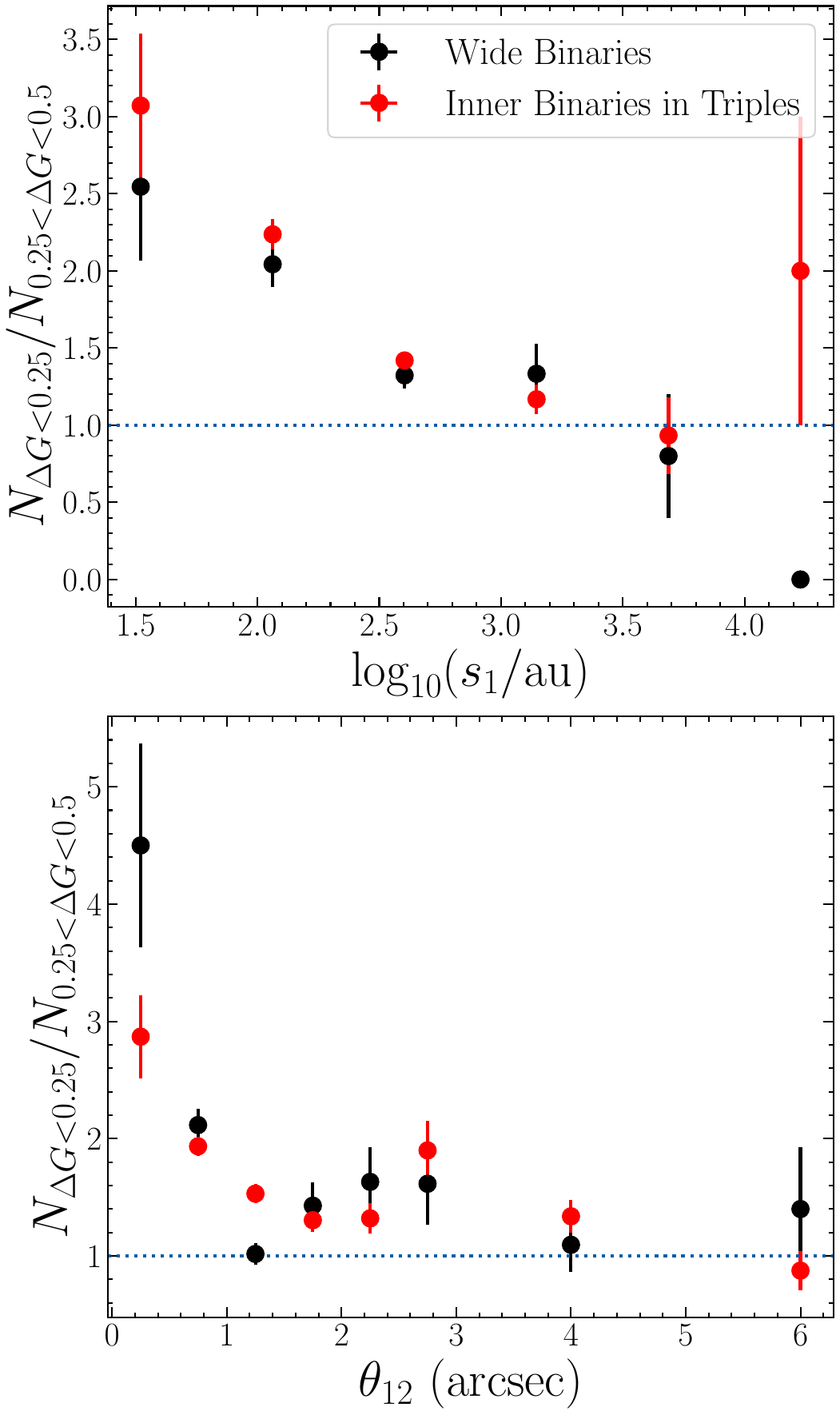}
    \caption{Twin properties of triple inner binaries compared to isolated wide binaries. The twin fraction of inner binaries in triples is displayed as a function of their physical separation (top) and angular separation (bottom). In red, we show the fraction of triples with $\Delta G < 0.25$ mag (twins) to those with $0.25 < \Delta G < 0.50$. In black, we show the same plots for the wide binaries at the same separations and distances, which have a well-established excess of twins \citep[][]{EB_19_twins}. 
 }\label{fig:twinfraction_sep_theta} 
\end{figure}

Following \citet{EB_19_twins}, we calculate a proxy for the twin fraction, defined as the ratio between the number of pairs with $\Delta G < 0.25$ mag (e.g., `twins') and inner binaries with similar magnitudes that are not quite twins ($0.25 < \Delta G < 0.50$). We plot this `twin fraction' in Figure \ref{fig:twinfraction_sep_theta}, as a function of both physical separation (top) and angular separation (bottom).
We compare the inner binaries of triples to our control sample of wide binaries (black), which have similar separation and distance distributions as the triple inner binaries (Appendix \ref{app:WB_control}). The upper panel divides our sample into bins of separations that are equally logarithmically spaced; the blue dotted line represents no twin excess ($N_{\Delta G < 0.25} / N_{0.25 < \Delta G  < 0.50} = 1$). The plot reveals a noticeable excess of twin inner binaries in triples that is comparable to the twin excess in wide binaries. For triples, the twin excess extends out to angular separations of $\theta\sim4''$, where observations are sensitive to a wide range of flux ratios. Furthermore, the twin phenomenon is present for inner binaries as wide as $\sim1000+$~au (top of Figure \ref{fig:twinfraction_sep_theta}), consistent with the wide binary population \citep{EB_19_twins}. 
Overall, the twin fraction evolves with separation in a manner that is strikingly similar to wide binaries at the same separation. This strong similarity suggests that triple inner binaries and isolated wide binaries likely share a common formation pathway.

One possible explanation for the binary twin excess at large separations ($\gsim100$~au) is that they originally formed at smaller separations ($s\lsim100$~au) through accretion in circumbinary disks and were subsequently widened by early dynamical interactions with other stars \citep{EB_19_twins, Hwang22_twinecc}. The fact that inner binaries in triples also exhibit a wide twin excess raises tension with this scenario.
Specifically, it is challenging for an interloping star to widen the inner binary without simultaneously disrupting the tertiary companion or destabilizing the entire triple. 

This tension suggests a few possible alternatives. One is that many wide binaries initially formed with tertiaries and indeed widened through this mechanism, but most wide tertiaries were unbound during this dynamical encounter, leaving behind only the surviving subset of stable triples we observe today. If so, the relative abundance of wide binaries and resolved triples could offer valuable constraints on the efficiency of dynamical processing in young clusters. Alternatively, some tertiaries may have formed later or been dynamically captured after the inner binary had already widened. A future study comparing wide binaries and resolved triples could offer new constraints on dynamical scattering and twin formation.

\subsection{Mutual Inclinations}\label{subsec:mutual_inclinations}

\begin{figure*}
    \centering
    \includegraphics[width=0.97
    \textwidth]{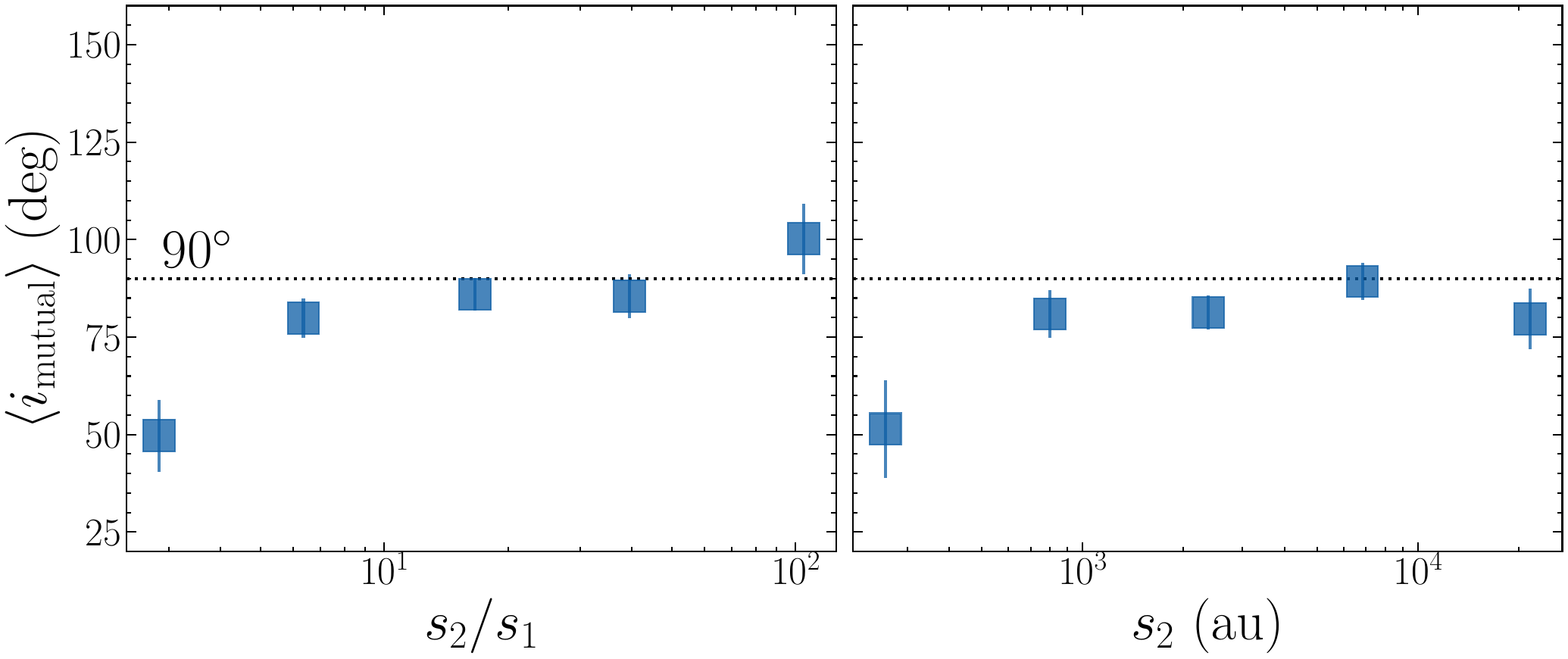}
    \caption{Mean mutual inclinations of triples as a function of separation ratio (left) and tertiary separation (right). Each point shows the inferred mean mutual inclination, $\langle i_{\rm mut} \rangle$ (Equation \ref{eq:mut_incl}), for triples in the given bin with binomial errors included. Only triples interior to 125 pc are included ($N\sim1200$), and the dotted line at $90^\circ$ indicates the expected value for randomly oriented inner and outer orbits. Wide triples are consistent with isotropic inclinations while more compact triples tend towards mutual orbital alignment.
 }\label{fig:mutual_inclinations} 
\end{figure*} 

The mutual inclination ($i_{\rm mut}$) in hierarchical triples represents the angle between the angular momentum vectors of the inner and outer orbits. For example, a mutual inclination of $0^\circ$ represents mutually aligned orbits (all in the same plane) while $i_{\rm mut}=90^\circ$ corresponds to a tertiary orbiting perpendicular to the inner binary's orbital plane.
Constraining $i_{\rm mut}$ is critical for understanding the dynamical evolution of triples; the efficiency of EKL, for example, depends sensitively on mutual inclination \citep[e.g.,][]{Naoz2016}. For resolved triples, the long orbital periods make it difficult to constrain individual inclinations. However, the average mutual inclination of a statistical triple population can be inferred using only astrometry. 

From the precise positions and parallaxes reported by {\it Gaia}, we calculate the direction of relative motion between the tertiary and inner binary. By studying the direction of motion of the tertiary relative to the inner binary (the ``relative motion sense''), we determine whether the tertiary exhibits a prograde or retrograde orbit with respect to the inner binary. The relative motion sense can therefore serve as a statistical proxy for mutual orbital alignment \citep{Sterzik2002, Tokovinin22_resolvedtriples}. 
For each subsystem, we define an angle $\gamma$ between the position and velocity vectors, which characterizes the direction of motion relative to the projected line joining the stars. The position vector for the outer orbit connects the tertiary to the center of mass of the inner binary.
This angle $\gamma$ lies in the range $[-180^\circ, +180^\circ]$, where positive values indicate counterclockwise (direct) motion and negative values indicate clockwise (retrograde) motion on the sky.

We assign to each subsystem a ``relative motion sense'' $\mathcal{S}$, which is the sign of the two-dimensional cross product between the position and velocity vectors. This yields $\mathcal{S} = +1$ for counterclockwise motion and $\mathcal{S} = -1$ for clockwise motion. Comparing the signs of $\mathcal{S}_{\mathrm{in}}$ and $\mathcal{S}_{\mathrm{out}}$ across the sample allows us to quantify whether the inner and outer orbits tend to rotate in the same (co-rotating) or opposite (counter-rotating) directions, providing a statistical constraint on mutual orbital alignment.
For each triple, we then compare the signs of the inner and outer subsystems. If $\mathcal{S}_{\mathrm{in}} = \mathcal{S}_{\mathrm{out}}$, the inner and outer orbits revolve in the same sense; otherwise, they have the opposite sense.

To quantify the overall tendency towards alignment across our sample, we follow \citet{Tokovinin22_resolvedtriples} in defining the sign correlation ($C$) as 
\begin{equation}
C = \frac{N_+ - N_-}{N_+ + N_-} \ ,
\end{equation}
where $N_+$ is the number of triples with the same motion sense in both subsystems, and $N_-$ is the number with opposite senses. This statistic can take on values between $+1$ (perfect alignment) and $-1$ (perfect anti-alignment), where $C=0$ indicates no preferred orientation. We assume errors in $C$ are binomial. Finally, the mean mutual inclination $\langle i_{\rm mut} \rangle$ can be estimated from \citep{Worley67}
\begin{equation}\label{eq:mut_incl}
    \langle i_{\rm mut} \rangle = 90^\circ (1 - C) .
\end{equation}
This relation assumes random orientations, which is reasonable for resolved triples.  Even if observational biases skew this assumption, such biases are symmetric with respect to rotation direction, so the sign correlation, $C$, remains an operational diagnostic of relative orbital alignment \citep{Sterzik2002, Tokovinin22_resolvedtriples}. While contrived configurations, such as exactly half prograde and half retrograde coplanar systems, can also yield $\langle i_{\rm mut} \rangle = 90^\circ$, such scenarios are not physically motivated.

In Figure~\ref{fig:mutual_inclinations}, we show the mean mutual inclination, $\langle i_{\rm mut} \rangle$, for triples within $125$~pc as a function of the separation ratio $s_2/s_1$ (left panel) and outer separation $s_2$ (right panel). The $125$~pc sample includes $\sim$1,200 triples, ensuring a sufficient number of systems in each bin (with the smallest bin containing $\sim$50 triples). We restrict to $125$~pc because at larger distances, proper motion errors can become comparable to or larger than the orbital velocities of wide tertiaries, making such inclination dominated by noise. Within $125$~pc, almost all tertiaries ($99.5\%$) have proper motion errors smaller than orbital motions, with the median proper motion error being $29$ times smaller than the outer orbital velocity.

In the left panel of Figure~\ref{fig:mutual_inclinations}, the triples are divided into five equally log-spaced bins in $s_2/s_1$, ranging from $1$ to $200$. 
The plot reveals that more compact systems (lower $s_2/s_1\lsim 10$) tend toward alignment, while wider triples ($s_2/s_1 \gtrsim 10$) show mutual inclinations consistent with random orientations. A similar pattern is observed in the right panel of Figure~\ref{fig:mutual_inclinations}: systems with smaller $s_2$ tend to be more aligned, whereas those with wider tertiaries are consistent with isotropy. At the smallest $s_2$ ($\lsim1$~au), tight stellar triples show evidence for mutual orbital alignment \citep[e.g.,][]{Bashi24}, which is a natural extension of our data.

\section{Results}\label{sec:results}
One of the largest uncertainties in modeling triple star populations is the initial conditions and intrinsic distributions. Generating a realistic synthetic triple population requires assumptions about the initial masses, separations, and eccentricities, all of which are poorly constrained in triples. In this section, we use our volume-limited sample of resolved triples to constrain the intrinsic distribution of masses, mass ratios, orbital periods, and eccentricities of triple star systems. We then use our findings to develop an observationally-grounded prescription for sampling triple parameters. 

\subsection{ Masses}\label{subsec:triple_masses}

\begin{figure*}
    \centering
    \includegraphics[width=0.95\textwidth]{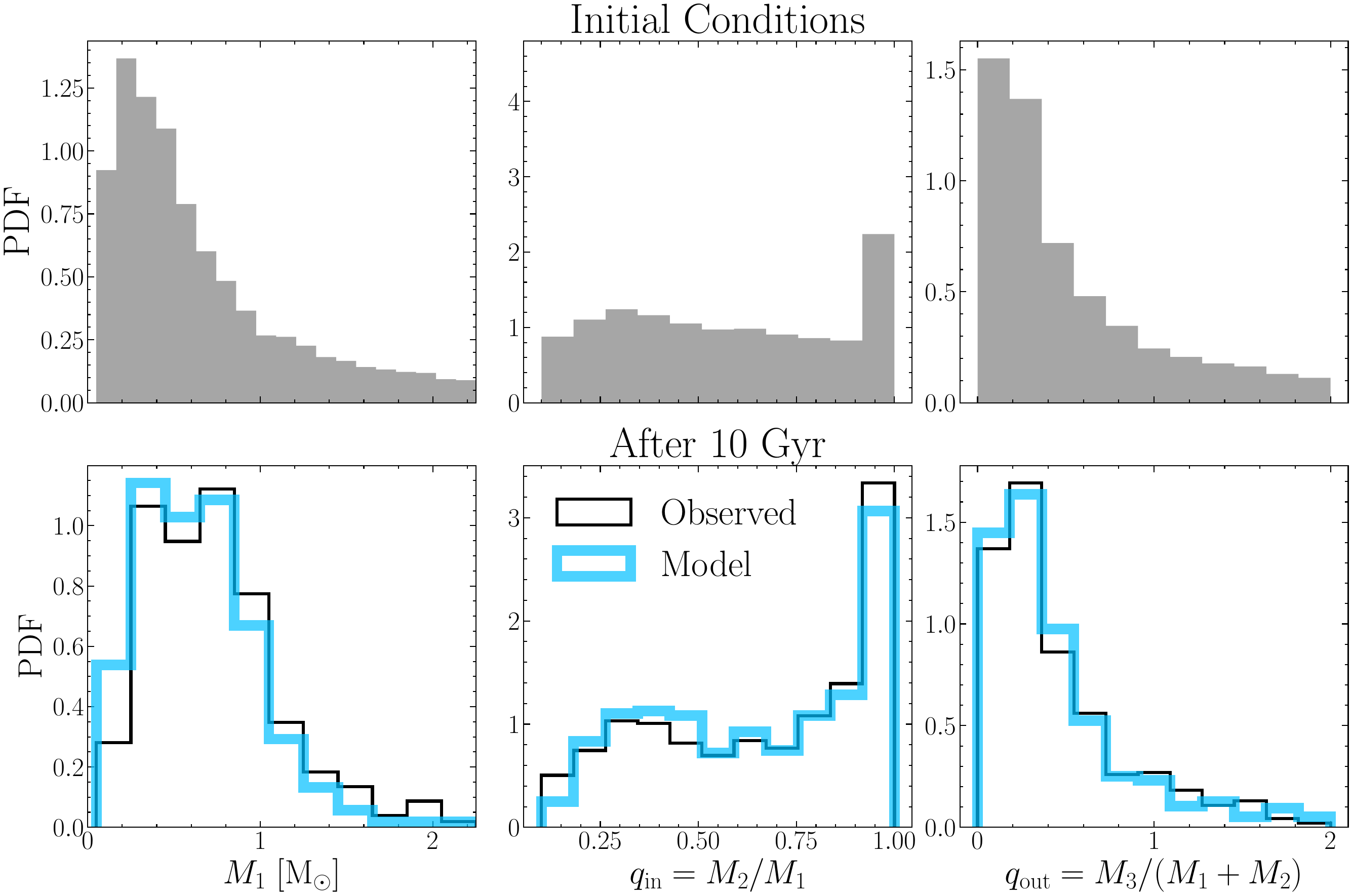}
    \caption{ The origin of triple masses. {\it Top:} The underlying distributions for initial triple primary masses ($M_1$), inner binary mass ratios ($q_{\rm in}$), and outer binary mass ratios ($q_{\rm out}$). {\it Bottom:} Assuming this intrinsic triple model, we show the results of $10$~Gyr evolution with constant star formation (blue). We subject these models to the {\it Gaia} selection function, keeping only those that are resolved and bright enough for detection at $t=10$~Gyr. The modeled population is compared to observed resolved triples within $100$~pc sample (black).
    The modeled population is broadly consistent with observed triples. }
    \label{fig:sampling_masses}
\end{figure*}

Our goal is to develop a physically motivated prescription for sampling masses of stars in triples in a manner that reproduces observations.
Starting from a main-sequence population with a Kroupa mass distribution, we sample binaries, then select a subset of these binaries to be triples. 

As \citet{Moe17} demonstrate, the distributions of primary
masses $M_1$, mass ratios $q$, orbital periods $P$, eccentricities $e$, and multiplicity fractions are covariant. Therefore, to ensure a widely applicable and physical prescription, we jointly sample the binary parameters using the multi-dimensional distributions of \citet{Moe17}.
We execute this using the ${\tt COSMIC}$ code's ${\tt multi\_dim}$ sampler \citep{COSMIC} with some modifications to make it more realistic. Firstly, we modify the ${\tt multi
\_dim.py}$ module in ${\tt COSMIC}$ to sample primary masses from a Kroupa initial mass function (IMF) instead of the default open-source version, which assumes a more top-heavy primary mass function. 
We also modify the code to incorporate updated M-dwarf multiplicity statistics from \citet{Winters19}, replacing the default assumption of zero binary fraction at $0.08~{\rm M_\odot}$\footnote{
The modified ${\tt multi\_dim.py}$ is provided on GitHub \url{https://github.com/cheyanneshariat/gaia_triples/tree/main/Data}.}, providing a population of both singles and binaries.
The binary population has primary masses ($M_1$), mass ratios ($q_{\rm in}=M_2/M_1$), orbital periods ($P_{\rm in}$), and eccentricities ($e_{\rm in}$) consistent with the observed distributions and binary fractions \citep{Moe17}. Note that binaries sampled through this scheme will have an excess of short-period ($P_{\rm in}<20$~days) circular binaries due to tidal evolution after their formation \citet{Moe17}. Therefore, when using this prescription to sample {\it initial} conditions for a triple population (close to formation), we restrict $P_{\rm in}>20$~days.

We generate triples by assigning an outer tertiary to a subset of the binary population, where the triple probability is a function of the binary's primary mass $M_1$. Triple fractions are adopted from \citet{Offner23}, where we use the fraction of triple and higher-order multiples among all multiple systems. In the terminology of \citet{Offner23}, this fraction is THF/MF, where THF is the triple/high-order fraction and MF is the multiplicity fraction (fraction of primaries with at least one companion).
For example, $\sim10\%$, $\sim25\%$, and $\sim60\%$ of binaries with primary masses $0.1~{\rm M_\odot},1~{\rm M_\odot}$, and $10~{\rm M_\odot}$, respectively, have a tertiary companion. For each triple, we sample the outer mass ratio $q_{\rm out}$ from a power-law distribution with slope $\gamma = -1.4$ and compute the tertiary mass $M_3 = q_{\rm out} \times (M_1 + M_2)$. We also explore alternative prescriptions, such as randomly drawing $M_3$ from the IMF, but find such models are inconsistent with the observed data, suggesting the tertiary mass is not entirely independent of the inner binary. Since the inner binary orbital period ($P_{\rm in}$) is already determined, we only sample the outer binary period ($P_{\rm out}$) from the log-normal distribution of binary stars \citep{DM91,Raghavan2010,Winters19}, keeping only tertiaries that are dynamically stable (see next section).
The top panel of Figure \ref{fig:sampling_masses} shows the intrinsic distributions of $M_1$, $q_{\rm in}$, and $q_{\rm out}$ at the zero-age main-sequence for triples, according to our model.

We now aim to test this model against observed main-sequence triples. After drawing initial conditions with the above prescription, we only retain triples that (a) have all three stars on the main sequence after $10$~Gyr of constant star formation, (b) would be entirely resolved by {\it Gaia} (Appendix~\ref{app:ang_res_criteria}), and (c) have all three components brighter than the {\it Gaia} detection threshold ($G \lesssim 20.5$; \citealt{Gaia_Collab21b}). For systems within $100$ pc, which we will compare this model against, the latter brightness cut corresponds to a minimum stellar mass of $\sim0.1~{\rm M_\odot}$. The main-sequence lifetime is determined by interpolating MIST isochrones \citep{Choi16}. 

The bottom panel of Figure \ref{fig:sampling_masses} plots the resulting masses and mass ratios from our mock triple population (blue) and compares them to observed distributions in our $100$~pc resolved triples sample (black). Both the masses and mass ratios in the mock population are broadly consistent with observed {\it Gaia} triples. In fact, the observed $M_1$ distribution, which deviates significantly from any canonical 
single-star IMF, is recovered. This supports the notion that the primary mass ($M_1$) in triples originates from a Kroupa IMF convolved with the triple fraction. The inner binary mass ratios ($q_{\rm in}$) are also consistent, advocating that triple inner binaries follow the same mass ratio distributions (and perhaps formation mechanisms) as isolated binaries. 
Lastly, the outer binary mass ratio, $q_{\rm out}$, is reproduced well with our simple power-law formulation.

From the initial distributions (top panel), most low mass-ratio ($q\lsim0.25$) systems are removed by the resolvability criteria, since their large flux contrasts make them unresolved at short separations. Also, a fraction of stars with $M \gtrsim 1.5$~M$_\odot$ evolved off the main-sequence before $10$~Gyr and therefore did not enter the observed sample. In our modeling, we do not dynamically evolve the triples, meaning three-body interactions combined with stellar evolution, which are known to tighten or merge $\sim20$--$30\%$ of triples over $\sim10$~Gyr \citep[e.g.,][]{Naoz2014, Toonen20, Shariat23, Shariat25Merge}, are not included. While this should, in principle, alter the mass ratio distributions, the fact that our models reproduce the observed masses well (bottom panel of Figure \ref{fig:sampling_masses}) without dynamics included might suggest that EKL-driven tightening or mergers have only a modest dependence on the triple stellar masses. 
This result is not surprising because EKL leads to a merger more frequently if the masses of the inner binary are different \citep[e.g.,][]{Naoz2014}.

\subsection{Orbital Periods}\label{subsec:triple_separations}

\begin{figure*}
    \centering
    \includegraphics[width=0.75\textwidth]{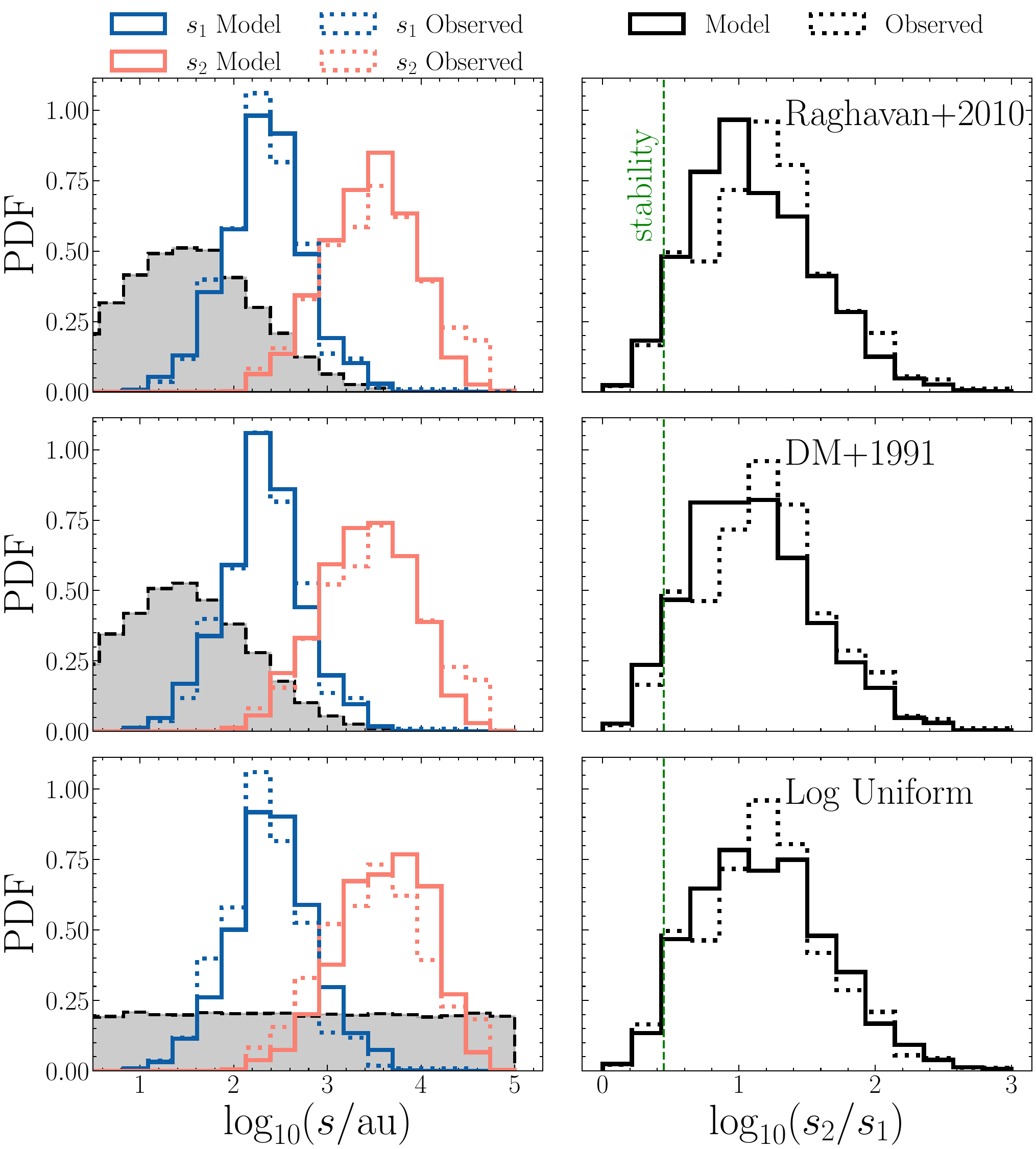}  
    \caption{The origin of triple separations. 
    The outer separations ($s_2$) of resolved triples are drawn from various parent distributions (gray histograms), whereas the inner separations ($s_1$) all follow \citet{Moe17}.
    We show the resulting mock distributions from these models (solid lines) and compare to observed triples (dotted lines). {\it Left}: Distributions of $s_1$ (blue) and $s_2$ (red).
    {\it Right}: Distributions of separation ratios ($s_2/s_1$).
    The top $2$ rows draw tertiary separations from the log-normal distributions \citet{Raghavan2010} and \citet{DM91}. The bottom row assumes a log-uniform distribution for $s_2$ between $10$ and $50,000$ au.
    The green line shows the threshold for dynamical instability ($s_2/s_1 = 2.8$). The log-normal models provide the best match with observations. These models contain only dynamically stable triples, yet still reproduce the observed fraction of systems with $s_2/s_1 < 2.8$, suggesting they are the result of projection effects. 
    } \label{fig:sampled_s1s2_trip} 
\end{figure*}

While separation distributions are constrained for main-sequence binaries at similar separations and masses \citep[e.g.,][]{DM91, Raghavan2010, Moe17, EB21_widebin}, only modest constraints exist for triples \citep[e.g.,][]{Tokovinin14b}. In the previous section, we jointly sample $M_1$, $e_{\rm in}$, $q_{\rm in}$, and $P_{\rm in}$ to get the inner binary parameters. In this section, we constrain the intrinsic distribution of the tertiary's orbital period, $P_{\rm out}$.

We test three different parent distributions. The first is from \citet{DM91}, a log-normal distribution with mean $\overline{\log_{10}(P)} = 4.8$ and standard deviation $\sigma_{\log_{10}(P)} = 2.3$, where $P$ is in days. The second, from \citet{Raghavan2010}, is also log-normal with $\overline{\log_{10}(P)} = 5.03$ and $\sigma_{\log_{10}(P)} = 2.28$. These two log-normal distributions are broadly similar, with the \citet{Raghavan2010} distribution yielding slightly more wide binaries. For both of these models, if the primary mass is below $0.6~{\rm M_\odot}$, we instead draw periods from a log-normal distribution with $\overline{\log_{10}(a)} = 1.3$ and $\sigma_{\log_{10}(a)} = 1.16$, consistent with the closer orbits of M-dwarf binaries \citep{Fischer92,Duchene13,Winters19}. While the true period distribution likely varies smoothly with mass, this discrete treatment offers a reasonable approximation given current limitations.
The third distribution we test is log-uniform from $10 < s/{\rm au} < 50,000$. We convert all sampled periods to projected separations (which {\it Gaia} measures) assuming Keplerian orbits and isotropic orientations (see Appendix \ref{app:projected_separations}).

For each triple, we sample $P_{\rm out}$ and check if it leads to a dynamically stable triple; otherwise, it is resampled.
Dynamical stability is defined by two criteria that determine whether the triple is (a) hierarchical and (b) long-term stable. To test hierarchy, we adopt the hierarchical parameter $\epsilon$, which is effectively the coefficient of the octupole term in the three-body Hamiltonian \citep[e.g.,][]{Naoz2013sec}
\begin{equation}\label{eq:eps_crit}
    \epsilon = \frac{a_{\rm in}}{a_{\rm out}}\frac{e_{\rm out}}{1-e_{\rm out}^2} < 0.1 \ .
\end{equation} 
For long-term stability, we apply the criteria from \citet{MA2001}:
\begin{equation}\label{eq:MA_stability_crit}
    \frac{a_{\rm out}}{a_{\rm in}}>2.8 \left(1+\frac{M_3}{M_1+M_2}\right)^\frac{2}{5} \frac{(1+e_{\rm out})^\frac{2}{5}}{(1-e_{\rm out})^\frac{6}{5}} \left(1-\frac{0.3i}{180^\circ}\right) \ .
\end{equation}
Here $a_{\rm in}$ ($a_{\rm out}$) is the semi-major axis of the inner (outer) orbit, $e_{\rm in}$ ($e_{\rm out}$) are the eccentricities of the inner (outer) orbit, $i$ is the mutual inclination in degrees, and $M_i$ are the masses, with $1,2,3$ corresponding to the primary, secondary, and tertiary, respectively.
We note that a deviation from strict hierarchy does not necessarily mean that a triple becomes instantaneously unstable  \citep{Grishin17,Mushkin20, Bhaskar21, Toonen2022, Zhang23,Weldon24}. While most of the triples in our observed sample are stable according to these criteria (e.g., Figure \ref{fig:summary_plots}), some of the triples have {\it projected} separations that are seemingly unstable. Later in this section, we check whether these visually unstable triples are purely due to projection effects.
 
These stability criteria also depend on masses, eccentricities, and mutual inclination. Masses and inner eccentricities are determined according to Section \ref{subsec:triple_masses}, and we assume thermal outer eccentricities and isotropic mutual inclinations.
If the system is not dynamically stable (according to Equations \ref{eq:eps_crit} and \ref{eq:MA_stability_crit}), then we re-sample $P_{\rm out}$ and $e_{\rm out}$ until these conditions are met. For comparison to observations, we only consider triples where all three stars are on the main sequence after $10$~Gyr with a constant star formation rate. Lastly, we mock-observe these systems using {\it Gaia} selection function (Appendix \ref{app:ang_res_criteria}), keeping only triples where all three components are resolved and brighter than the {\it Gaia} detection threshold (see Section \ref{subsec:triple_masses}). Note that the effective angular resolution at which two stars can be resolved also depends on their distance, which we sample from the $100$ pc catalog. 
One last bias that we account for in our model is the truncation of the widest triples ($s > 30{,}000$~au) due to both perturbations and selection effects. We estimate this using the observed fall-off in {\it Gaia} wide binaries (Appendix \ref{app:missing_binaries}), which reduces the number of triples with wide tertiaries as a function of their outer separation.

Figure \ref{fig:sampled_s1s2_trip} shows the results of sampling from different period distributions and compares them to observations. In the left panel, gray histograms show the parent distribution used to sample $P_{\rm out}$, including \citet{Raghavan2010}, \citet{DM91}, and log-uniform. Dotted lines show the observed $s_1$ (blue) and $s_2$ (red) of $100$~pc resolved triples, and the solid lines show the mock population.
Both the \citet{Raghavan2010} and \citet{DM91} log-normal period distributions are consistent with triple observations. Specifically, they reproduce both the observed $s_1$ and $s_2$ separations from $\sim50 - 50,000$~au and the $s_2/s_1$ distributions relatively well (right column). Importantly, these models predict virtually the same fraction of systems with projected separation ratios $s_2/s_1<2.8$ as observed in our catalog. This supports the notion that such {\it visually} unstable triples arise from projection effects, and that nearly all Galactic triples are consistent with being hierarchical and dynamically stable.

In contrast, the log-uniform model over-predicts the number of wide inner and outer separations relative to observations. This arises because the log-uniform distribution samples wide separations (larger $s_2$) more often than the log-normal distributions, allowing wider inner binaries to exist (larger $s_1$). Such a log-uniform model has also been shown to be inconsistent with observed wide binaries \citep[e.g.,][]{EB18}. 

Using the same framework as above, we test whether drawing both $P_{\rm in}$ and $P_{\rm out}$ simultaneously and retaining only stable combinations can reproduce the observed distributions. This differs from the default assumption, which only resamples $P_{\rm out}$. We find that this approach biases the $s_1$ distribution toward smaller values, in tension with observations. This suggests that fixing the inner binary and sampling the outer orbit until stability is achieved yields a more realistic triple population.

We also test whether the simple stability criterion, $P_{\rm out}/P_{\rm in} > 5$, can reproduce observed triple separations. Interestingly, this criterion produces significantly more triples with $s_2/s_1 < 2.8$ than is observed, indicating that slightly more involved criteria (Equation \ref{eq:eps_crit} and \ref{eq:MA_stability_crit}) are required to describe triple stability. 

Recursively sampling tertiaries from the binary log-normal distributions reproduces triple separations extremely well (Figure \ref{fig:sampled_s1s2_trip}). One interpretation of this result is that tertiaries form at a wide range of separations about inner binaries, governed by the intrinsic binary log-normal distribution. Then, only those that are long-term stable remain in the field. This suggests that both wide binaries and triples form through similar star-formation mechanisms in their nascent environments \citep[][]{Tokovinin22_resolvedtriples}. Wide multiples in young stellar populations show evidence of hierarchical collapse being a dominant mechanism \citep[][]{Joncour17}, which is a natural top-down process that can account for these observations.

\subsection{Eccentricities}\label{subsec:triple_ecc}

\begin{figure*}
    \centering
\includegraphics[width=0.95\textwidth]{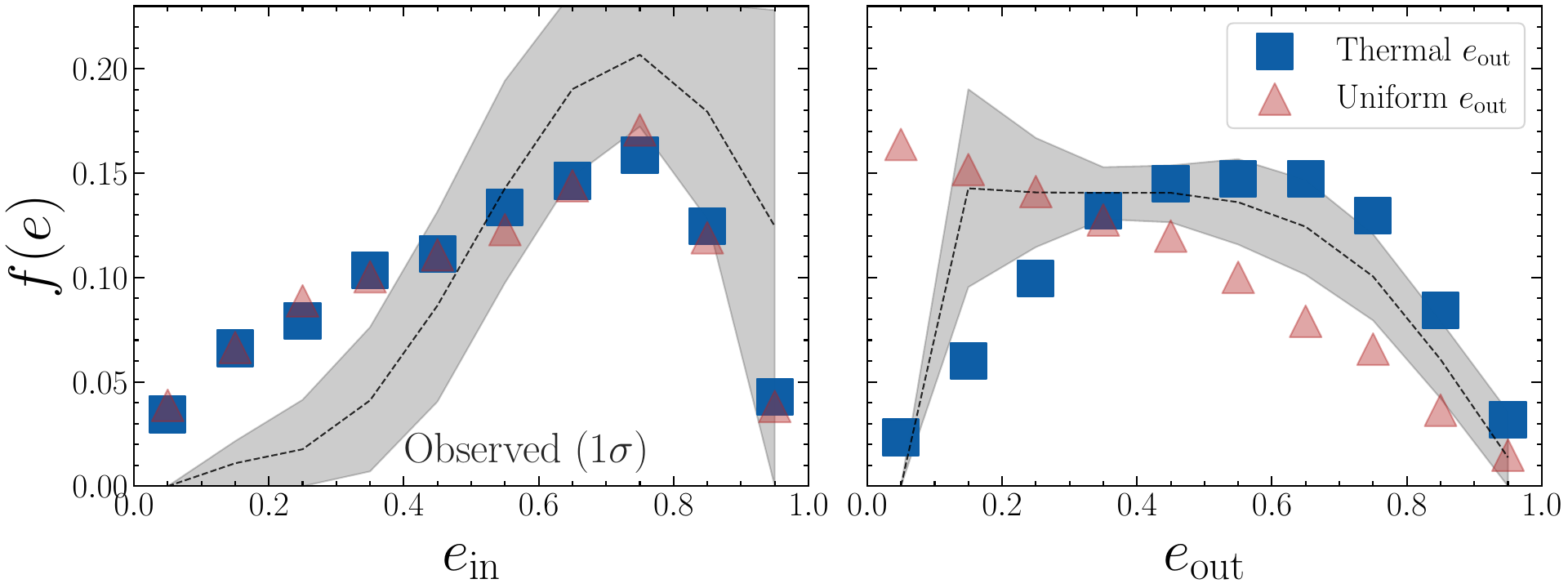}  
    \caption{The origin of triple eccentricities. Observed eccentricity distributions for the inner (left) and outer (right) orbits of resolved triples are shown, with the mean (black dashed line) and $1\sigma$ uncertainty (gray band). These are compared to two models that sample outer eccentricities ($e_{\rm out}$) from either a thermal (blue squares) or uniform (red triangles) distribution. Both models sample inner eccentricities ($e_{\rm in}$) following \citet{Moe17} and are subject to dynamical stability constraints.
    } \label{fig:ecc_model} 
\end{figure*}

The distribution of outer eccentricities ($e_{\rm out}$) in triple systems is essential for modeling their dynamical evolution and constraining formation pathways. However, measuring the eccentricity of wide systems is challenging due to their long orbital periods. In our model, the inner eccentricity ($e_{\rm in}$) is jointly sampled with $M_1$, $q_{\rm in}$, and $P_{\rm in}$ following the prescription of \citet{Moe17} (see Section~\ref{subsec:triple_masses}). The resulting $e_{\rm in}$ distribution is roughly thermal for long orbital periods ($\log(P/{\rm days})>1.5$) and shifts to smaller values for closer binaries \citep{Moe17}. The eccentricity of the outer orbit, on the other hand, is much less constrained.

While individual eccentricities are difficult to measure, the eccentricity distribution of a population, $f(e)$, can be estimated using astrometry. \citet{Tokovinin16_ecc} develop a method that statistically reconstructs the eccentricity distributions of wide binaries using their proper motions and relative positions \citep[see][for details]{Tokovinin16_ecc,Tokovinin20_ecc}. Here, we apply their method on our $100$~pc resolved triple sample to determine $f(e)$ for both inner and outer orbits.
Overall, we find that inner orbits are relatively eccentric with a mean $\langle e_{\rm in} \rangle \approx 0.69$, while outer orbits are more moderate, $\langle e_{\rm out} \rangle \approx 0.47$.
\citet{Tokovinin22_resolvedtriples} also apply this method to their resolved triples sample, finding that $\langle e_{\rm in} \rangle \approx 0.66\pm0.02$ and $\langle e_{\rm out} \rangle \approx 0.54\pm0.02$, which is consistent to with our values. Here, we examine the origin of these observed eccentricities. 

We test both a uniform and thermal intrinsic $e_{\rm out}$ distributions by sampling from each, applying the selection function, and comparing the resulting $e_{\rm out}$ to observations. If a sampled triple violates dynamical stability (Equations \ref{eq:eps_crit} and \ref{eq:MA_stability_crit}), both $e_{\rm out}$ and $P_{\rm out}$ are resampled until a stable configuration is found. This approach effectively assumes that the inner binary parameters are set earlier in the formation process, while the tertiary may form (or settle) across a wide range of orbital configurations, with only the stable ones surviving to be observed in the field today.

Figure \ref{fig:ecc_model} plots the eccentricity distribution $f(e)$ for the inner (left) and outer orbit (right) of observed triples in our resolved sample (black). We compare this distribution to the models that assume a uniform $e_{\rm out}$ (red triangles) and thermal $e_{\rm out}$ (blue squares). Both models assume $e_{\rm in}$ follows \citet{Moe17}, so any differences are the result of enforcing stability. For the observed distribution, $1\sigma$ uncertainties are derived using bootstrap \citep[][]{Tokovinin16_ecc} and shaded in gray.

A thermal distribution of $e_{\rm out}$ matches observations of resolved triples better than a uniform distribution.
The uniform distribution becomes weighted towards smaller $e_{\rm out}$ after imposing stability, while the observed wide tertiaries show a dearth of small $e_{\rm out}$. On the other hand, the thermal $e_{\rm out}$ distribution exhibits a decline at high eccentricities due to dynamical stability constraints, resulting in a peak around $e_{\rm out} \approx 0.5-0.6$ and only a small fraction of systems with $e_{\rm out} \gsim 0.8$ \citep[in agreement with][]{Tokovinin22_resolvedtriples}. The robust dropoff at large $e_{\rm out}$ further emphasizes the need for considering $e_{\rm out}$ in triple stability criteria, as also highlighted in Section \ref{subsec:triple_separations}.

 The $e_{\rm in}$ distribution peaks slightly higher, near $0.7$, and similarly aligns well with the observed curve. The model underpredicts the fraction of systems with $e_{\rm in} > 0.9$ and slightly overpredicts the fraction with $0.1 < e_{\rm in} < 0.4$, likely because it does not account for dynamical evolution that can drive inner binaries to high eccentricities \citep[e.g., EKL;][]{Naoz2016}. Dynamics also contributes to the dearth of high $e_{\rm out}$ triples, since those triples exhibit the strongest EKL, causing the inner binary to tighten or merge \citep[e.g.,][]{Naoz2014,Shariat25Merge}

\subsection{How to Sample Triple Parameters}\label{subsec:how_to_sample}
The agreement between our models and the observed distributions of triple-star masses (Figure~\ref{fig:sampling_masses}), orbital separations (Figure~\ref{fig:sampled_s1s2_trip}), and eccentricities (Figure~\ref{fig:ecc_model}) highlights the need to re-evaluate triple population modeling. For instance, most previous models of triple populations assume that the $q_{\rm in}$ and $q_{\rm out}$ are drawn from the same underlying distribution, often taken to be uniform. However, observations show that this assumption does not hold true for the resolved triple population.

To allow for easy population modeling in an observationally-consistent way, we provide a simple Python tool on GitHub\footnote{\url{https://github.com/cheyanneshariat/gaia_triples}} that generates mock triple systems by sampling masses, orbital periods, and eccentricities from our empirically validated scheme. This tool can be applied to sample initial conditions for a triple population and can also generate a complete mock stellar population, with singles, binaries, and triples. We summarize the prescription for creating synthetic stellar populations (including triples) below:

\begin{enumerate}
    \item \textbf{Draw stellar masses:} Sample stellar masses from a Kroupa IMF to represent single stars in the population.
    \item \textbf{Add binaries:} Generate a binary population with jointly-sampled multiplicity fractions, primary masses ($M_1$), mass ratios ($q=M_2/M_1$), orbital periods ($P_{\rm in}$), and eccentricities ($e_{\rm in}$) following \citet{Moe17}. 

    \item \textbf{Assign triples:} For each binary, assign a probability that the system has a tertiary companion (i.e., is a triple), given by the triple fraction as a function of $M_1$.

    \item \textbf{Sample outer mass ratio:} For triples, assign the tertiary mass as $ M_3 = q_{\rm out} (M_1 + M_2) $, where the outer mass ratio ($ q_{\rm out} $) is sampled from a power law with logarithmic slope $\gamma=-1.4$.

    \item \textbf{Sample outer period and eccentricity:} 
     For triple outer orbits, sample a period ($P_{\rm out}$) from the log-normal distribution of binary stars and eccentricity ($e_{\rm out}$) from a thermal distribution. If the triple fails the hierarchy or dynamical stability criteria (Equations~\ref{eq:eps_crit}, \ref{eq:MA_stability_crit}), redraw until these conditions are met. 

\end{enumerate}

The above sampling procedure reproduces key properties of the observed triple population (masses, separations, eccentricities, multiplicity statistics) while generating a realistic distribution of singles and binaries as well. In a follow-up paper, we apply these initial conditions to evolve a synthetic triple population using dynamical simulations with stellar evolution.

Although our sampling approach performs well across a range of separations and stellar masses, we do not claim that it is a universal prescription. For example, massive triples ($M \gtrsim 8~{\rm M}_\odot$) and unresolved (compact) triples are underrepresented in these observations. However, our inner binary sampling follows empirical prescriptions that were calibrated using massive stars and short-period systems \citep{Moe17}, allowing the sampling scheme to be sensibly applied to the full triple population. 

The above sampling scheme does not explicitly draw mutual inclinations, which are less constrained by observations, especially for unresolved triples. For the wide triples in our sample, mutual inclinations are broadly consistent with being isotropic (Figure~\ref{fig:mutual_inclinations}).  
Compact triples with $P_{\rm out} \lesssim 1000$~days show a tendency for moderate outer eccentricities and aligned orbits \citep[e.g.,][]{Bashi24}, although such systems are likely rare among the overall population \citep{Tokovinin14b}. In our sample of wide systems, the more compact triples also start to show a mild preference for mutual orbital alignment (Figure \ref{fig:mutual_inclinations}). Nonetheless, triples with $s_2 \gtrsim 100$~au are generally consistent with isotropic mutual inclinations, which is the default assumption in our model.

\subsection{Completeness}\label{subsec:completeness}
\begin{figure}
    \centering
    \includegraphics[width=1.0\columnwidth]{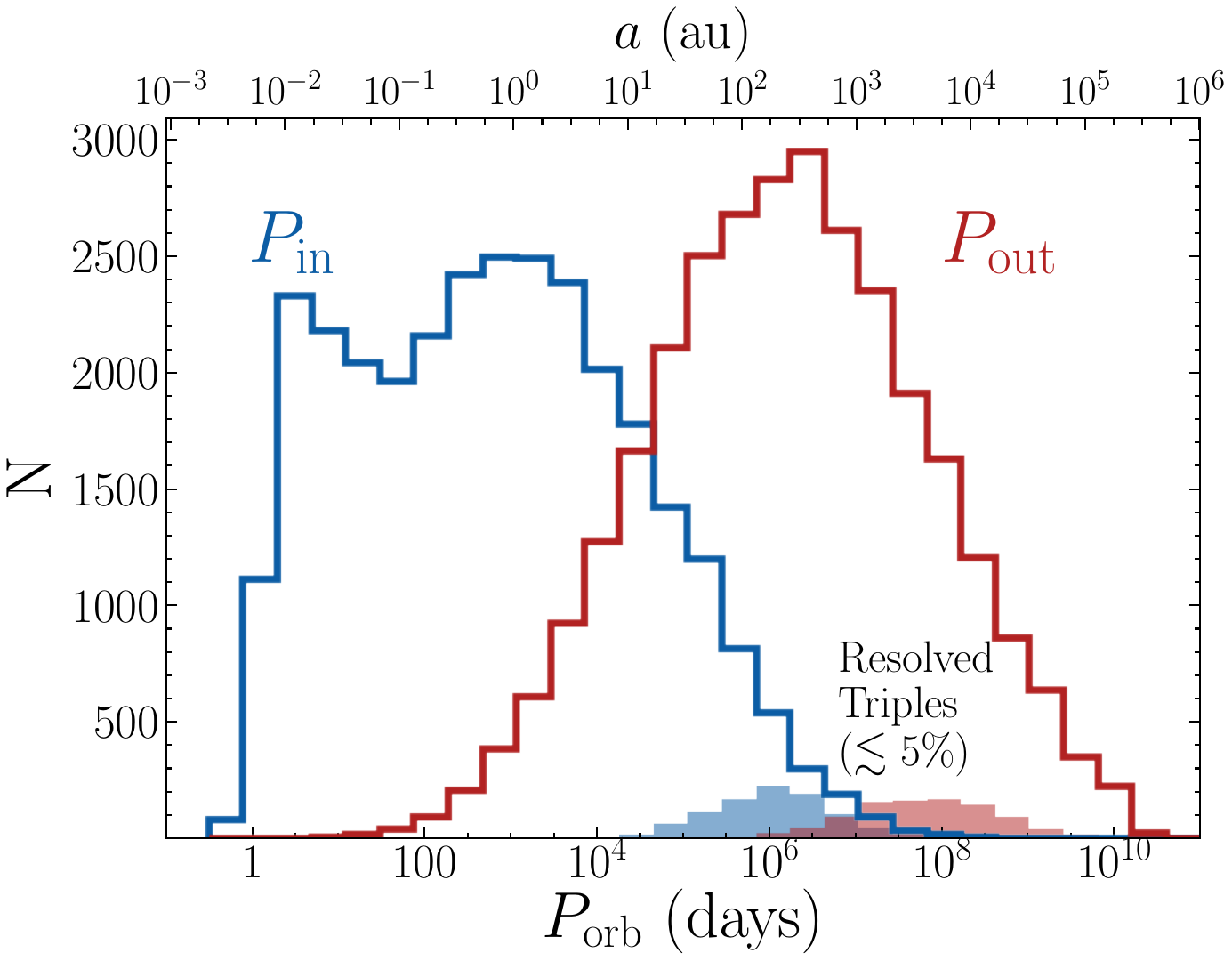}
    \caption{ Completeness of the $100$~pc resolved triple sample. The solid curves show the intrinsic distribution of the inner (blue) and outer (red) orbital periods for $30,000$ triples (roughly the number expected within $100$~pc). The shaded histograms show the subset that are fully spatially resolved. Resolved triples comprise only $\lsim5\%$ of the full triple population within $100$~pc.
    } \label{fig:period_completeness} 
\end{figure}

We adopt our intrinsic triple model from the previous section to estimate the completeness of our resolved triples sample. This approach relies on assumptions about the intrinsic distribution of triple masses and orbital periods. Among these, the incompleteness is most sensitive to the assumed period distribution, which especially disfavors low-mass stars because they have relatively smaller periods \citep[e.g.,][]{Winters19}. 

Even in a nearby sample (e.g., $d<100$~pc), only a small fraction of all triples are fully spatially resolved by {\it Gaia}. Figure \ref{fig:period_completeness} illustrates this, where we show the intrinsic period distributions for a mock population of $30,000$ hierarchical triples sampled using our prescription (Section \ref{subsec:how_to_sample}). Among those with all three masses greater than $0.1~{\rm M_\odot}$ ($\sim21,000$), only $620$ triples are completely resolved and on the main sequence, corresponding to a completeness fraction of $\sim3\%$. The $100$~pc volume is also expected to host $\sim30,000$ triples, assuming a stellar density $n_\star=0.1~{\rm pc^{-3}}$ and total triple fraction $0.1$. Our observed sample contains $650-700$ resolved main-sequence triples, which is consistent with our completeness estimates.

\subsection{Triple Fraction}\label{subsec:triple_fraction}
\begin{figure*}
    \centering
    \includegraphics[width=0.99
    \textwidth]{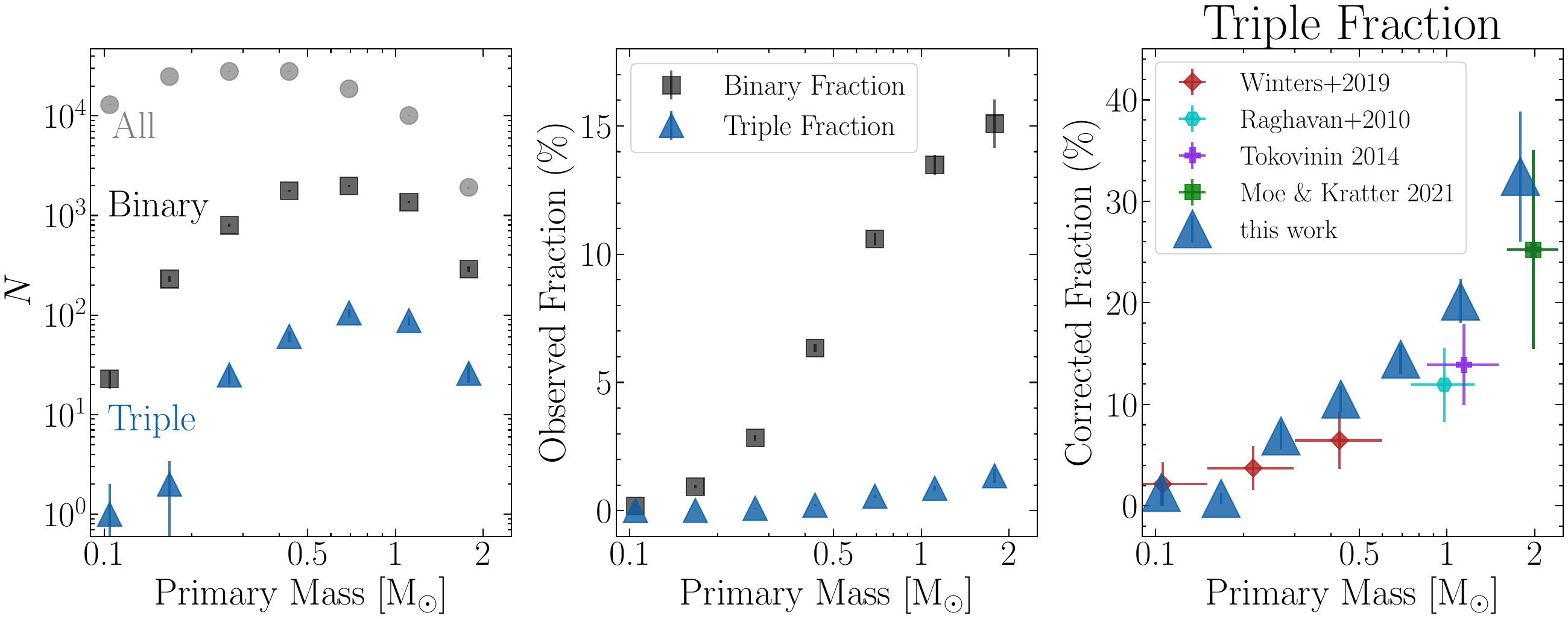}
    \caption{Triple fraction as a function of primary mass. {\bf Left:} The number of total stars (gray circles), in wide binaries (black squares) and in wide triples (blue triangles) in the {\it Gaia} $100$~pc sample that pass initial quality cuts (Section \ref{sec:creating_catalog}). {\bf Middle:} Observed fraction of stars in binaries (black squares) and triples (blue triangles) with companions wider than $100$~au. {\bf Right:} Completeness-corrected triple fractions (blue triangles) compared to reported fractions from  \citet{Winters19}, \citet{Raghavan2010}, \citet{Tokovinin14b}, and \citet{MoeKratter21}. Poisson errors are displayed for each point, and reported errors are displayed for the comparison samples. 
 }\label{fig:triple_fraction} 
\end{figure*}

For stars closer than $100$~pc, {\it Gaia} resolves almost all stellar companions with separations beyond $100$~au. Therefore, our triple sample is nearly complete for inner binaries with $d<100$~pc and $s_{\rm in}>100$~au, allowing us to faithfully estimate the fraction of stars with wide ($s_{\rm in}>100$~au) companions. For all main-sequence stars within the $100$~pc volume, we calculate the observed fraction of wide triples and binaries, defined by companions with $s>100$~au. Then, assuming triple masses and separations from Section \ref{subsec:how_to_sample}, we evaluate the completeness fraction as a function of $M_1$, from which we back out the intrinsic triple fraction.

Following \citet{Offner23}, we define the binary fraction, $f_{\rm binary} = N_{\rm Binary} / N_{\rm All}$, where $N_{\rm Binary}$ is the number of binaries with physical separations greater than $100$~au and $N_{\rm All} = N_{\rm Single} + N_{\rm Binary} + N_{\rm Triple}+...$ is the total number of systems within 100 pc that pass the initial {\it Gaia} query cuts (Section \ref{sec:creating_catalog}). An additional cut of ${\tt astrometric\_sigma5d\_max}<1$ is applied all singles, binaries, and triples within $100$~pc to remove sources with bad astrometry \citep[e.g.,][]{Gaia_Collab21b}.   
The triple fraction is defined as $f_{\rm triples} = N_{\rm Triples} / N_{\rm All}$, where $N_{\rm Triples}$ is the total number of triples with $s_1>100$~au.
$N_{\rm Binary}$ does not include systems that are in the resolved triples catalog, such that it is strictly a binary fraction.

In the left panel of Figure \ref{fig:triple_fraction}, we plot the total number of stars ($N_{\rm All}$), binaries ($N_{\rm Binary}$), and triples ($N_{\rm Triple}$), as a function of primary mass ($M_1$). For triples, $M_1$ is the most massive star in the inner binary. We consider only stars with primary mass above $0.1~{\rm M_\odot}$ since the {\it Gaia} $100$ pc sample has low completeness below this mass \citep{Gaia_Collab21b}. We also restrict to masses below $2.5~{\rm M_\odot}$, since there are only a few main-sequence stars in the $100$~pc sample with such high masses. 

In the middle panel of Figure \ref{fig:triple_fraction}, we plot the observed wide triple/binary fraction as a function of primary mass. At the lowest primary masses ($\lsim0.5~{\rm M_\odot}$), a minority ($\lsim6\%$) of systems have wide stellar companions, consistent with previous work \citep[e.g.,][]{Allen07,Winters19}. Both the observed wide triple and binary fraction increases monotonically with primary mass, and for $1~{\rm M_\odot}$ primaries, $\sim15\%$ are in wide binaries and $\sim7\%$ are in resolved triples. These fractions only measure wide, resolved companions. However, the majority of binaries and triples in this $100$~pc sample will reside in unresolved systems. We account for these unresolved systems by estimating the completeness of our catalog for each primary mass bin and correcting these observed fractions accordingly.

We assume that the underlying period distributions follow the scheme developed in Section \ref{subsec:triple_separations}, which reproduces observations. Under this assumption, the outer orbit is derived from log-normal period distributions: \citet{Winters19} for $M_1 < 0.6~{\rm M_\odot}$ and \citet{DM91} for all others. The inner orbit is drawn from \citet{Moe17}. 
Then we sample masses according to the procedure in Section \ref{subsec:triple_masses}, derive separations, and determine the fraction of systems that are resolved by {\it Gaia} and would therefore make it into our sample (see Appendix \ref{app:ang_res_criteria} for resolvability criteria). We use this intrinsic model to estimate the completeness fraction, the fraction of triples with three resolved components, for each primary mass bin. We then correct the observed triple fraction for each primary mass by the completeness fraction to derive the intrinsic triple fraction. 

The right panel of Figure \ref{fig:triple_fraction} plots the intrinsic triple fraction as a function of primary mass from our sample, after correcting for incompleteness. We compare to triple fractions reported by previous multiplicity studies of M dwarfs \citep[$M_1<0.6~{\rm M_\odot}$;][]{Winters19}, FGK dwarfs \citep[$0.75<M_1/{\rm M_\odot}<1.5$;][]{Raghavan2010,Tokovinin14b}, and A-type stars \citep[$1.6<M_1/{\rm M_\odot}<2.4$;][]{MoeKratter21}. Our results generally agree with previous volume-limited surveys, but predict a slightly larger triple fraction for FGK primaries.
The plot shows that low mass stars $M_1\lsim0.5~{\rm M_\odot}$ have typical triple fractions of $2-5\%$, while $2~{\rm M_\odot}$ stars have a $35\%$ triple fraction. The triple fraction strictly increases with primary mass. We note that the highest mass bin contains only a few triples within the $100$~pc sample, so the estimated triple fractions contain larger uncertainties.

\subsection{Comparison to Wide Binaries}\label{subsec:comparison_to_binary}

\begin{figure*}
    \centering
    \includegraphics[width=0.95\textwidth]{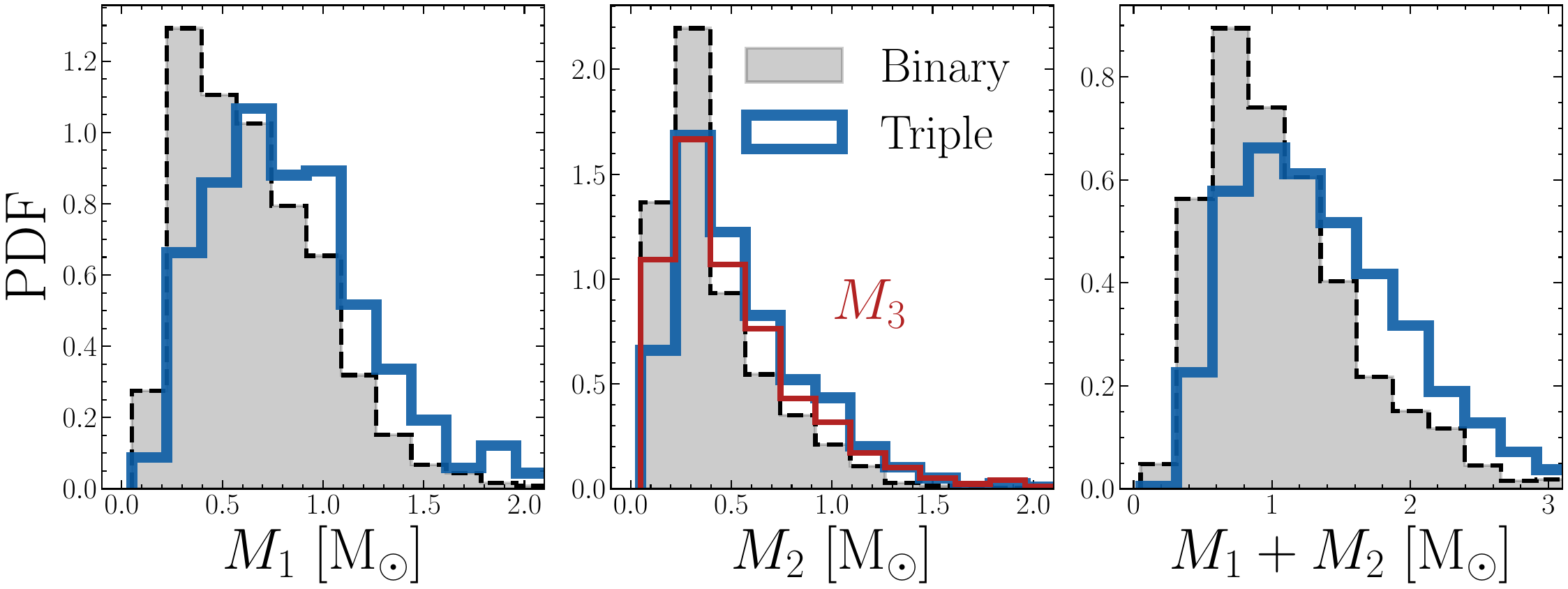}\\[0.4em]
    \includegraphics[width=0.95\textwidth]{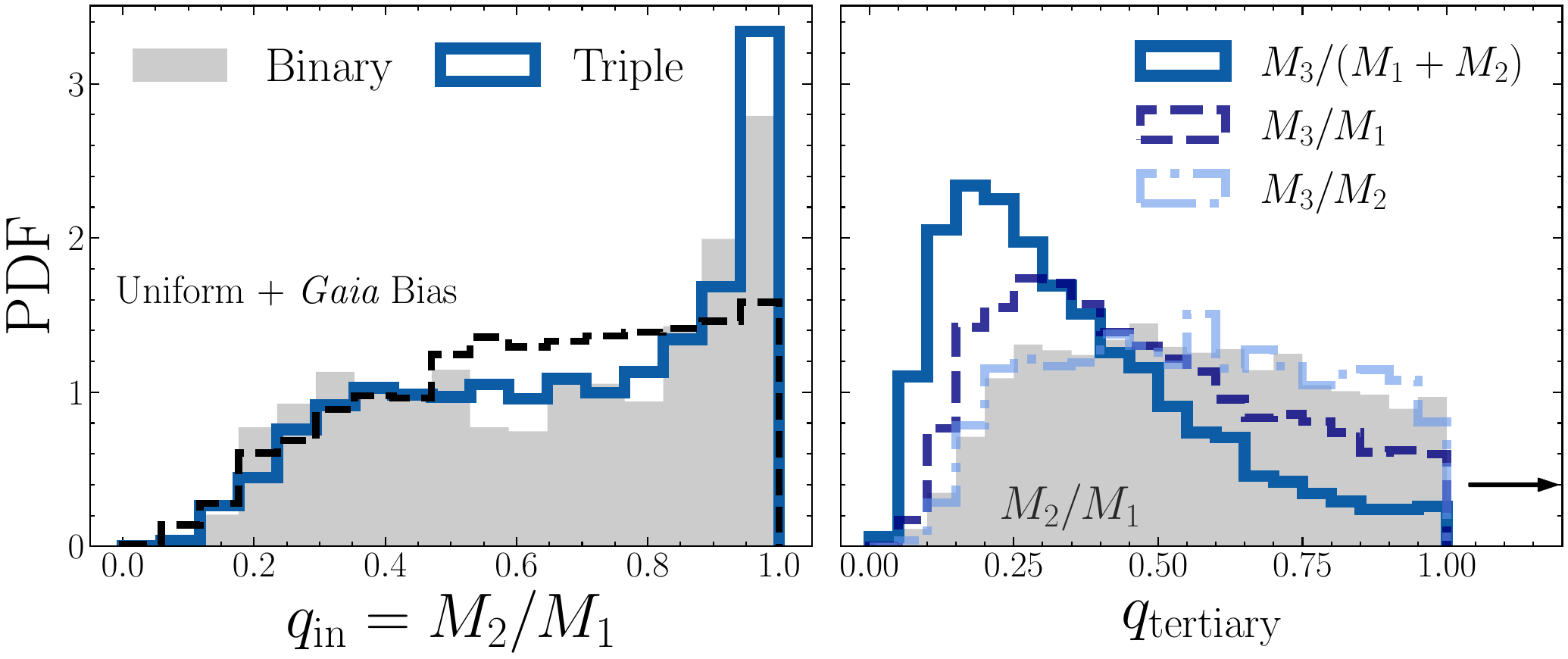}
    \caption{ Comparing observed resolved triples to isolated wide binaries.
    The wide binary control sample is designed to have the same separations and distances to the triples.
    \textit{Top:} Comparing individual stellar masses between binaries (gray) and triples (blue).
    On the left, we compare the primary masses ($M_1$). In the middle, we compare the secondary ($M_2$) and tertiary ($M_3$, red) triple masses to the secondary in wide binaries. On the right, we compare the total mass of triple inner binaries to the total mass of wide binaries.  
    \textit{Bottom:} Comparing mass ratios between binaries (gray) and triples (blue). On the left, we compare the inner binary mass ratios ($q_{\rm in} = M_2/M_1$). For reference, we plot a uniform intrinsic mass ratio distribution subjected to {\it Gaia}'s selection effects (dashed black curve). On the right, we plot tertiary-to-inner mass ratios in triples shown in various forms ($M_3/(M_1+M_2)$, $M_3/M_1$, $M_3/M_2$), compared to the binary mass ratio distribution. Here, the binary sample is different from the previous one, where it is now controlled by the separations/distances of triple outer binaries.
    }\label{fig:binary_triple_comp} 

\end{figure*}

Since the advent of {\it Gaia}, wide binaries in the solar neighborhood have been studied extensively \citep[e.g.,][]{EB18, EB21_widebin, Tian20, Hartman20}, whereas wide triples have received comparatively less attention \citep[with the exception of][]{Tokovinin22_resolvedtriples}.
In this section, we compare the properties of isolated wide binaries (i.e., without resolved tertiaries) to inner binaries of resolved triples. Such a comparison tests whether a triple environment, either through its dynamical influence or formation history, leads to different binary properties.
To ensure a fair comparison, we construct a control sample of wide binaries matched in separation and distance distributions to the triples (Appendix \ref{app:WB_control}), ensuring similar observational biases. As a result, any differences in observable properties, such as mass distributions, can be attributed to intrinsic differences between the populations.

In Figure \ref{fig:binary_triple_comp}, we compare the masses (top) and mass ratios (bottom) between resolved binaries and resolved triples. 
In the top left panel, we compare the primary masses $M_1$ in binaries (gray) to triple inner binaries (blue).
In the top middle panel, we compare the secondary masses $M_2$ in binaries (gray) to the secondary ($M_2$, blue) and tertiary ($M_3$, red) masses in triples.
In the top right panel, we compare the total mass of binaries to the total mass of triple inner binaries.
The median total mass of triples is $1.65$~M$_\odot$ and the median mass of the inner binary is $1.1$~M$_\odot$.  In the 100 pc sample, which is less susceptible to systematic biases, our sample contains $423$ triple main-sequence systems and $703$ triples total. Among the main sequence triples, $26\%$ of them have a tertiary that is the most massive star, and $11\%$ of the time the tertiary is more massive than the inner binary altogether. However, $50\%$ of the time the tertiary is more massive than one of the stars in the inner binary. These numbers are broadly consistent with the \citet{Tokovinin22_resolvedtriples} $100$~pc sample.

Figure \ref{fig:binary_triple_comp} also confirms that stars in triples are systematically more massive than those in wide binaries \citep[e.g.,][and references therein]{Offner23}. This trend reflects the well-established increase in multiplicity with stellar mass (e.g., Figure \ref{fig:triple_fraction}). Physically, this may result from enhanced accretion during formation or dynamical capture favoring higher-mass stars \citep[e.g.,][]{Clark21}. The primary mass in triples has a median of $0.7$~M$_\odot$, compared to $0.53$~M$_\odot$ in binaries (left panel). Similarly, both $M_2$ and $M_3$ in triples have medians of $\sim0.4$~M$_\odot$, exceeding the binary $M_2$ median of $0.33$~M$_\odot$ (middle panel). As a result, the total inner binary mass in triples also exceeds that of wide binaries (right panel).

The bottom row of Figure \ref{fig:binary_triple_comp} compares the mass ratio distributions between binaries and triples. The left panel compares $q_{\rm in}=M_2/M_1$ to binary mass ratios. We also plot the expected curve for a uniform mass ratio distribution subject to the {\it Gaia} selection function (black dashed curve). For this uniform model, periods are chosen from the log-normal distribution \citep[][]{DM91} and distances are drawn from the observed sample. While intrinsically uniform, this curve truncates at low-$q$ because high contrast pairs are more difficult to resolve at fixed angular separation. Overall, both regular binaries and triple inner binaries show nearly the same mass ratio distributions. They are both mostly consistent with a uniform distribution for $q\lsim0.9$, but show a stark twin excess at $q>0.9$. 

The similarity between the mass ratios of wide binaries and triple inner binaries suggests that both populations share a common formation pathway, despite the presence of a tertiary in triples. One common pathway could be hierarchical fragmentation within a single collapsing core that produces bound multiple systems, with dynamical interactions leaving behind a mixture of wide binaries, triples, and higher-order hierarchies \citep[e.g.,][]{Thomasson24}. Observations of the Taurus star-forming regions show that wide hierarchical multiples ($s \gtrsim 1000$~au) outnumber isolated wide binaries, supporting this hierarchical formation scenario \citep[e.g.,][]{Joncour17}.

The bottom right panel of Figure \ref{fig:binary_triple_comp} plots the outer mass ratio $q_{\rm out} = M_3/(M_1+M_2)$ (blue) and compares it to the mass ratio of wide binaries. Note that these mass ratios can be greater than one. Since all of these mass ratios involve the tertiary, we devise our comparison sample of binaries to be those with separations matching the {\it outer} separation of triples, such that the selection biases are similar. 
The plot reveals that the mass ratio of the outer binary in triples, $q_{\rm out}$ (blue curve), is different from that of regular wide binaries (gray histogram). A common assumption in triple modeling is that the inner and outer binary mass ratios are drawn from the same distribution. This is not consistent with observations, and instead, observed triples are weighted towards smaller $q_{\rm out}$.
Assuming that the tertiary mass is drawn randomly, a $q_{\rm out}$ distribution weighted towards small $q_{\rm out}$, given that the inner binary contains two stars. However, the $q_{\rm out}$ distribution cannot be reproduced if the tertiary mass is drawn from an independent Kroupa IMF, but is more consistent with a $\gamma=-1.3$ power law (e.g., Figure \ref{fig:sampling_masses}).
The tendency for tertiaries to be of lower mass may reflect their formation history. Early dynamical unfolding and dynamical capture—two widely accepted mechanisms for forming wide multiples—favor lower-mass tertiaries, as they are more easily captured or dynamically widened during interactions \citep[e.g.,][and references therein]{Reipurth12}.

The other tertiary mass ratios, $M_3/M_2$ and $M_3/M_1$, have morphologically similar distributions to wide binaries for $q<1$, though $M_3/M_1$ the distribution shows a slight excess of lower mass ratios in triples. 

\section{Discussion }\label{sec:discussion}

\subsection{Comparison to Previous Samples }\label{subsec:tok_compare}

\subsubsection{Solar-Type Triples: Tokovinin 2014 }\label{subsubsec:tok_compare_2014}

\begin{figure*}
    \centering
    \includegraphics[width=0.9\textwidth]{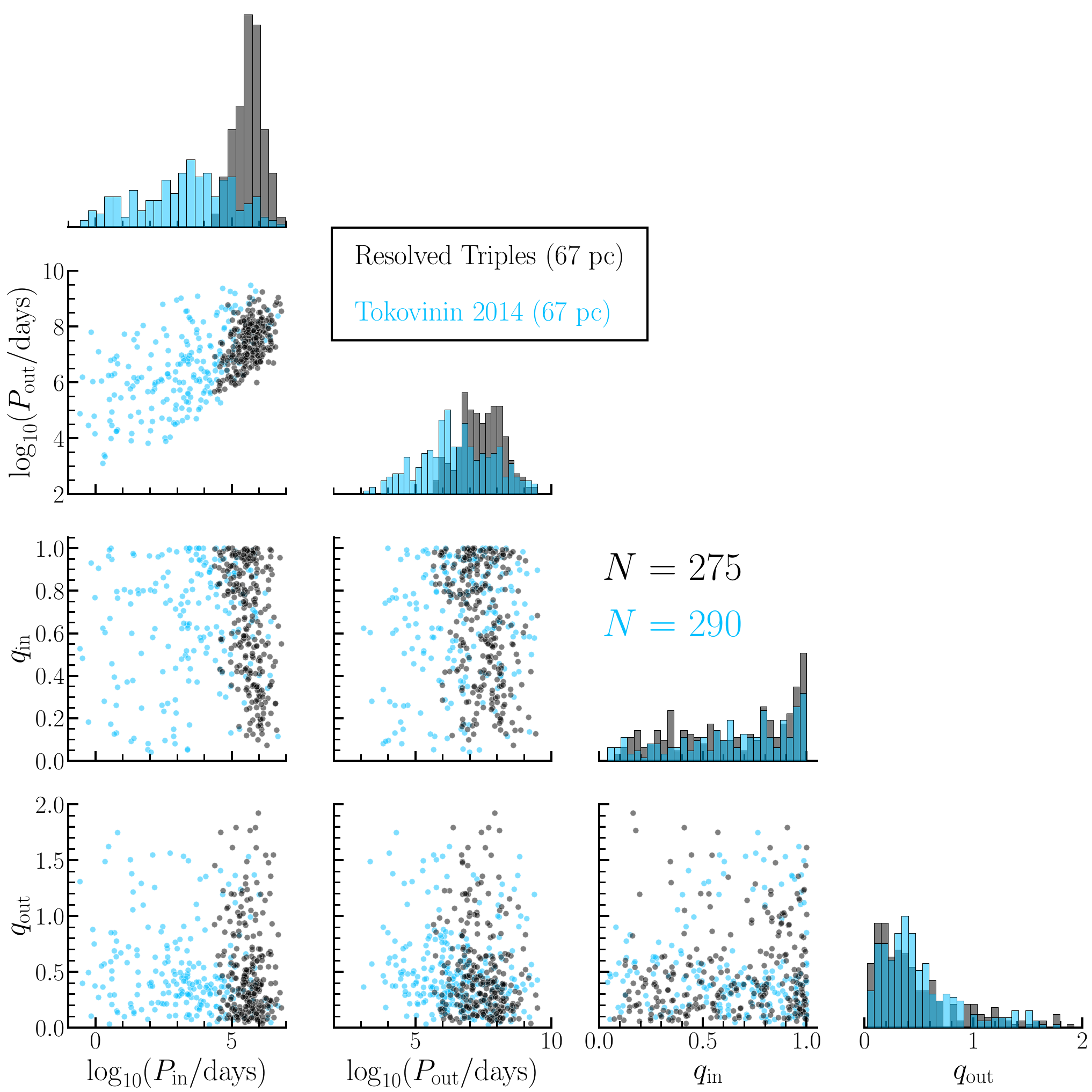}
    \caption{
    Comparison of our resolved triple systems (black) to solar-type triples from \citet{Tokovinin14b} (gray), both limited to $d<67$~pc. The corner plot shows the various parameter spaces spanned by $P_{\rm in}$, $P_{\rm out}$, $q_{\rm in}$, and $q_{\rm out}$. Our resolved sample places no restriction on spectral type, whereas the \citet{Tokovinin14a} catalog includes only F- and G-type primaries. Resolved triples make up a minority of the full triple population, containing systems with the largest inner and outer orbital periods, but spanning a similar range of mass ratios.
    }\label{fig:tok_compare_corner} 

\end{figure*}

\citet{Tokovinin14a} assembled a $>90\%$ complete sample of F- and G-type stars within $67$~pc of the Sun. Using this parent sample, \citet{Tokovinin14b} built a catalog of hierarchical multiples, including $290$ triples, using spectroscopy, astrometry, and imaging. While they focus on systems with solar-type stars, their triples span a broad range of orbital periods from $\sim1-10^{10}$~days. The completeness of detecting companions of the primary star is $80\%$, which decreases to $\sim30\%$ for detecting subsystems in the secondaries. 

Figure \ref{fig:tok_compare_corner} compares the parameter space of the resolved triples in our sample to solar-type triples from \citet{Tokovinin14b}.
The corner plot includes $P_{\rm in}$, $P_{\rm out}$, $q_{\rm in}$, and $q_{\rm out}$, where the diagonals show distributions and the off-diagonal components show scatter plots comparing the various parameters. To match \citet{Tokovinin14b}, we limit our resolved triples to the same volume limit ($d<67$~pc). Note that the \citet{Tokovinin14b} triples contain only triples with FG stars ($N=290$) while our sample places no restriction on the spectral type ($N=275$).  
As evidenced by the plot, resolved triples make up a minority of the entire triple population. The periods of resolved triples concentrate between $10^5 \lsim P_{\rm in/}{\rm days} \lsim 10^7$ and $10^6\lsim P_{\rm out/}{\rm days} \lsim 10^{10}$, placing them on the top right in $P_{\rm in} - P_{\rm out}$ space. However, resolved triples span the same broad range of inner and outer mass ratios as the more compact triple population.

If we limit our $67$~pc resolved triples sample to only F- and G-dwarf primaries, matching \citet{Tokovinin14b}, we find $98$ triples compared to the $\sim45$ fully resolved triples present in their sample. This reflects the fact that {\it Gaia} data revealed many tertiaries within the $67$~pc sample that were not previously known to be bound to their inner binaries.

\subsubsection{100 pc Resolved Triples: Tokovinin 2022 }\label{subsubsec:tok_compare_2022}

\citet{Tokovinin22_resolvedtriples} construct a catalog of resolved triple star systems within $100$~pc using {\it Gaia} astrometry, similar to this work. While their methods overlap in identifying common proper motion companions, there are several differences in scope and approach. Most notably, we extend our search volume out to $500$~pc, whereas \citet{Tokovinin22_resolvedtriples} restricts to $100$~pc. Increasing the search volume also increases the fraction of chance alignments relative to real triples. However, we explicitly calculate a chance alignment probability for each triple in our sample, allowing us to create a large catalog with high purity. To avoid major contamination from chance alignments, even within $100$~pc, \citet{Tokovinin22_resolvedtriples} impose a hard separation cutoff at $10^4$~au. We indeed confirm that below $10^4$~au, most companions are truly bound (e.g., see Appendix \ref{app:chance_alignment}), but do not impose any separation cut (beyond the initial $1$~pc limit).

Another key difference is that we do not impose any restrictions on stellar type or mass, allowing our catalog to include triples containing white dwarfs, red giants, and more massive main-sequence stars. In contrast, \citet{Tokovinin22_resolvedtriples} focus exclusively on triples with main-sequence primaries below $1.5~{\rm M_\odot}$. When restricted to main-sequence triples within $100$~pc, the two catalogs are broadly consistent: \citet{Tokovinin22_resolvedtriples} identifies $392$ systems, while our sample contains $423$. Applying similar mass and photometric cuts to our sample yields $390$ systems, showing good agreement. Without imposing any photometric cuts, our catalog contains $706$ triples within $100$~pc, reflecting the inclusion of evolved stars and those without measured ${\tt bp\_rp}$ colors. Expanding to the full distance limit of $500$~pc, we find a total of $9767$ resolved triples. We note that these numbers are based on our adopted cut of $R_{\rm triple}<0.1$ (i.e., less than a $10\%$ chance alignment probability); relaxing this threshold would increase the number of identified systems, but at the cost of introducing a larger fraction of spurious matches, particularly at wide separations (Appendix \ref{app:chance_alignment}).

\subsection{Exoplanets in Wide Triples}\label{subsec:exoplanets}

The multiplicity of planet-hosting stars can provide a new perspective on the history of planetary systems.
To identify planet-hosting stars in our catalog, we cross-match the resolved triples to the catalog of stellar exoplanet hosts from the NASA Exoplanet Archive\footnote{\url{http://exoplanetarchive.ipac.caltech.edu}}, using the version downloaded on February 24th, 2025. The Exoplanet Archive includes {\it Gaia} DR2 IDs for each stellar host. We match these DR2 IDs to DR3 IDs using the {\it Gaia} ${\tt dr2\_neighbourhood}$ catalog. The planet-hosting star could be any of the three stars in the triple, including one of the inner binary components or the tertiary star. For example, our closest star system Alpha Centauri, which is a triple, hosts a planet around the tertiary star \citep[e.g.,][]{Ang16}

In total, the match returns $14$ planetary systems with host stars that reside in a resolved triple. $10$ of these planets orbit the primary star in the inner binary, $2$ of them orbit the secondary star in the inner binary, and $2$ orbit the tertiary star. In Table \ref{tab:triple_planets} we list these planets along with their host name, host {\it Gaia} DR3 ID, planet name, and the triple component they orbit (primary, secondary, or tertiary).
Among these systems, most have been identified as being in triples previously \citep[][]{Cuntz22, Gonzalez-Payo24}, with the exception of two: K2-27 and K2-31. 

K2-27 is a solar-type ($\sim1$~M$_\odot$) star hosting a warm Neptune (K2-27b) with $M_{\rm p} = 30$~M$_\oplus$ in a $6.8$~day eccentric ($e=0.25$) orbit \citep[][]{Montet15, Van-Eylen16,Petigura17,Mayo18,Crossfield16}. Forming such a massive planet in situ is challenging, as it may require an unusually high solid density in the inner disk; a more commonly invoked scenario is through some form of planetary migration \citep[e.g.,][]{Baraffe06, Correia20}. For such a close orbit, the non-zero eccentricity of K2-27b might support that it migrated inward through high eccentricity migration where thermal atmospheric tides, evaporation of the atmosphere, or excitation from a distant companion acted against bodily tides to keep the orbit eccentric \citep[e.g.,][]{Correia20}.

K2-31b is a $1.8$~M$_{\rm Jup}$ Hot Jupiter on a circular grazing orbit of only $1.25$ days \citep[e.g.,][]{Crossfield16,Dai16,Grziwa16,Kokori23}. The planet's orbit is consistent with high eccentricity migration followed by tidal capture. The presence of this system in a triple further favors this scenario, where torques from the third star could have ignited the planet's dynamical evolution \citep[see e.g.,][]{Yang2025}.

Our cross-match also recovers WD 1856b, a $10$~M$_{\rm Jup}$ planet on a circular $1.4$-day orbit around its white dwarf host \citep{Vanderburg20}. The white dwarf is the tertiary component of a hierarchical triple, with an M-dwarf binary orbiting it $\sim1000$~au away. Its current orbit might have been achieved through a common envelope scenario \citep[e.g.,][]{Vanderburg20,Lagos21} or is likely the product of high-eccentricity tidal migration via the EKL mechanism \citep[e.g.,][]{Munoz20,Stephan21,OConnor21}.

Several of the planets in our triples catalog show evidence for dynamical formation histories, perhaps influenced by their triple-star environments. A larger sample of planet-hosting triples can be generated using direct imaging or utilizing {\it Gaia} metrics. Future work could also use this catalog to explore whether planets preferentially orbit any of the triple components (e.g., primary, secondary, or tertiary) or whether planets in triples differ from those in regular wide binaries.

\section{Conclusions}\label{sec:conclusions}

Using the precise astrometry provided by {\it Gaia}, we create the largest catalog of resolved triple star systems to date. The well-understood selection function and low contamination rate make the sample a powerful testbed for anchoring models of triple star formation and dynamics. In this paper, we leverage our understanding of the catalog's completeness to infer intrinsic population demographics of triple star systems. Our main conclusions are summarized below:
\begin{enumerate}
    \item{ {\it Resolved Triples Catalog:} We construct a sample of $9767$ resolved triple star systems within $500$ pc of the Sun. The triples contain main-sequence, red giant, and white dwarf components (Table \ref{tab:triple_types}), including a newly identified triple white dwarf (Figure \ref{fig:triple_images}). The triples span separations of $\sim10$ to $50,000$~au and component masses from $0.1{-}5~{\rm M_\odot}$ (Figure \ref{fig:summary_plots}). The complete catalog, along with the chance-alignment probability for each triple, is available online\footnote{\url{https://github.com/cheyanneshariat/gaia_triples/blob/main/Data/triples_catalog.csv}}.
    }
    
    \item{ {\it Twin Excess:} Inner binaries in triples show a statistically significant excess of near-equal-mass {\it twin} binaries (Figures \ref{fig:deltaG_all}) out to separations of $1000+$~au (Figure \ref{fig:twinfraction_sep_theta}). The twin properties of triple inner binaries are remarkably similar to isolated wide binaries, suggesting shared formation mechanisms between the two populations.}  

    \item{ {\it Mutual Inclinations:} Wide triples are consistent with isotropic mutual inclinations $i_{\rm mutual}$ between the inner and outer orbits, while compact triples show a slight preference towards mutual orbital alignment (Figure \ref{fig:mutual_inclinations}).
    }

    \item{ {\it Triple Inner Binaries:} The primary mass in triples ($M_1$) is reproduced from the Kroupa IMF conditioned on the triple fraction. The inner binary mass ratio ($q_{\rm in}=M_2/M_1$) orbital period ($P_{\rm in}$), and eccentricity ($e_{\rm in}$) are consistent with the distributions of binary stars (Figure \ref{fig:sampling_masses}, \ref{fig:sampled_s1s2_trip}, and \ref{fig:ecc_model}).}

    \item{ {\it Triple Outer Binaries:} The outer binary mass ratio distribution, $q_{\rm in}=M_3/(M_1+M_2)$, is reproduced assuming a power-law with logarithmic slope $\gamma=-1.4$. The tertiary's orbit is consistent with periods ($P_{\rm out}$) drawn from the binary log-normal distribution (Figure \ref{fig:sampled_s1s2_trip}) and eccentricities ($e_{\rm out}$) from a thermal distribution (Figure \ref{fig:ecc_model}), subject to dynamical stability.}    
    
    \item{ {\it Stability of Observed Triples:} Nearly all observed triples in our sample are consistent with being hierarchical and stable according to Equation \eqref{eq:eps_crit} and Equation \eqref{eq:MA_stability_crit}. 
    Mock-observing hierarchical triples recovers a fraction of visually unstable triples ($s_2/s_1<2.8$) similar to observations (Figure \ref{fig:sampled_s1s2_trip}), suggesting that most `visually' unstable triples are the result of projection effects. Notably, assuming the simpler stability criterion, $P_{\rm out}/P_{\rm in} > 5$, is inconsistent with the data.}

    \item{ {\it Triple Fraction:} After correcting for incompleteness, we estimate the triple fraction as a function of primary mass (Figure~\ref{fig:triple_fraction}). The triple fraction rises monotonically with $M_1$, from $\sim5\%$ for low-mass stars ($\lsim 0.5~{\rm M_\odot}$) to $\sim35\%$ for $2~{\rm M_\odot}$ stars. These rates are consistent with previous multiplicity studies \citep[e.g.,][]{Offner23}. }

    \item { {\it Sampling Triple Parameters:} We synthesize our results to develop a prescription for sampling synthetic triple populations\footnote{\url{https://github.com/cheyanneshariat/gaia_triples}}, anchored to observations.}

\end{enumerate}

\section{Acknowledgments}\label{acknowledgments}
We thank the anonymous referee for constructive feedback that improved the manuscript.
We thank Isaac Cheng and Tirth Surti for useful discussions.
C.S. is supported by the Joshua and Beth Friedman Foundation Fund. K.E. acknowledges support from NSF grant AST-2307232. S.N. acknowledges the partial support of NSF-BSF grant AST-2206428 and NASA XRP grant 80NSSC23K0262 as well as Howard and Astrid
Preston for their generous support.
The computations presented here were conducted in the Resnick High Performance Computing Center, a facility supported by Resnick Sustainability Institute at the California Institute of Technology
This work has made use of data from the European Space Agency (ESA) mission Gaia, processed by the Gaia Data Processing and Analysis Consortium (DPAC). 

\clearpage
\appendix 

\section{Chance Alignment Probability}\label{app:chance_alignment}

We employ $\mathcal{R}$ (Equation \ref{eq:r_chance}) to estimate the probability that a given triple is a chance alignment. Note that $\mathcal{R}$ for a triple represents the largest $\mathcal{R}_{ij}$ value of the pairs that make up the triple. We create an empirical chance alignment catalog to test the validity of interpreting $\mathcal{R}$ as a chance alignment probability. A sample of purely chance-aligned triples is created by shifting the {\it Gaia} source catalog by $0.5^\circ$ in declination and then searching for resolved triples among them. Any triples found in this shifted catalog are, by construction, purely chance alignments. Then, we use the KDE (Section \ref{sec:creating_catalog}) to derive $\mathcal{R}$ for the shifted catalog. We follow \citet{EB21_widebin} in applying a
leave-10\% out method, where the density for 10\% of binary candidates is evaluated using a KDE constructed from the other 90\%. 

Since the chance alignment probability $\mathcal{R}$ is computed for each binary pair individually, and each candidate triple consists of multiple such pairs, we define the chance alignment probability for the full triple system ($\mathcal{R}_{\rm triple}$) as the maximum $\mathcal{R}$ value among its constituent pairs. This conservative choice reflects the idea that the least secure pair dominates the confidence in the system being a true triple. For the shifted catalog, where triples are identified through two or more unrelated chance alignments involving a shared star, the likelihood of such a configuration arising randomly is effectively the product of two independent alignments. As such, we adopt $\mathcal{R}_{\rm triple}=\mathcal{R}^2$ as the relevant contamination statistic for triples identified in the shifted catalog.

\begin{figure*}[h]
    \centering
    \includegraphics[width=0.9\textwidth]{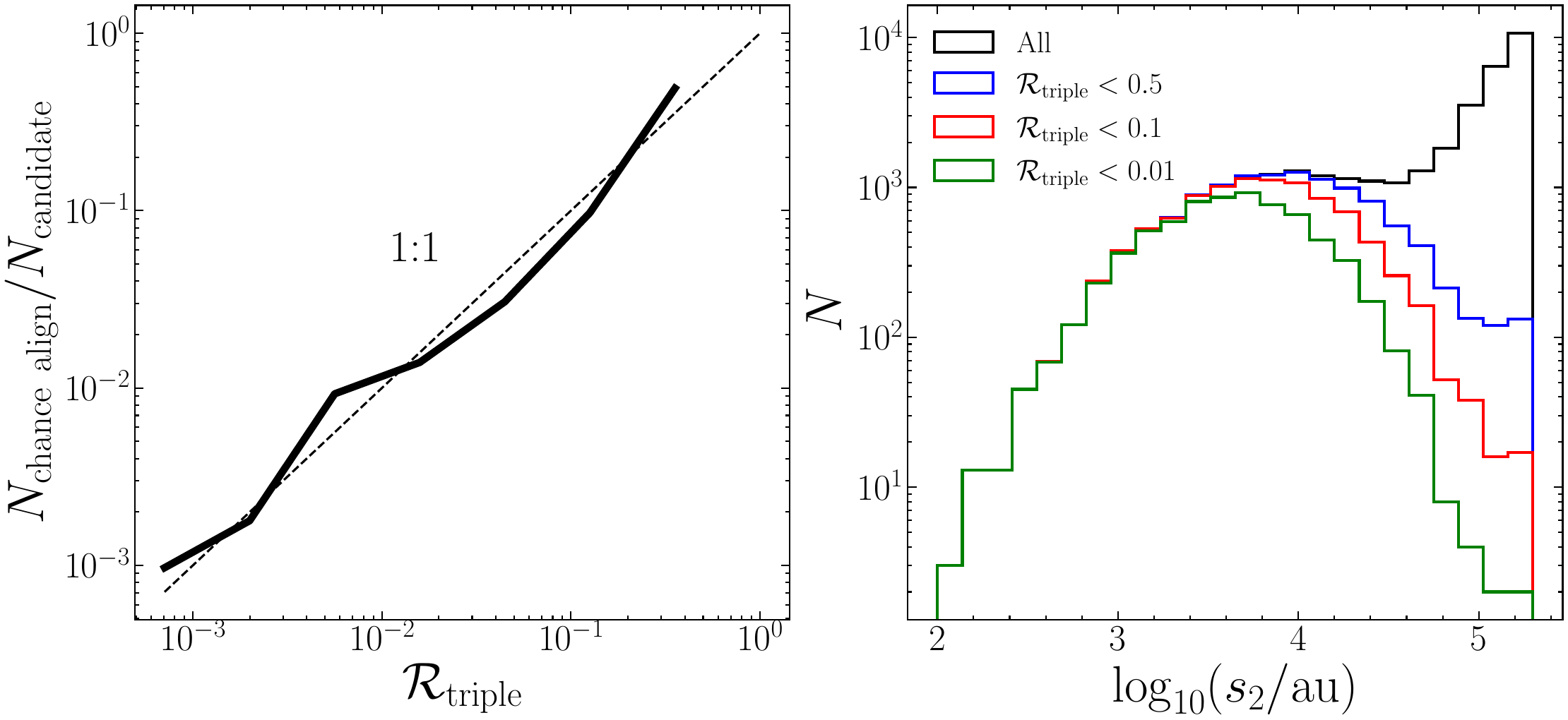}
    \caption{ The $\mathcal{R}_{\rm triple}$ statistics as a measure of chance alignment probability. {\it Left:} We plot the true number of chance alignments as a function of $\mathcal{R}_{\rm triple}$. $N_{\rm chance~align}/N_{\rm candidate}$ is the ratio between the number of chance-aligned triples (from the shifted catalog) with a given $\mathcal{R}_{\rm triple}$ compared to all triple candidates with that $\mathcal{R}_{\rm triple}$ value
    The dashed line also shows the $1:1$ line, illustrating that $\mathcal{R}_{\rm triple}$ is a solid estimate {\it Right:} Tertiary separations for different cuts on $\mathcal{R}_{\rm triple}$.
    }\label{fig:R_chance_plot} 
\end{figure*}

The left panel of Figure \ref{fig:R_chance_plot} plots the ratio between the number of the number of chance alignments
(from the shifted catalog) with a given $\mathcal{R}_{\rm triple}$ to the total number of
triple candidates with that $\mathcal{R}_{\rm triple}$: $N_{\rm chance~align}/N_{\rm candidate}$. If $\mathcal{R}_{\rm triple}$ was exactly the chance alignment probability, it would follow a $1:1$ line. $\mathcal{R}_{\rm triple}$ generally falls close to the $1:1$ line, making it a functioning definition of the chance alignment probability for triples.

In the right panel of Figure \ref{fig:R_chance_plot} we show the distribution of tertiary separations for different cuts on $\mathcal{R}_{\rm triple}$. As anticipated, chance alignments dominate at the largest separations ($s_2\gsim30,000$~au) while almost all sources with $s_2\lsim10,000$~au are real.

\section{Unmatched Pairs}\label{app:unmatched_pairs}
Our selection of resolved triples began by choosing, from a list of resolved pairs, those that share a common companion with other pairs. For triples, there exist three unique possible matches between the three possible pairings: 1-2, 1-3, and 2-3. Here, star 1 and star 2 are the stars that create the inner binary (closest pair), and the remaining, more distant tertiary is star 3. Among the $9767$ resolved triples, $6004$ had all three of these pairs match and satisfy all the cuts. For the remaining $3763$ only two pairs match in the triple, and the connection between the remaining two stars was inferred. For example, 1-2 and 2-3 were identified as confidently bound pairs by our procedure, but not 1-3. The triple will still contain all three stars, despite 1-3 not matching. The reason for the missing pairs is because of the extra orbital motion of the inner components, making them evade the proper motion cut (Equation \ref{eq:proper_motion_cut}). In this section, we explore the consistency of these `missing' pairs with the other stars in the triple to ensure high quality for the triple sample.


 The missing pair was 1-2, 1-3, 2-3 in $97$, $1635$, $2031$ of the triples, respectively. For each of the missing pairs, we calculate the uncertainty-normalized parallax difference between the components as  $\Delta \varpi / \sigma_{\Delta \varpi} = (\varpi_i - \varpi_j)/\sqrt{\sigma_{\varpi,i}^2 + \sigma_{\varpi,j}^2}$, where $i$,$j\in$ (1, 2, 3), corresponding to a triple component. The 
standard deviation in $\Delta \varpi / \sigma_{\Delta \varpi}$ for the three different types of missing pairs (1-2, 1-3, 2-3) are $2.3$, $2.7$, and $1.95$, respectively. The parallaxes of these pairs are consistent within $\approx2$–$3\sigma$, indicating that they are likely physically bound. The proper motions, however, differ more than expected because the orbital motion of the inner binary was not accounted for when searching for pairs, causing them to be missed by the simple binary selection criteria of Equation \eqref{eq:proper_motion_cut}. The missing pairs could be recovered if a looser proper motion cut were used, but this would come at the expense of more false positives.

\section{Resolvability Criteria}\label{app:ang_res_criteria}
Nearby sources in {\it Gaia} are resolved as distinct detections if their angular separation exceeds the angular resolution limit, $\theta_{\rm min}$, which depends on the difference in apparent magnitude between the two stars ($\Delta G = |G_1 - G_2|$). For example, high-contrast pairs with small angular separations are outshone by their primary and thereby remain unresolved. \citet{EB_review} presents {\it Gaia}'s sensitivity to companions as a function of angular separation ($\theta$) and $\Delta G$ in their Figure 2. We adopt their `no cuts' sensitivity curve, since we apply no additional selection to the parent sample beyond the basic {\it Gaia} query in Section~\ref{sec:creating_catalog}. The sensitivity curve continuously increases from $0$ at small $\theta$ to $1$ at large $\theta$. We therefore model the angular resolution criteria continuously by linearly interpolating the curves provided in \citet{EB_review}. As an example, $90\%$ of low contrast pairs ($\Delta G \lsim 4$) are detected with an effective angular resolution of $\theta_{\rm min}\approx1''$. On the other hand, high contrast pairs ($\Delta G \approx 10 $) are only $90\%$ detected at wide angular separations ($\theta \gsim 8''$).

For each triple, we assign a distance ($d$) according to our comparison sample of interest (often the $100$~pc sample). Then, we calculate the angular separation of the inner ($\theta_1 = s_1/d$) and outer ($\theta_2 = s_2/d$) binary. We also calculate the $\Delta G$ between all three stars from their masses (see Section \ref{subsec:triple_masses}). Then, for all pairs of stars in the triple, we use their $\theta$ and $\Delta G$ to identify {\it Gaia}'s sensitivity to that pair, which defines its detection probability. Each pair is retained according to this probability. For a triple to be completely resolved, we require that all stars pass this detection threshold.  Furthermore, we enforce that all three stars have $G \lesssim 20.5$ at their distance to ensure that they are bright enough for {\it Gaia} detection in the first place (\citealt{Gaia_Collab21b}).

\section{Wide Binary Comparison Sample}\label{app:WB_control}
\begin{figure*}
    \centering
    \includegraphics[width=0.99\textwidth]{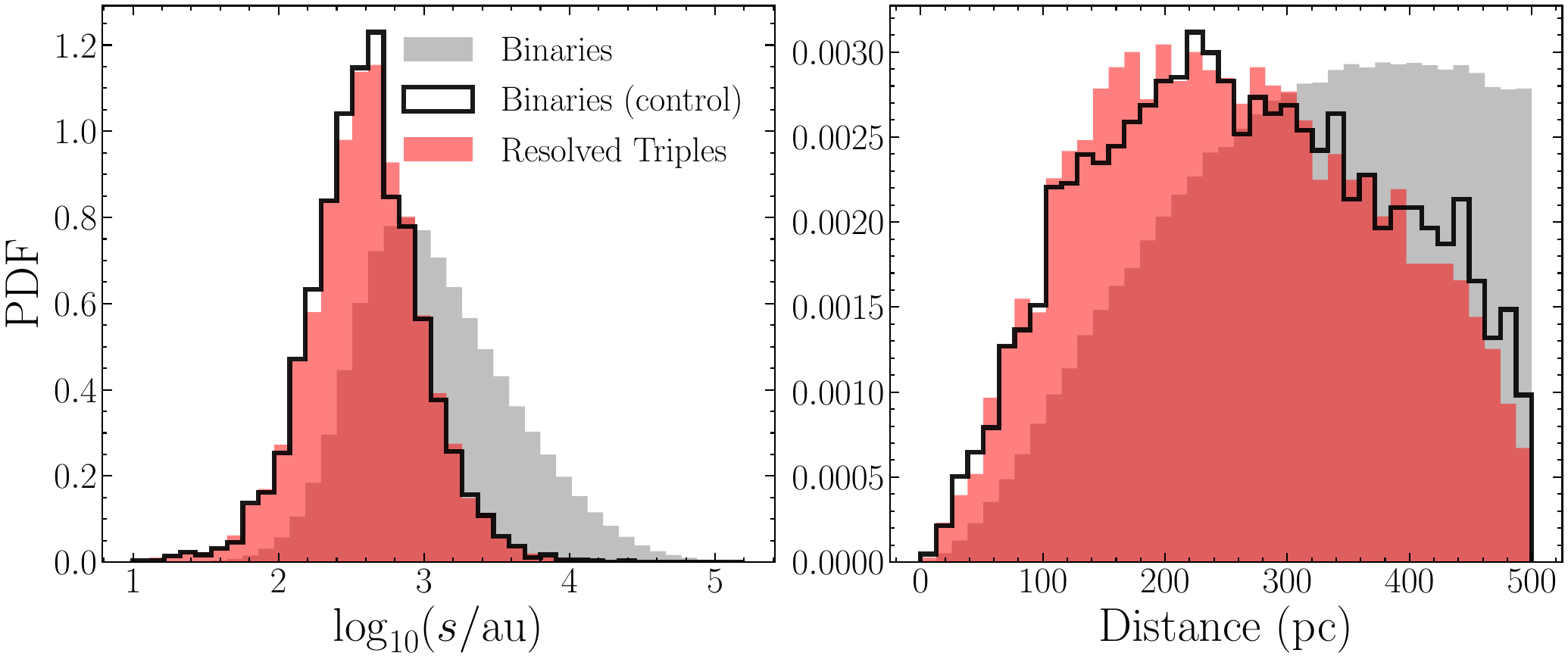}
    \caption{A wide binary control sample with similar separations and distances to triple inner binaries. On the left, we show the separation triple inner binaries (red), all wide binaries (gray), and the control sample of wide binaries (black). On the right, we show the distance distribution of the same population. The wide binary controls ample is a subset of all wide binaries where the separation and distance distribution is similar to triple inner binaries. We compare the wide binary control sample (black curves) to triples in Section \ref{subsec:comparison_to_binary}.  }
    \label{fig:WB_control_sample}
\end{figure*}

In Section \ref{subsec:comparison_to_binary}, we compare properties of the inner and outer binary of triples to regular wide binaries. This exercise allows us to identify the role of a tertiary, or triple-star evolution, in altering the present-day properties of triples. However, the inner binaries of triples have inherently different separations and distances that regular wide binaries in the field (e.g., Figure \ref{fig:summary_plots}). Therefore, we construct two samples of wide binaries with similar separations and distances to the triples, such that these samples are subject to similar systematic biases as the triples. The first sample is controlled to have the same separations and distances as the inner binaries in our 500 pc sample, while the second is controlled to have the same separations and distances as the outer binaries. 

The process of developing binary control samples can be visualized in Figure \ref{fig:WB_control_sample}. Here, we plot the separations on the left and distances on the right. The gray distributions are the wide-binary separations (left) and distances (right). Over 10,000 iterations, we sample a separation from the triple distribution (red) and select one wide binary with a separation within $5$ au of this chosen triple separation. This creates a sample of wide binaries with a nearly identical separation distribution to the triples, as shown with the black curve on the left of Figure \ref{fig:WB_control_sample}. Then, we repeat this process for triple distances, which filters the control to also have similar distances to the triples (black curve in the right panel). After applying this procedure, we now have a sample of $3254$ binaries with the same underlying separations and distances to the triples, allowing us to consistently compare the two and find differences that are not purely systematic biases.  Importantly, the distribution of angular separations $\theta=s/d$ are similar as well.

\section{Projected Separations from Orbital Elements}\label{app:projected_separations}

The projected separation of an orbit is dependent on the inclination of the orbit with respect to the line of sight and the phase of the bodies in the orbit at any given time. To convert the semi-major axes ($a_1,a_2$) of the triple orbits to projected separations ($s_1,s_2$), we statistically sample on-sky inclinations (from an isotropic distribution) and orbital phases for both the inner and outer orbit, from which we derive projected separations. To sample orbital phases, we sample the mean anomaly from a uniform distribution between $0$ and $2\pi$. Together with the mean anomaly ($M$) and eccentricity of the orbit, we numerically solve for the eccentric anomaly ($E$) using
\begin{equation}
    M = E - e  \sin(E).
\end{equation}
Lastly, we convert the eccentric anomaly to true anomaly ($\nu$) using
\begin{equation}
    \tan{(\nu/2)} =  \sqrt{\frac{1+e}{1-e}}\tan{(E/2)}.
\end{equation}
Now we can calculate the physical separation between the two stars in a Keplerian orbit using  
\begin{equation}
r = \frac{a(1 - e^2)}{1 + e\cos{(\nu)}}.
\end{equation}
This physical separation will change depending on the on-sky inclination of the orbit ($i$) and the argument of periapsis ($\omega$, chosen from a uniform distribution) with the following relation
\begin{equation}
s = r \sqrt{\cos^2{(\nu + \omega)}  \sin^2{(i)} + \sin^2{(\nu + \omega)}},
\end{equation}
This procedure is performed for both the inner and outer orbit $5$ times each, and the median separation is taken to be the observed one. This method of statistically sampling orientations and phases allows us to estimate an $s_1$ and $s_2$ for our simulated triple sample, allowing for a comparison with the $s_1$ and $s_2$ measured by {\it Gaia} observations. 

\section{Missing Wide Companions}\label{app:missing_binaries}

In Section \ref{subsec:triple_separations}, we discuss methods of sampling the tertiary separations for triples using a model. In doing so, we need to consider the truncation of systems at separations above $\sim30,000$~au. Two main effects decrease the number of observed companions at wide separations: (i) strict chance alignment cuts, which remove some real systems with $s_2\gsim 30,000$~au (Section \ref{sec:creating_catalog}) and (ii) perturbations in the Galactic field, such as stellar flybys \citep[e.g.,][]{Kaib2014, Michaely2016} and galactic tides \citep[e.g.,][]{Jiang10, Grishin22}, which unbind a fraction of wide multiples, including triples \citep[e.g.,][]{Shariat23}.

\begin{figure*}
    \centering
    \includegraphics[width=0.5\textwidth]{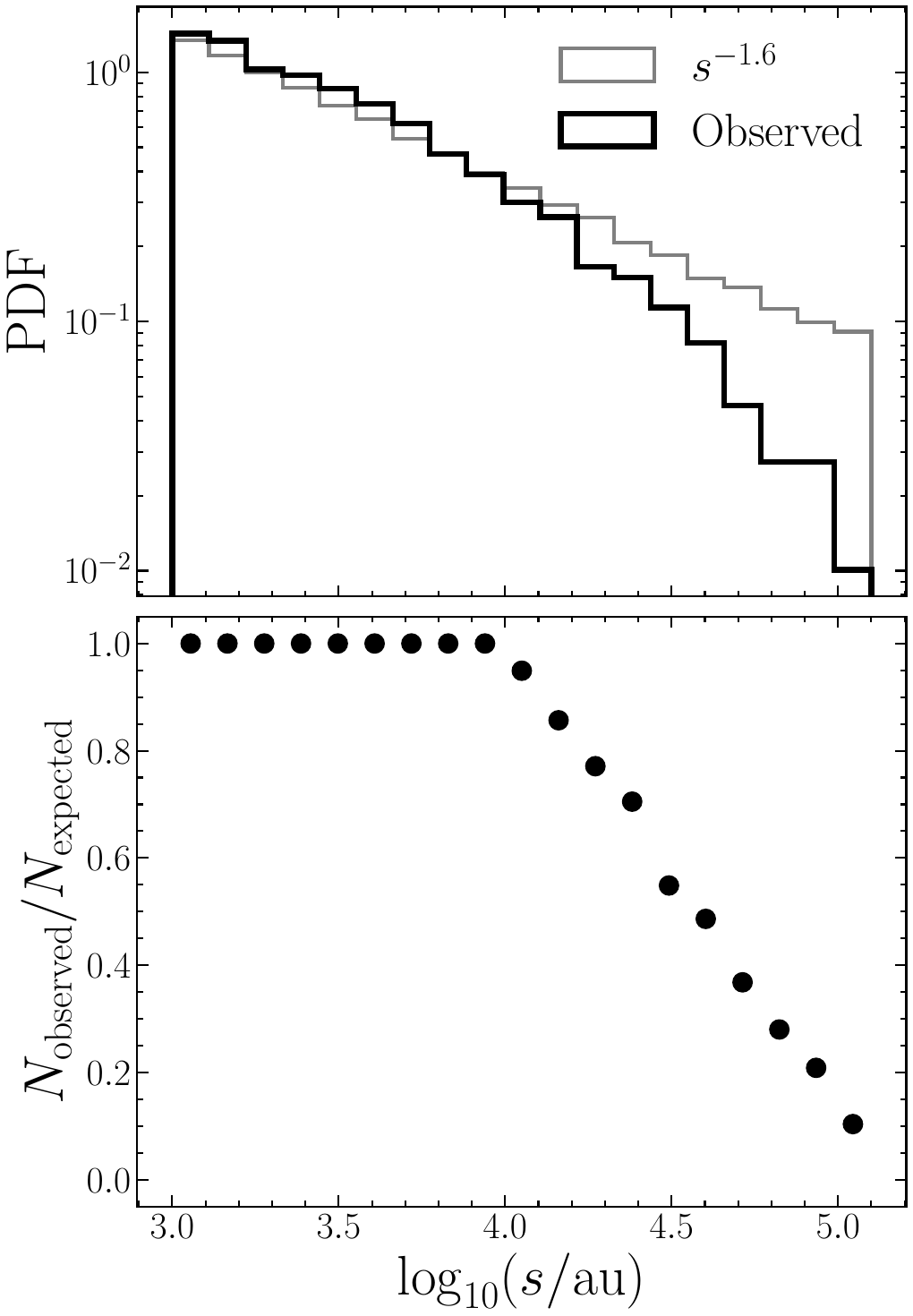}
    \caption{The absence of wide stellar companions. The top panel shows the observed separation distribution of wide binaries within $300$~pc (black) compared to a $s^{-1.6}$ power-law. The bottom shows the ratio between these two histograms at each separation bin. Above $\sim10,000$~au, the separation distributions drop below the expected power law due to the combined effect of perturbations unbinding wide companions and chance alignment cuts falsely removing real systems. }
    \label{fig:missing_binaries}
\end{figure*}

We estimate the combined effect of these using the {\it Gaia} wide binary population. We take the $d<300$~pc wide binaries from the catalog of \citep{EB21_widebin}, which applies a similar chance alignment statistic as this work. Then, we assume an intrinsic separation distribution of ${\rm d}N/{\rm d}s\propto s^{-1.6}$ for the wide binaries, motivated by observations \citep{Andrews17, EB18}. Lastly, we compare this to the observed wide binary separations (top panel of Figure \ref{fig:missing_binaries}). We find that the observed separation distribution follows the power law closely until $\sim30,000$~au, after which it declines steeply. Assuming the $s^{-1.6}$ as the intrinsic distribution, the drop-off can be explained by the combined effects of chance alignment filtering and dynamical disruption in the Galactic environment.
The ratio between the number of observed wide binaries and the number expected from an $s^{-1.6}$ power law estimates the completeness at these separations. Below $\sim30,000$~au, this fraction is $1$ (all binaries are recovered) but drops to $0.2$ at $s\sim50,000$~au.
Since wide tertiary companions are subject to the same forces as wide binaries, this approach offers a reasonable correction for the truncation of wide tertiaries in our sample.

\section{Exoplanets in Triples}\label{app:exoplanets}

After cross-matching our triple catalog to the exoplanet archive (Section \ref{subsec:exoplanets}), we identify planets that orbit stars that reside in resolved triples. In Table \ref{tab:triple_planets} we list these planets, their host star's {\it Gaia} source IDs, and which specific star in the triple they orbit (primary, secondary, or tertiary).

\startlongtable
\begin{deluxetable*}{cccc}
\tablecaption{Planets in Stellar Triples\label{tab:triple_planets}}
\tablehead{
\colhead{Host Name} &
\colhead{Gaia DR3} &
\colhead{Planet} &
\colhead{Triple} \\
\colhead{} &
\colhead{Source ID} &
\colhead{Name} &
\colhead{Component} 
}
\startdata
HD 18143    & 116037204451525376  & HD 18143 b    & Star 1           \\
HD 18143    & 116037204451525376  & HD 18143 c    & Star 1           \\
TOI-1768    & 852857845015534080  & TOI-1768.01   & Star 1           \\
TOI-4336 A  & 6113245033656232448 & TOI-4336 A b  & Star 1           \\
HD 153557   & 1408029509584967168 & HD 153557 b   & Star 2           \\
HD 153557   & 1408029509584967168 & HD 153557 c   & Star 2           \\
HD 153557   & 1408029509584967168 & HD 153557 d   & Star 2           \\
HIP 65 A    & 4923860051276772608 & HIP 65 A b    & Star 2           \\
K2-27       & 3798552815560689792 & K2-27 b       & Star 2           \\
K2-31       & 6050191241556876672 & K2-31 b       & Star 2           \\
16 Cyg B    & 2135550755683407232 & 16 Cyg B b    & Tertiary         \\
HD 40979    & 961428192989499904  & HD 40979 b    & Tertiary         \\
HD 95544    & 1133573746387113856 & HD 95544 b    & Tertiary         \\
TOI-470     & 2912264564319611136 & TOI-470 b     & Tertiary         \\
TOI-833     & 5250780970316845696 & TOI-833 b     & Tertiary         \\
V1298 Tau   & 51886335968692480   & V1298 Tau b   & Tertiary         \\
V1298 Tau   & 51886335968692480   & V1298 Tau c   & Tertiary         \\
V1298 Tau   & 51886335968692480   & V1298 Tau d   & Tertiary         \\
V1298 Tau   & 51886335968692480   & V1298 Tau e   & Tertiary         \\
WD 1856+534 & 2146576589564898688 & WD 1856+534 b & Tertiary        
\enddata
\vspace{1mm}
\end{deluxetable*}

\section{Resolved Quadruples}\label{app:resolved_quads}
In our initial graph search for resolved multiples, we recover $\sim3000$ resolved quadruples, $314$ of which have a high bound probability $\geq90\%$. 
These candidate quadruples are hierarchically structured, typically in either a 2+2 or 2+1+1 configuration.
Though rare, such wide hierarchical quadruples may serve as key case studies for theories of clustered star formation and the dynamical unfolding of early multiple systems.
We do not analyze these systems in any detail, but provide a machine readable table of resolved quadruples, along with basic properties and chance alignment probabilities for each system.

\begin{deluxetable*}{cccccc}
\centering
\tablecaption{Resolved Quadruples Catalog\label{tab:quad_preview}}
\tablehead{
\colhead{Star 1 ID} & 
\colhead{Star 2 ID} & 
\colhead{Star 3 ID} & 
\colhead{Star 4 ID} &
\colhead{\ldots} &
\colhead{\ldots}
}
\startdata
5912556128336434816	& 5912556128347850624	& 5912556849890951168	& 5912556849902360192 & \ldots & \ldots \\
2202734042876873472	& 2202734042886828160	& 2202746755990117248	& 2202746755990117376 & \ldots & \ldots \\
\ldots & \ldots & \ldots & \ldots & \ldots & \ldots \\
5515192381247219328	& 5515192385540764672	& 5515192449968480896	& 5515192454269353728 & \ldots & \ldots \\
\enddata
\tablecomments{Example rows the resolved quadruples sample. Additional columns (e.g., separations, photometry, astrometry) are omitted for brevity.}
\end{deluxetable*}

\section{Higher-Order Multiples}\label{app:higher_order_mults}

Stellar systems with more than three stars, such as quadruples, offer an important window into the most complex outcomes of star formation. Their architectures, for example, can help constrain models of hierarchical star formation \citep[e.g.,][]{Tokovinin14a, Tokovinin14b, Offner23}. 

In addition to these systems, some triples in our catalog host unresolved inner subsystems, raising their true multiplicity above three. To identify such systems, we cross-match our resolved triples with existing {\it Gaia} binary catalogs, including the non-single star (NSS) catalog and the eclipsing binary catalog \citep[][]{gaia_ecl}. These embedded subsystems—ranging from $\sim1$~au binaries to contact systems—are of particular interest since their formation may be dominated by Kozai cycles with tidal friction \citep[e.g.,][]{Fabrycky07,Naoz2014,Borkovits16_keplertriples, Toonen20}.
We first search for nested binaries among our resolved triples sample by matching to the {\it Gaia} Eclipsing Binary Catalog \citep{gaia_ecl}. This catalog contains $2,184,477$ sources with apparent G-band brightnesses ranging from a few mag down to $20$th mag across the entire sky. Along with the sources, \citet{gaia_ecl} publishes orbital periods (most between $0.2-5$~days), light curve model parameters, and a ${\tt global\_ranking}$ metric that quantifies the strength of the model fit. Only sources with ${\tt global\_ranking}>0.4$ are published, and by inspecting their ZTF light curves of several, most are visually eclipsing.

Matching these to our triples yields $142$ eclipsing binary (EB) candidates that are in resolved triples, comprising $1.5\%$ of our high-confidence sample ($\mathcal{R} < 0.1$). Extending this to a lower confidence sample of triples with $\mathcal{R} < 0.5$ adds $15$ extra matches. Among the $142$ good matches, $109$ belong to one of the components in the inner binary and $33$ belong to the tertiary. Assuming that the components are true eclipsing binaries, this would make these $142$ systems hierarchical quadruple star systems in a `2+1+1' (inner EB) or `1+1+2' (tertiary EB) configuration. Future work is warranted to investigate the eclipsing binary fraction in these systems compared to random pairing in field stars or even wide binaries. Some evidence already suggests an excess of double eclipsing binaries compared to random pairings of field stars \citep[][]{Fezenko22}. Furthermore, studying the relative frequency between EBs that are in the inner binary or tertiary may reveal clues to the early dynamical environments in protostellar clusters.

To find nested binaries with larger orbital periods and that are not necessarily eclipsing, we match our resolved triples to {\it Gaia}'s non-single-star (NSS) catalog \citep{GAIADR3}. We focus on systems with astrometric two-body orbital solutions, where both the period and eccentricity are determined. Our match results in $185$ NSS binaries that reside in resolved triples. Half ($92$) have an `SB1' solution, $43$ have an `Orbital' solution, $22$ have an `AstroSpectroSB1' solution, $19$ have `SB2' solutions, and the rest ($9$) are eclipsing binaries which overlap with the previous cross-match. The median {\tt significance} is $30$ and the median ${\tt goodness\_of\_fit}$ is $1.6$. Most ($111$) of these NSS binaries match the primary component of the triple's inner binary, and $63$ match the tertiary. We note that $52$ triples have one component with an NSS acceleration solution, a large number have components with ${\tt RUWE}>1.4$ (potentially indicating binarity). 

We provide a machine-readable table of higher order multiples with known orbital solution from the {\it Gaia} eclipsing or NSS catalogs. In total, it includes $347$ resolved triples with at least one system (multiplicity at least 4) and $6$ where two components are themselves unresolved binaries (multiplicity at least 5).
A comprehensive census of higher-order multiples, combining both resolved and unresolved components, will be useful for building a complete picture of stellar multiplicity. Note that the cut on proper motion difference (Equation \ref{eq:deltamu_criteria} and \ref{eq:mu_orbit} removes a significant fraction of unresolved binaries from the catalog, biasing those that did make it into our sample.

\begin{deluxetable*}{cccccccc}
\centering
\tablecaption{Triple Candidates with Unresolved Subsystems \label{tab:triples_higher_order}}
\tablehead{
\colhead{Primary ID} & 
\colhead{Secondary ID} & 
\colhead{Tertiary ID} & 
\colhead{$s_1$ (au)} & 
\colhead{$s_2$ (au)} & 
\colhead{Matched Stars} & 
\colhead{Solution Type} 
& \colhead{N Matched}
}
\startdata
2852562732196504320 & 2852562732195048064 & 2852562766556242816 & 373.4 & 6523.8 & Primary, Tertiary & eclipsing, Orbital & 2 \\
6185430896893625472 & 6185430931253573376 & 6185430995676548616 & 811.1 & 14609.7 & Primary, Tertiary & SB1, SB1 & 2 \\
\ldots & \ldots & \ldots & \ldots & \ldots & \ldots & \ldots & \ldots \\
4530301305194578176 & 4530301306175249096 & 4530300545702749440 & 189.0 & 1425.0 & Primary & SB1 & 1 \\
\enddata
\tablecomments{Only the first two rows and last row are shown for brevity. Solution types come from Gaia DR3 catalogs of eclipsing binaries and non-single stars. `N Matched' is the number of triple components that are unresolved binaries: for example, a value of $2$ means the system has $5$ stars total. Most ($347/353$) of these triples have only one subsystem.}
\end{deluxetable*}

\bibliography{references}

\begin{thebibliography}{}
\expandafter\ifx\csname natexlab\endcsname\relax\def\natexlab#1{#1}\fi
\providecommand{\url}[1]{\href{#1}{#1}}
\providecommand{\dodoi}[1]{doi:~\href{http://doi.org/#1}{\nolinkurl{#1}}}
\providecommand{\doeprint}[1]{\href{http://ascl.net/#1}{\nolinkurl{http://ascl.net/#1}}}
\providecommand{\doarXiv}[1]{\href{https://arxiv.org/abs/#1}{\nolinkurl{https://arxiv.org/abs/#1}}}

\bibitem[{{Allen} {et~al.}(2007){Allen}, {Koerner}, {McElwain}, {Cruz}, \& {Reid}}]{Allen07}
{Allen}, P.~R., {Koerner}, D.~W., {McElwain}, M.~W., {Cruz}, K.~L., \& {Reid}, I.~N. 2007, \aj, 133, 971, \dodoi{10.1086/510346}

\bibitem[{{Andrews} {et~al.}(2017){Andrews}, {Chanam{\'e}}, \& {Ag{\"u}eros}}]{Andrews17}
{Andrews}, J.~J., {Chanam{\'e}}, J., \& {Ag{\"u}eros}, M.~A. 2017, \mnras, 472, 675, \dodoi{10.1093/mnras/stx2000}

\bibitem[{{Anglada-Escud{\'e}} {et~al.}(2016){Anglada-Escud{\'e}}, {Amado}, {Barnes}, {Berdi{\~n}as}, {Butler}, {Coleman}, {de La Cueva}, {Dreizler}, {Endl}, {Giesers}, {Jeffers}, {Jenkins}, {Jones}, {Kiraga}, {K{\"u}rster}, {L{\'o}pez-Gonz{\'a}lez}, {Marvin}, {Morales}, {Morin}, {Nelson}, {Ortiz}, {Ofir}, {Paardekooper}, {Reiners}, {Rodr{\'\i}guez}, {Rodr{\'\i}guez-L{\'o}pez}, {Sarmiento}, {Strachan}, {Tsapras}, {Tuomi}, \& {Zechmeister}}]{Ang16}
{Anglada-Escud{\'e}}, G., {Amado}, P.~J., {Barnes}, J., {et~al.} 2016, \nat, 536, 437, \dodoi{10.1038/nature19106}

\bibitem[{{Baraffe} {et~al.}(2006){Baraffe}, {Alibert}, {Chabrier}, \& {Benz}}]{Baraffe06}
{Baraffe}, I., {Alibert}, Y., {Chabrier}, G., \& {Benz}, W. 2006, \aap, 450, 1221, \dodoi{10.1051/0004-6361:20054040}

\bibitem[{{Bashi} \& {Tokovinin}(2024)}]{Bashi24}
{Bashi}, D., \& {Tokovinin}, A. 2024, \aap, 692, A247, \dodoi{10.1051/0004-6361/202452637}

\bibitem[{{Bate}(2000)}]{Bate00}
{Bate}, M.~R. 2000, \mnras, 314, 33, \dodoi{10.1046/j.1365-8711.2000.03333.x}

\bibitem[{{Bate}(2012)}]{Bate12}
---. 2012, \mnras, 419, 3115, \dodoi{10.1111/j.1365-2966.2011.19955.x}

\bibitem[{{Bate} \& {Bonnell}(1997)}]{Bate97}
{Bate}, M.~R., \& {Bonnell}, I.~A. 1997, \mnras, 285, 33, \dodoi{10.1093/mnras/285.1.33}

\bibitem[{{Bhaskar} {et~al.}(2021){Bhaskar}, {Li}, {Hadden}, {Payne}, \& {Holman}}]{Bhaskar21}
{Bhaskar}, H., {Li}, G., {Hadden}, S., {Payne}, M.~J., \& {Holman}, M.~J. 2021, \aj, 161, 48, \dodoi{10.3847/1538-3881/abcbfc}

\bibitem[{{Binney} \& {Tremaine}(2008)}]{Binney08}
{Binney}, J., \& {Tremaine}, S. 2008, {Galactic Dynamics: Second Edition}

\bibitem[{{Borkovits} {et~al.}(2016){Borkovits}, {Hajdu}, {Sztakovics}, {Rappaport}, {Levine}, {B{\'\i}r{\'o}}, \& {Klagyivik}}]{Borkovits16_keplertriples}
{Borkovits}, T., {Hajdu}, T., {Sztakovics}, J., {et~al.} 2016, \mnras, 455, 4136, \dodoi{10.1093/mnras/stv2530}

\bibitem[{{Breivik} {et~al.}(2020){Breivik}, {Coughlin}, {Zevin}, {Rodriguez}, {Kremer}, {Ye}, {Andrews}, {Kurkowski}, {Digman}, {Larson}, \& {Rasio}}]{COSMIC}
{Breivik}, K., {Coughlin}, S., {Zevin}, M., {et~al.} 2020, \apj, 898, 71, \dodoi{10.3847/1538-4357/ab9d85}

\bibitem[{{Choi} {et~al.}(2016){Choi}, {Dotter}, {Conroy}, {Cantiello}, {Paxton}, \& {Johnson}}]{Choi16}
{Choi}, J., {Dotter}, A., {Conroy}, C., {et~al.} 2016, \apj, 823, 102, \dodoi{10.3847/0004-637X/823/2/102}

\bibitem[{{Clark} \& {Whitworth}(2021)}]{Clark21}
{Clark}, P.~C., \& {Whitworth}, A.~P. 2021, \mnras, 500, 1697, \dodoi{10.1093/mnras/staa3176}

\bibitem[{{Correia} {et~al.}(2020){Correia}, {Bourrier}, \& {Delisle}}]{Correia20}
{Correia}, A.~C.~M., {Bourrier}, V., \& {Delisle}, J.~B. 2020, \aap, 635, A37, \dodoi{10.1051/0004-6361/201936967}

\bibitem[{{Crossfield} {et~al.}(2016){Crossfield}, {Ciardi}, {Petigura}, {Sinukoff}, {Schlieder}, {Howard}, {Beichman}, {Isaacson}, {Dressing}, {Christiansen}, {Fulton}, {L{\'e}pine}, {Weiss}, {Hirsch}, {Livingston}, {Baranec}, {Law}, {Riddle}, {Ziegler}, {Howell}, {Horch}, {Everett}, {Teske}, {Martinez}, {Obermeier}, {Benneke}, {Scott}, {Deacon}, {Aller}, {Hansen}, {Mancini}, {Ciceri}, {Brahm}, {Jord{\'a}n}, {Knutson}, {Henning}, {Bonnefoy}, {Liu}, {Crepp}, {Lothringer}, {Hinz}, {Bailey}, {Skemer}, \& {Defrere}}]{Crossfield16}
{Crossfield}, I. J.~M., {Ciardi}, D.~R., {Petigura}, E.~A., {et~al.} 2016, \apjs, 226, 7, \dodoi{10.3847/0067-0049/226/1/7}

\bibitem[{{Cuntz} {et~al.}(2022){Cuntz}, {Luke}, {Millard}, {Boyle}, \& {Patel}}]{Cuntz22}
{Cuntz}, M., {Luke}, G.~E., {Millard}, M.~J., {Boyle}, L., \& {Patel}, S.~D. 2022, \apjs, 263, 33, \dodoi{10.3847/1538-4365/ac9302}

\bibitem[{{Dai} {et~al.}(2016){Dai}, {Winn}, {Albrecht}, {Arriagada}, {Bieryla}, {Butler}, {Crane}, {Hirano}, {Johnson}, {Kiilerich}, {Latham}, {Narita}, {Nowak}, {Palle}, {Ribas}, {Rogers}, {Sanchis-Ojeda}, {Shectman}, {Teske}, {Thompson}, {Van Eylen}, {Vanderburg}, {Wittenmyer}, \& {Yu}}]{Dai16}
{Dai}, F., {Winn}, J.~N., {Albrecht}, S., {et~al.} 2016, \apj, 823, 115, \dodoi{10.3847/0004-637X/823/2/115}

\bibitem[{{Duch{\^e}ne} \& {Kraus}(2013)}]{Duchene13}
{Duch{\^e}ne}, G., \& {Kraus}, A. 2013, \araa, 51, 269, \dodoi{10.1146/annurev-astro-081710-102602}

\bibitem[{{Duquennoy} \& {Mayor}(1991)}]{DM91}
{Duquennoy}, A., \& {Mayor}, M. 1991, \aap, 248, 485

\bibitem[{{El-Badry}(2024)}]{EB_review}
{El-Badry}, K. 2024, \nar, 98, 101694, \dodoi{10.1016/j.newar.2024.101694}

\bibitem[{El-Badry \& Rix(2018)}]{EB18}
El-Badry, K., \& Rix, H.-W. 2018, Monthly Notices of the Royal Astronomical Society, 480, 4884, \dodoi{10.1093/mnras/sty2186}

\bibitem[{{El-Badry} \& {Rix}(2019)}]{EB19_metal}
{El-Badry}, K., \& {Rix}, H.-W. 2019, \mnras, 482, L139, \dodoi{10.1093/mnrasl/sly206}

\bibitem[{{El-Badry} {et~al.}(2021){El-Badry}, {Rix}, \& {Heintz}}]{EB21_widebin}
{El-Badry}, K., {Rix}, H.-W., \& {Heintz}, T.~M. 2021, \mnras, 506, 2269, \dodoi{10.1093/mnras/stab323}

\bibitem[{{El-Badry} {et~al.}(2019){El-Badry}, {Rix}, {Tian}, {Duch{\^e}ne}, \& {Moe}}]{EB_19_twins}
{El-Badry}, K., {Rix}, H.-W., {Tian}, H., {Duch{\^e}ne}, G., \& {Moe}, M. 2019, \mnras, 489, 5822, \dodoi{10.1093/mnras/stz2480}

\bibitem[{{Fabrycky} \& {Tremaine}(2007)}]{Fabrycky07}
{Fabrycky}, D., \& {Tremaine}, S. 2007, \apj, 669, 1298, \dodoi{10.1086/521702}

\bibitem[{{Fezenko} {et~al.}(2022){Fezenko}, {Hwang}, \& {Zakamska}}]{Fezenko22}
{Fezenko}, G.~B., {Hwang}, H.-C., \& {Zakamska}, N.~L. 2022, \mnras, 511, 3881, \dodoi{10.1093/mnras/stac309}

\bibitem[{{Fischer} \& {Marcy}(1992)}]{Fischer92}
{Fischer}, D.~A., \& {Marcy}, G.~W. 1992, \apj, 396, 178, \dodoi{10.1086/171708}

\bibitem[{{Gaia Collaboration} {et~al.}(2016){Gaia Collaboration}, {Prusti}, {de Bruijne}, {Brown}, {Vallenari}, {Babusiaux}, {Bailer-Jones}, {Bastian}, {Biermann}, {Evans}, {Eyer}, {Jansen}, {Jordi}, {Klioner}, {Lammers}, {Lindegren}, {Luri}, {Mignard}, {Milligan}, {Panem}, {Poinsignon}, {Pourbaix}, {Randich}, {Sarri}, {Sartoretti}, {Siddiqui}, {Soubiran}, {Valette}, {van Leeuwen}, {Walton}, {Aerts}, {Arenou}, {Cropper}, {Drimmel}, {H{\o}g}, {Katz}, {Lattanzi}, {O'Mullane}, {Grebel}, {Holland}, {Huc}, {Passot}, {Bramante}, {Cacciari}, {Casta{\~n}eda}, {Chaoul}, {Cheek}, {De Angeli}, {Fabricius}, {Guerra}, {Hern{\'a}ndez}, {Jean-Antoine-Piccolo}, {Masana}, {Messineo}, {Mowlavi}, {Nienartowicz}, {Ord{\'o}{\~n}ez-Blanco}, {Panuzzo}, {Portell}, {Richards}, {Riello}, {Seabroke}, {Tanga}, {Th{\'e}venin}, {Torra}, {Els}, {Gracia-Abril}, {Comoretto}, {Garcia-Reinaldos}, {Lock}, {Mercier}, {Altmann}, {Andrae}, {Astraatmadja}, {Bellas-Velidis}, {Benson}, {Berthier}, {Blomme}, {Busso}, {Carry}, {Cellino}, {Clementini},
  {Cowell}, {Creevey}, {Cuypers}, {Davidson}, {De Ridder}, {de Torres}, {Delchambre}, {Dell'Oro}, {Ducourant}, {Fr{\'e}mat}, {Garc{\'\i}a-Torres}, {Gosset}, {Halbwachs}, {Hambly}, {Harrison}, {Hauser}, {Hestroffer}, {Hodgkin}, {Huckle}, {Hutton}, {Jasniewicz}, {Jordan}, {Kontizas}, {Korn}, {Lanzafame}, {Manteiga}, {Moitinho}, {Muinonen}, {Osinde}, {Pancino}, {Pauwels}, {Petit}, {Recio-Blanco}, {Robin}, {Sarro}, {Siopis}, {Smith}, {Smith}, {Sozzetti}, {Thuillot}, {van Reeven}, {Viala}, {Abbas}, {Abreu Aramburu}, {Accart}, {Aguado}, {Allan}, {Allasia}, {Altavilla}, {{\'A}lvarez}, {Alves}, {Anderson}, {Andrei}, {Anglada Varela}, {Antiche}, {Antoja}, {Ant{\'o}n}, {Arcay}, {Atzei}, {Ayache}, {Bach}, {Baker}, {Balaguer-N{\'u}{\~n}ez}, {Barache}, {Barata}, {Barbier}, {Barblan}, {Baroni}, {Barrado y Navascu{\'e}s}, {Barros}, {Barstow}, {Becciani}, {Bellazzini}, {Bellei}, {Bello Garc{\'\i}a}, {Belokurov}, {Bendjoya}, {Berihuete}, {Bianchi}, {Bienaym{\'e}}, {Billebaud}, {Blagorodnova}, {Blanco-Cuaresma}, {Boch},
  {Bombrun}, {Borrachero}, {Bouquillon}, {Bourda}, {Bouy}, {Bragaglia}, {Breddels}, {Brouillet}, {Br{\"u}semeister}, {Bucciarelli}, {Budnik}, {Burgess}, {Burgon}, {Burlacu}, {Busonero}, {Buzzi}, {Caffau}, {Cambras}, {Campbell}, {Cancelliere}, {Cantat-Gaudin}, {Carlucci}, {Carrasco}, {Castellani}, {Charlot}, {Charnas}, {Charvet}, {Chassat}, {Chiavassa}, {Clotet}, {Cocozza}, {Collins}, {Collins}, {Costigan}, {Crifo}, {Cross}, {Crosta}, {Crowley}, {Dafonte}, {Damerdji}, {Dapergolas}, {David}, {David}, {De Cat}, {de Felice}, {de Laverny}, {De Luise}, {De March}, {de Martino}, {de Souza}, {Debosscher}, {del Pozo}, {Delbo}, {Delgado}, {Delgado}, {di Marco}, {Di Matteo}, {Diakite}, {Distefano}, {Dolding}, {Dos Anjos}, {Drazinos}, {Dur{\'a}n}, {Dzigan}, {Ecale}, {Edvardsson}, {Enke}, {Erdmann}, {Escolar}, {Espina}, {Evans}, {Eynard Bontemps}, {Fabre}, {Fabrizio}, {Faigler}, {Falc{\~a}o}, {Farr{\`a}s Casas}, {Faye}, {Federici}, {Fedorets}, {Fern{\'a}ndez-Hern{\'a}ndez}, {Fernique}, {Fienga}, {Figueras}, {Filippi},
  {Findeisen}, {Fonti}, {Fouesneau}, {Fraile}, {Fraser}, {Fuchs}, {Furnell}, {Gai}, {Galleti}, {Galluccio}, {Garabato}, {Garc{\'\i}a-Sedano}, {Gar{\'e}}, {Garofalo}, {Garralda}, {Gavras}, {Gerssen}, {Geyer}, {Gilmore}, {Girona}, {Giuffrida}, {Gomes}, {Gonz{\'a}lez-Marcos}, {Gonz{\'a}lez-N{\'u}{\~n}ez}, {Gonz{\'a}lez-Vidal}, {Granvik}, {Guerrier}, {Guillout}, {Guiraud}, {G{\'u}rpide}, {Guti{\'e}rrez-S{\'a}nchez}, {Guy}, {Haigron}, {Hatzidimitriou}, {Haywood}, {Heiter}, {Helmi}, {Hobbs}, {Hofmann}, {Holl}, {Holland}, {Hunt}, {Hypki}, {Icardi}, {Irwin}, {Jevardat de Fombelle}, {Jofr{\'e}}, {Jonker}, {Jorissen}, {Julbe}, {Karampelas}, {Kochoska}, {Kohley}, {Kolenberg}, {Kontizas}, {Koposov}, {Kordopatis}, {Koubsky}, {Kowalczyk}, {Krone-Martins}, {Kudryashova}, {Kull}, {Bachchan}, {Lacoste-Seris}, {Lanza}, {Lavigne}, {Le Poncin-Lafitte}, {Lebreton}, {Lebzelter}, {Leccia}, {Leclerc}, {Lecoeur-Taibi}, {Lemaitre}, {Lenhardt}, {Leroux}, {Liao}, {Licata}, {Lindstr{\o}m}, {Lister}, {Livanou}, {Lobel}, {L{\"o}ffler},
  {L{\'o}pez}, {Lopez-Lozano}, {Lorenz}, {Loureiro}, {MacDonald}, {Magalh{\~a}es Fernandes}, {Managau}, {Mann}, {Mantelet}, {Marchal}, {Marchant}, {Marconi}, {Marie}, {Marinoni}, {Marrese}, {Marschalk{\'o}}, {Marshall}, {Mart{\'\i}n-Fleitas}, {Martino}, {Mary}, {Matijevi{\v{c}}}, {Mazeh}, {McMillan}, {Messina}, {Mestre}, {Michalik}, {Millar}, {Miranda}, {Molina}, {Molinaro}, {Molinaro}, {Moln{\'a}r}, {Moniez}, {Montegriffo}, {Monteiro}, {Mor}, {Mora}, {Morbidelli}, {Morel}, {Morgenthaler}, {Morley}, {Morris}, {Mulone}, {Muraveva}, {Musella}, {Narbonne}, {Nelemans}, {Nicastro}, {Noval}, {Ord{\'e}novic}, {Ordieres-Mer{\'e}}, {Osborne}, {Pagani}, {Pagano}, {Pailler}, {Palacin}, {Palaversa}, {Parsons}, {Paulsen}, {Pecoraro}, {Pedrosa}, {Pentik{\"a}inen}, {Pereira}, {Pichon}, {Piersimoni}, {Pineau}, {Plachy}, {Plum}, {Poujoulet}, {Pr{\v{s}}a}, {Pulone}, {Ragaini}, {Rago}, {Rambaux}, {Ramos-Lerate}, {Ranalli}, {Rauw}, {Read}, {Regibo}, {Renk}, {Reyl{\'e}}, {Ribeiro}, {Rimoldini}, {Ripepi}, {Riva}, {Rixon},
  {Roelens}, {Romero-G{\'o}mez}, {Rowell}, {Royer}, {Rudolph}, {Ruiz-Dern}, {Sadowski}, {Sagrist{\`a} Sell{\'e}s}, {Sahlmann}, {Salgado}, {Salguero}, {Sarasso}, {Savietto}, {Schnorhk}, {Schultheis}, {Sciacca}, {Segol}, {Segovia}, {Segransan}, {Serpell}, {Shih}, {Smareglia}, {Smart}, {Smith}, {Solano}, {Solitro}, {Sordo}, {Soria Nieto}, {Souchay}, {Spagna}, {Spoto}, {Stampa}, {Steele}, {Steidelm{\"u}ller}, {Stephenson}, {Stoev}, {Suess}, {S{\"u}veges}, {Surdej}, {Szabados}, {Szegedi-Elek}, {Tapiador}, {Taris}, {Tauran}, {Taylor}, {Teixeira}, {Terrett}, {Tingley}, {Trager}, {Turon}, {Ulla}, {Utrilla}, {Valentini}, {van Elteren}, {Van Hemelryck}, {van Leeuwen}, {Varadi}, {Vecchiato}, {Veljanoski}, {Via}, {Vicente}, {Vogt}, {Voss}, {Votruba}, {Voutsinas}, {Walmsley}, {Weiler}, {Weingrill}, {Werner}, {Wevers}, {Whitehead}, {Wyrzykowski}, {Yoldas}, {{\v{Z}}erjal}, {Zucker}, {Zurbach}, {Zwitter}, {Alecu}, {Allen}, {Allende Prieto}, {Amorim}, {Anglada-Escud{\'e}}, {Arsenijevic}, {Azaz}, {Balm}, {Beck}, {Bernstein},
  {Bigot}, {Bijaoui}, {Blasco}, {Bonfigli}, {Bono}, {Boudreault}, {Bressan}, {Brown}, {Brunet}, {Bunclark}, {Buonanno}, {Butkevich}, {Carret}, {Carrion}, {Chemin}, {Ch{\'e}reau}, {Corcione}, {Darmigny}, {de Boer}, {de Teodoro}, {de Zeeuw}, {Delle Luche}, {Domingues}, {Dubath}, {Fodor}, {Fr{\'e}zouls}, {Fries}, {Fustes}, {Fyfe}, {Gallardo}, {Gallegos}, {Gardiol}, {Gebran}, {Gomboc}, {G{\'o}mez}, {Grux}, {Gueguen}, {Heyrovsky}, {Hoar}, {Iannicola}, {Isasi Parache}, {Janotto}, {Joliet}, {Jonckheere}, {Keil}, {Kim}, {Klagyivik}, {Klar}, {Knude}, {Kochukhov}, {Kolka}, {Kos}, {Kutka}, {Lainey}, {LeBouquin}, {Liu}, {Loreggia}, {Makarov}, {Marseille}, {Martayan}, {Martinez-Rubi}, {Massart}, {Meynadier}, {Mignot}, {Munari}, {Nguyen}, {Nordlander}, {Ocvirk}, {O'Flaherty}, {Olias Sanz}, {Ortiz}, {Osorio}, {Oszkiewicz}, {Ouzounis}, {Palmer}, {Park}, {Pasquato}, {Peltzer}, {Peralta}, {P{\'e}turaud}, {Pieniluoma}, {Pigozzi}, {Poels}, {Prat}, {Prod'homme}, {Raison}, {Rebordao}, {Risquez}, {Rocca-Volmerange}, {Rosen},
  {Ruiz-Fuertes}, {Russo}, {Sembay}, {Serraller Vizcaino}, {Short}, {Siebert}, {Silva}, {Sinachopoulos}, {Slezak}, {Soffel}, {Sosnowska}, {Strai{\v{z}}ys}, {ter Linden}, {Terrell}, {Theil}, {Tiede}, {Troisi}, {Tsalmantza}, {Tur}, {Vaccari}, {Vachier}, {Valles}, {Van Hamme}, {Veltz}, {Virtanen}, {Wallut}, {Wichmann}, {Wilkinson}, {Ziaeepour}, \& {Zschocke}}]{Gaia_Collab}
{Gaia Collaboration}, {Prusti}, T., {de Bruijne}, J.~H.~J., {et~al.} 2016, \aap, 595, A1, \dodoi{10.1051/0004-6361/201629272}

\bibitem[{{Gaia Collaboration} {et~al.}(2021{\natexlab{a}}){Gaia Collaboration}, {Smart}, {Sarro}, {Rybizki}, {Reyl{\'e}}, {Robin}, {Hambly}, {Abbas}, {Barstow}, {de Bruijne}, {Bucciarelli}, {Carrasco}, {Cooper}, {Hodgkin}, {Masana}, {Michalik}, {Sahlmann}, {Sozzetti}, {Brown}, {Vallenari}, {Prusti}, {Babusiaux}, {Biermann}, {Creevey}, {Evans}, {Eyer}, {Hutton}, {Jansen}, {Jordi}, {Klioner}, {Lammers}, {Lindegren}, {Luri}, {Mignard}, {Panem}, {Pourbaix}, {Randich}, {Sartoretti}, {Soubiran}, {Walton}, {Arenou}, {Bailer-Jones}, {Bastian}, {Cropper}, {Drimmel}, {Katz}, {Lattanzi}, {van Leeuwen}, {Bakker}, {Casta{\~n}eda}, {De Angeli}, {Ducourant}, {Fabricius}, {Fouesneau}, {Fr{\'e}mat}, {Guerra}, {Guerrier}, {Guiraud}, {Jean-Antoine Piccolo}, {Messineo}, {Mowlavi}, {Nicolas}, {Nienartowicz}, {Pailler}, {Panuzzo}, {Riclet}, {Roux}, {Seabroke}, {Sordo}, {Tanga}, {Th{\'e}venin}, {Gracia-Abril}, {Portell}, {Teyssier}, {Altmann}, {Andrae}, {Bellas-Velidis}, {Benson}, {Berthier}, {Blomme}, {Brugaletta}, {Burgess},
  {Busso}, {Carry}, {Cellino}, {Cheek}, {Clementini}, {Damerdji}, {Davidson}, {Delchambre}, {Dell'Oro}, {Fern{\'a}ndez-Hern{\'a}ndez}, {Galluccio}, {Garc{\'\i}a-Lario}, {Garcia-Reinaldos}, {Gonz{\'a}lez-N{\'u}{\~n}ez}, {Gosset}, {Haigron}, {Halbwachs}, {Harrison}, {Hatzidimitriou}, {Heiter}, {Hern{\'a}ndez}, {Hestroffer}, {Holl}, {Jan{\ss}en}, {Jevardat de Fombelle}, {Jordan}, {Krone-Martins}, {Lanzafame}, {L{\"o}ffler}, {Lorca}, {Manteiga}, {Marchal}, {Marrese}, {Moitinho}, {Mora}, {Muinonen}, {Osborne}, {Pancino}, {Pauwels}, {Recio-Blanco}, {Richards}, {Riello}, {Rimoldini}, {Roegiers}, {Siopis}, {Smith}, {Ulla}, {Utrilla}, {van Leeuwen}, {van Reeven}, {Abreu Aramburu}, {Accart}, {Aerts}, {Aguado}, {Ajaj}, {Altavilla}, {{\'A}lvarez}, {{\'A}lvarez Cid-Fuentes}, {Alves}, {Anderson}, {Anglada Varela}, {Antoja}, {Audard}, {Baines}, {Baker}, {Balaguer-N{\'u}{\~n}ez}, {Balbinot}, {Balog}, {Barache}, {Barbato}, {Barros}, {Bartolom{\'e}}, {Bassilana}, {Bauchet}, {Baudesson-Stella}, {Becciani}, {Bellazzini},
  {Bernet}, {Bertone}, {Bianchi}, {Blanco-Cuaresma}, {Boch}, {Bombrun}, {Bossini}, {Bouquillon}, {Bragaglia}, {Bramante}, {Breedt}, {Bressan}, {Brouillet}, {Burlacu}, {Busonero}, {Butkevich}, {Buzzi}, {Caffau}, {Cancelliere}, {C{\'a}novas}, {Cantat-Gaudin}, {Carballo}, {Carlucci}, {Carnerero}, {Casamiquela}, {Castellani}, {Castro-Ginard}, {Castro Sampol}, {Chaoul}, {Charlot}, {Chemin}, {Chiavassa}, {Cioni}, {Comoretto}, {Cornez}, {Cowell}, {Crifo}, {Crosta}, {Crowley}, {Dafonte}, \& {Dapergolas}}]{Gaia_Collab21b}
{Gaia Collaboration}, {Smart}, R.~L., {Sarro}, L.~M., {et~al.} 2021{\natexlab{a}}, \aap, 649, A6, \dodoi{10.1051/0004-6361/202039498}

\bibitem[{{Gaia Collaboration} {et~al.}(2021{\natexlab{b}}){Gaia Collaboration}, {Brown}, {Vallenari}, {Prusti}, {de Bruijne}, {Babusiaux}, {Biermann}, {Creevey}, {Evans}, {Eyer}, {Hutton}, {Jansen}, {Jordi}, {Klioner}, {Lammers}, {Lindegren}, {Luri}, {Mignard}, {Panem}, {Pourbaix}, {Randich}, {Sartoretti}, {Soubiran}, {Walton}, {Arenou}, {Bailer-Jones}, {Bastian}, {Cropper}, {Drimmel}, {Katz}, {Lattanzi}, {van Leeuwen}, {Bakker}, {Cacciari}, {Casta{\~n}eda}, {De Angeli}, {Ducourant}, {Fabricius}, {Fouesneau}, {Fr{\'e}mat}, {Guerra}, {Guerrier}, {Guiraud}, {Jean-Antoine Piccolo}, {Masana}, {Messineo}, {Mowlavi}, {Nicolas}, {Nienartowicz}, {Pailler}, {Panuzzo}, {Riclet}, {Roux}, {Seabroke}, {Sordo}, {Tanga}, {Th{\'e}venin}, {Gracia-Abril}, {Portell}, {Teyssier}, {Altmann}, {Andrae}, {Bellas-Velidis}, {Benson}, {Berthier}, {Blomme}, {Brugaletta}, {Burgess}, {Busso}, {Carry}, {Cellino}, {Cheek}, {Clementini}, {Damerdji}, {Davidson}, {Delchambre}, {Dell'Oro}, {Fern{\'a}ndez-Hern{\'a}ndez}, {Galluccio},
  {Garc{\'\i}a-Lario}, {Garcia-Reinaldos}, {Gonz{\'a}lez-N{\'u}{\~n}ez}, {Gosset}, {Haigron}, {Halbwachs}, {Hambly}, {Harrison}, {Hatzidimitriou}, {Heiter}, {Hern{\'a}ndez}, {Hestroffer}, {Hodgkin}, {Holl}, {Jan{\ss}en}, {Jevardat de Fombelle}, {Jordan}, {Krone-Martins}, {Lanzafame}, {L{\"o}ffler}, {Lorca}, {Manteiga}, {Marchal}, {Marrese}, {Moitinho}, {Mora}, {Muinonen}, {Osborne}, {Pancino}, {Pauwels}, {Petit}, {Recio-Blanco}, {Richards}, {Riello}, {Rimoldini}, {Robin}, {Roegiers}, {Rybizki}, {Sarro}, {Siopis}, {Smith}, {Sozzetti}, {Ulla}, {Utrilla}, {van Leeuwen}, {van Reeven}, {Abbas}, {Abreu Aramburu}, {Accart}, {Aerts}, {Aguado}, {Ajaj}, {Altavilla}, {{\'A}lvarez}, {{\'A}lvarez Cid-Fuentes}, {Alves}, {Anderson}, {Anglada Varela}, {Antoja}, {Audard}, {Baines}, {Baker}, {Balaguer-N{\'u}{\~n}ez}, {Balbinot}, {Balog}, {Barache}, {Barbato}, {Barros}, {Barstow}, {Bartolom{\'e}}, {Bassilana}, {Bauchet}, {Baudesson-Stella}, {Becciani}, {Bellazzini}, {Bernet}, {Bertone}, {Bianchi}, {Blanco-Cuaresma}, {Boch},
  {Bombrun}, {Bossini}, {Bouquillon}, {Bragaglia}, {Bramante}, {Breedt}, {Bressan}, {Brouillet}, {Bucciarelli}, {Burlacu}, {Busonero}, {Butkevich}, {Buzzi}, {Caffau}, {Cancelliere}, {C{\'a}novas}, {Cantat-Gaudin}, {Carballo}, {Carlucci}, {Carnerero}, {Carrasco}, {Casamiquela}, {Castellani}, {Castro-Ginard}, {Castro Sampol}, {Chaoul}, {Charlot}, {Chemin}, {Chiavassa}, {Cioni}, {Comoretto}, {Cooper}, {Cornez}, {Cowell}, {Crifo}, {Crosta}, {Crowley}, {Dafonte}, {Dapergolas}, {David}, {David}, {de Laverny}, {De Luise}, {De March}, {De Ridder}, {de Souza}, {de Teodoro}, {de Torres}, {del Peloso}, {del Pozo}, {Delbo}, {Delgado}, {Delgado}, {Delisle}, {Di Matteo}, {Diakite}, {Diener}, {Distefano}, {Dolding}, {Eappachen}, {Edvardsson}, {Enke}, {Esquej}, {Fabre}, {Fabrizio}, {Faigler}, {Fedorets}, {Fernique}, {Fienga}, {Figueras}, {Fouron}, {Fragkoudi}, {Fraile}, {Franke}, {Gai}, {Garabato}, {Garcia-Gutierrez}, {Garc{\'\i}a-Torres}, {Garofalo}, {Gavras}, {Gerlach}, {Geyer}, {Giacobbe}, {Gilmore}, {Girona},
  {Giuffrida}, {Gomel}, {Gomez}, {Gonzalez-Santamaria}, {Gonz{\'a}lez-Vidal}, {Granvik}, {Guti{\'e}rrez-S{\'a}nchez}, {Guy}, {Hauser}, {Haywood}, {Helmi}, {Hidalgo}, {Hilger}, {H{\l}adczuk}, {Hobbs}, {Holland}, {Huckle}, {Jasniewicz}, {Jonker}, {Juaristi Campillo}, {Julbe}, {Karbevska}, {Kervella}, {Khanna}, {Kochoska}, {Kontizas}, {Kordopatis}, {Korn}, {Kostrzewa-Rutkowska}, {Kruszy{\'n}ska}, {Lambert}, {Lanza}, {Lasne}, {Le Campion}, {Le Fustec}, {Lebreton}, {Lebzelter}, {Leccia}, {Leclerc}, {Lecoeur-Taibi}, {Liao}, {Licata}, {Lindstr{\o}m}, {Lister}, {Livanou}, {Lobel}, {Madrero Pardo}, {Managau}, {Mann}, {Marchant}, {Marconi}, {Marcos Santos}, {Marinoni}, {Marocco}, {Marshall}, {Martin Polo}, {Mart{\'\i}n-Fleitas}, {Masip}, {Massari}, {Mastrobuono-Battisti}, {Mazeh}, {McMillan}, {Messina}, {Michalik}, {Millar}, {Mints}, {Molina}, {Molinaro}, {Moln{\'a}r}, {Montegriffo}, {Mor}, {Morbidelli}, {Morel}, {Morris}, {Mulone}, {Munoz}, {Muraveva}, {Murphy}, {Musella}, {Noval}, {Ord{\'e}novic}, {Orr{\`u}},
  {Osinde}, {Pagani}, {Pagano}, {Palaversa}, {Palicio}, {Panahi}, {Pawlak}, {Pe{\~n}alosa Esteller}, {Penttil{\"a}}, {Piersimoni}, {Pineau}, {Plachy}, {Plum}, {Poggio}, {Poretti}, {Poujoulet}, {Pr{\v{s}}a}, {Pulone}, {Racero}, {Ragaini}, {Rainer}, {Raiteri}, {Rambaux}, {Ramos}, {Ramos-Lerate}, {Re Fiorentin}, {Regibo}, {Reyl{\'e}}, {Ripepi}, {Riva}, {Rixon}, {Robichon}, {Robin}, {Roelens}, {Rohrbasser}, {Romero-G{\'o}mez}, {Rowell}, {Royer}, {Rybicki}, {Sadowski}, {Sagrist{\`a} Sell{\'e}s}, {Sahlmann}, {Salgado}, {Salguero}, {Samaras}, {Sanchez Gimenez}, {Sanna}, {Santove{\~n}a}, {Sarasso}, {Schultheis}, {Sciacca}, {Segol}, {Segovia}, {S{\'e}gransan}, {Semeux}, {Shahaf}, {Siddiqui}, {Siebert}, {Siltala}, {Slezak}, {Smart}, {Solano}, {Solitro}, {Souami}, {Souchay}, {Spagna}, {Spoto}, {Steele}, {Steidelm{\"u}ller}, {Stephenson}, {S{\"u}veges}, {Szabados}, {Szegedi-Elek}, {Taris}, {Tauran}, {Taylor}, {Teixeira}, {Thuillot}, {Tonello}, {Torra}, {Torra}, {Turon}, {Unger}, {Vaillant}, {van Dillen}, {Vanel},
  {Vecchiato}, {Viala}, {Vicente}, {Voutsinas}, {Weiler}, {Wevers}, {Wyrzykowski}, {Yoldas}, {Yvard}, {Zhao}, {Zorec}, {Zucker}, {Zurbach}, \& {Zwitter}}]{GAIADR3}
{Gaia Collaboration}, {Brown}, A.~G.~A., {Vallenari}, A., {et~al.} 2021{\natexlab{b}}, \aap, 649, A1, \dodoi{10.1051/0004-6361/202039657}

\bibitem[{{Gao} {et~al.}(2023){Gao}, {Toonen}, \& {Leigh}}]{Gao23}
{Gao}, Y., {Toonen}, S., \& {Leigh}, N. 2023, \mnras, 518, 526, \dodoi{10.1093/mnras/stac3068}

\bibitem[{{Gonz{\'a}lez-Payo} {et~al.}(2024){Gonz{\'a}lez-Payo}, {Caballero}, {Gorgas}, {Cort{\'e}s-Contreras}, {G{\'a}lvez-Ortiz}, \& {Cifuentes}}]{Gonzalez-Payo24}
{Gonz{\'a}lez-Payo}, J., {Caballero}, J.~A., {Gorgas}, J., {et~al.} 2024, \aap, 689, A302, \dodoi{10.1051/0004-6361/202450048}

\bibitem[{{Grishin} \& {Perets}(2022)}]{Grishin22}
{Grishin}, E., \& {Perets}, H.~B. 2022, \mnras, 512, 4993, \dodoi{10.1093/mnras/stac706}

\bibitem[{{Grishin} {et~al.}(2017){Grishin}, {Perets}, {Zenati}, \& {Michaely}}]{Grishin17}
{Grishin}, E., {Perets}, H.~B., {Zenati}, Y., \& {Michaely}, E. 2017, \mnras, 466, 276, \dodoi{10.1093/mnras/stw3096}

\bibitem[{{Grziwa} {et~al.}(2016){Grziwa}, {Gandolfi}, {Csizmadia}, {Fridlund}, {Parviainen}, {Deeg}, {Cabrera}, {Djupvik}, {Albrecht}, {Palle}, {P{\"a}tzold}, {B{\'e}jar}, {Prieto-Arranz}, {Eigm{\"u}ller}, {Erikson}, {Fynbo}, {Guenther}, {Hatzes}, {Kiilerich}, {Korth}, {Kuutma}, {Monta{\~n}{\'e}s-Rodr{\'\i}guez}, {Nespral}, {Nowak}, {Rauer}, {Saario}, {Sebastian}, \& {Slumstrup}}]{Grziwa16}
{Grziwa}, S., {Gandolfi}, D., {Csizmadia}, S., {et~al.} 2016, \aj, 152, 132, \dodoi{10.3847/0004-6256/152/5/132}

\bibitem[{{Guszejnov} {et~al.}(2017){Guszejnov}, {Hopkins}, \& {Krumholz}}]{Guszejnov17}
{Guszejnov}, D., {Hopkins}, P.~F., \& {Krumholz}, M.~R. 2017, \mnras, 468, 4093, \dodoi{10.1093/mnras/stx725}

\bibitem[{{Hamers} {et~al.}(2022){Hamers}, {Glanz}, \& {Neunteufel}}]{Hamers22}
{Hamers}, A.~S., {Glanz}, H., \& {Neunteufel}, P. 2022, \apjs, 259, 25, \dodoi{10.3847/1538-4365/ac49e7}

\bibitem[{{Hamilton}(2022)}]{Hamilton22_widebinecc}
{Hamilton}, C. 2022, \apjl, 929, L29, \dodoi{10.3847/2041-8213/ac6600}

\bibitem[{{Hartman} \& {L{\'e}pine}(2020)}]{Hartman20}
{Hartman}, Z.~D., \& {L{\'e}pine}, S. 2020, \apjs, 247, 66, \dodoi{10.3847/1538-4365/ab79a6}

\bibitem[{{Heintz} {et~al.}(2022){Heintz}, {Hermes}, {El-Badry}, {Walsh}, {van Saders}, {Fields}, \& {Koester}}]{Heintz22}
{Heintz}, T.~M., {Hermes}, J.~J., {El-Badry}, K., {et~al.} 2022, \apj, 934, 148, \dodoi{10.3847/1538-4357/ac78d9}

\bibitem[{{Heintz} {et~al.}(2024){Heintz}, {Hermes}, {Tremblay}, {Baya Ould Rouis}, {Redding}, {Kaiser}, \& {van Saders}}]{Heintz24}
{Heintz}, T.~M., {Hermes}, J.~J., {Tremblay}, P.~E., {et~al.} 2024, arXiv e-prints, arXiv:2405.02423.
\newblock \doarXiv{2405.02423}

\bibitem[{{Hwang} {et~al.}(2022{\natexlab{a}}){Hwang}, {El-Badry}, {Rix}, {Hamilton}, {Ting}, \& {Zakamska}}]{Hwang22_twinecc}
{Hwang}, H.-C., {El-Badry}, K., {Rix}, H.-W., {et~al.} 2022{\natexlab{a}}, \apjl, 933, L32, \dodoi{10.3847/2041-8213/ac7c70}

\bibitem[{{Hwang} {et~al.}(2021){Hwang}, {Ting}, {Schlaufman}, {Zakamska}, \& {Wyse}}]{Hwang21_metal}
{Hwang}, H.-C., {Ting}, Y.-S., {Schlaufman}, K.~C., {Zakamska}, N.~L., \& {Wyse}, R. F.~G. 2021, \mnras, 501, 4329, \dodoi{10.1093/mnras/staa3854}

\bibitem[{{Hwang} {et~al.}(2022{\natexlab{b}}){Hwang}, {Ting}, \& {Zakamska}}]{Hwang22_ecc}
{Hwang}, H.-C., {Ting}, Y.-S., \& {Zakamska}, N.~L. 2022{\natexlab{b}}, \mnras, 512, 3383, \dodoi{10.1093/mnras/stac675}

\bibitem[{{Jiang} \& {Tremaine}(2010)}]{Jiang10}
{Jiang}, Y.-F., \& {Tremaine}, S. 2010, \mnras, 401, 977, \dodoi{10.1111/j.1365-2966.2009.15744.x}

\bibitem[{{Joncour} {et~al.}(2017){Joncour}, {Duch{\^e}ne}, \& {Moraux}}]{Joncour17}
{Joncour}, I., {Duch{\^e}ne}, G., \& {Moraux}, E. 2017, \aap, 599, A14, \dodoi{10.1051/0004-6361/201629398}

\bibitem[{{Kaib} \& {Raymond}(2014)}]{Kaib2014}
{Kaib}, N.~A., \& {Raymond}, S.~N. 2014, \apj, 782, 60, \dodoi{10.1088/0004-637X/782/2/60}

\bibitem[{{Katz} \& {Dong}(2012)}]{Katz12}
{Katz}, B., \& {Dong}, S. 2012, arXiv e-prints, arXiv:1211.4584, \dodoi{10.48550/arXiv.1211.4584}

\bibitem[{{Knigge} {et~al.}(2022){Knigge}, {Toonen}, \& {Boekholt}}]{Knigge22}
{Knigge}, C., {Toonen}, S., \& {Boekholt}, T.~C.~N. 2022, \mnras, 514, 1895, \dodoi{10.1093/mnras/stac1336}

\bibitem[{{Kokori} {et~al.}(2023){Kokori}, {Tsiaras}, {Edwards}, {Jones}, {Pantelidou}, {Tinetti}, {Bewersdorff}, {Iliadou}, {Jongen}, {Lekkas}, {Nastasi}, {Poultourtzidis}, {Sidiropoulos}, {Walter}, {W{\"u}nsche}, {Abraham}, {Agnihotri}, {Albanesi}, {Arce-Mansego}, {Arnot}, {Audejean}, {Aumasson}, {Bachschmidt}, {Baj}, {Barroy}, {Belinski}, {Bennett}, {Benni}, {Bernacki}, {Betti}, {Biagini}, {Bosch}, {Brandebourg}, {Br{\'a}t}, {Bretton}, {Brincat}, {Brouillard}, {Bruzas}, {Bruzzone}, {Buckland}, {Cal{\'o}}, {Campos}, {Carre{\~n}o}, {Carrion Rodrigo}, {Casali}, {Casalnuovo}, {Cataneo}, {Chang}, {Changeat}, {Chowdhury}, {Ciantini}, {Cilluffo}, {Coliac}, {Conzo}, {Correa}, {Coulon}, {Crouzet}, {Crow}, {Curtis}, {Daniel}, {Dauchet}, {Dawes}, {Deldem}, {Deligeorgopoulos}, {Dransfield}, {Dymock}, {Eenm{\"a}e}, {Esseiva}, {Evans}, {Falco}, {Farf{\'a}n}, {Fern{\'a}ndez-Laj{\'u}s}, {Ferratfiat}, {Ferreira}, {Ferretti}, {Fio{\l}ka}, {Fowler}, {Futcher}, {Gabellini}, {Gainey}, {Gaitan}, {Gajdo{\v{s}}},
  {Garc{\'\i}a-S{\'a}nchez}, {Garlitz}, {Gillier}, {Gison}, {Gonzales}, {Gorshanov}, {Grau Horta}, {Grivas}, {Guerra}, {Guillot}, {Haswell}, {Haymes}, {Hentunen}, {Hills}, {Hose}, {Humbert}, {Hurter}, {Hynek}, {Irzyk}, {Jacobsen}, {Jannetta}, {Johnson}, {J{\'o}{\'z}wik-Wabik}, {Kaeouach}, {Kang}, {Kiiskinen}, {Kim}, {Kivila}, {Koch}, {Kolb}, {Ku{\v{c}}{\'a}kov{\'a}}, {Lai}, {Laloum}, {Lasota}, {Lewis}, {Liakos}, {Libotte}, {Lomoz}, {Lopresti}, {Majewski}, {Malcher}, {Mallonn}, {Mannucci}, {Marchini}, {Mari}, {Marino}, {Marino}, {Mario}, {Marquette}, {Mart{\'\i}nez-Bravo}, {Ma{\v{s}}ek}, {Matassa}, {Michel}, {Michelet}, {Miller}, {Miny}, {Molina}, {Mollier}, {Monteleone}, {Montigiani}, {Morales-Aimar}, {Mortari}, {Morvan}, {Mugnai}, {Murawski}, {Naponiello}, {Naudin}, {Naves}, {N{\'e}el}, {Neito}, {Neveu}, {Noschese}, {{\"O}{\u{g}}men}, {Ohshima}, {Orbanic}, {Pace}, {Pantacchini}, {Paschalis}, {Pereira}, {Peretto}, {Perroud}, {Phillips}, {Pintr}, {Pioppa}, {Plazas}, {Poelarends}, {Popowicz}, {Purcell},
  {Quinn}, {Raetz}, {Rees}, {Regembal}, {Rocchetto}, {Rocci}, {Rockenbauer}, {Roth}, {Rousselot}, {Rubia}, {Ruocco}, {Russo}, {Salisbury}, {Salvaggio}, {Santos}, {Savage}, {Scaggiante}, {Sedita}, {Shadick}, {Silva}, {Sioulas}, {{\v{S}}koln{\'\i}k}, {Smith}, {Smolka}, {Solmaz}, {Stanbury}, {Stouraitis}, {Tan}, {Theusner}, \& {Thurston}}]{Kokori23}
{Kokori}, A., {Tsiaras}, A., {Edwards}, B., {et~al.} 2023, \apjs, 265, 4, \dodoi{10.3847/1538-4365/ac9da4}

\bibitem[{{Kounkel} {et~al.}(2019){Kounkel}, {Covey}, {Moe}, {Kratter}, {Su{\'a}rez}, {Stassun}, {Rom{\'a}n-Z{\'u}{\~n}iga}, {Hernandez}, {Kim}, {Pe{\~n}a Ram{\'\i}rez}, {Roman-Lopes}, {Stringfellow}, {Jaehnig}, {Borissova}, {Tofflemire}, {Krolikowski}, {Rizzuto}, {Kraus}, {Badenes}, {Longa-Pe{\~n}a}, {G{\'o}mez Maqueo Chew}, {Barba}, {Nidever}, {Brown}, {De Lee}, {Pan}, {Bizyaev}, {Oravetz}, \& {Oravetz}}]{Kounkel19}
{Kounkel}, M., {Covey}, K., {Moe}, M., {et~al.} 2019, \aj, 157, 196, \dodoi{10.3847/1538-3881/ab13b1}

\bibitem[{{Kouwenhoven} {et~al.}(2010){Kouwenhoven}, {Goodwin}, {Parker}, {Davies}, {Malmberg}, \& {Kroupa}}]{Kouwenhoven10}
{Kouwenhoven}, M.~B.~N., {Goodwin}, S.~P., {Parker}, R.~J., {et~al.} 2010, \mnras, 404, 1835, \dodoi{10.1111/j.1365-2966.2010.16399.x}

\bibitem[{{Kozai}(1962)}]{Kozai1962}
{Kozai}, Y. 1962, \aj, 67, 591, \dodoi{10.1086/108790}

\bibitem[{{Kratter} \& {Lodato}(2016)}]{Kratter16}
{Kratter}, K., \& {Lodato}, G. 2016, \araa, 54, 271, \dodoi{10.1146/annurev-astro-081915-023307}

\bibitem[{{Kroupa}(1995{\natexlab{a}})}]{Kroupa95a}
{Kroupa}, P. 1995{\natexlab{a}}, \mnras, 277, 1491, \dodoi{10.1093/mnras/277.4.1491}

\bibitem[{{Kroupa}(1995{\natexlab{b}})}]{Kroupa95b}
---. 1995{\natexlab{b}}, \mnras, 277, 1507, \dodoi{10.1093/mnras/277.4.1507}

\bibitem[{{Kummer} {et~al.}(2023){Kummer}, {Toonen}, \& {de Koter}}]{Kummer23}
{Kummer}, F., {Toonen}, S., \& {de Koter}, A. 2023, arXiv e-prints, arXiv:2306.09400, \dodoi{10.48550/arXiv.2306.09400}

\bibitem[{{Lagos} {et~al.}(2021){Lagos}, {Schreiber}, {Zorotovic}, {G{\"a}nsicke}, {Ronco}, \& {Hamers}}]{Lagos21}
{Lagos}, F., {Schreiber}, M.~R., {Zorotovic}, M., {et~al.} 2021, \mnras, 501, 676, \dodoi{10.1093/mnras/staa3703}

\bibitem[{{Laos} {et~al.}(2020){Laos}, {Stassun}, \& {Mathieu}}]{Laos20}
{Laos}, E., {Stassun}, K.~G., \& {Mathieu}, R.~D. 2020, \apj, 902, 107, \dodoi{10.3847/1538-4357/abb3fe}

\bibitem[{{Lee} {et~al.}(2019){Lee}, {Offner}, {Kratter}, {Smullen}, \& {Li}}]{Lee19}
{Lee}, A.~T., {Offner}, S. S.~R., {Kratter}, K.~M., {Smullen}, R.~A., \& {Li}, P.~S. 2019, \apj, 887, 232, \dodoi{10.3847/1538-4357/ab584b}

\bibitem[{{Leiner} {et~al.}(2025){Leiner}, {Gosnell}, {Geller}, {Sun}, {Mathieu}, \& {Sills}}]{Leiner25}
{Leiner}, E.~M., {Gosnell}, N.~M., {Geller}, A.~M., {et~al.} 2025, \apjl, 979, L1, \dodoi{10.3847/2041-8213/ad9d0c}

\bibitem[{{Lidov}(1962)}]{Lidov1962}
{Lidov}, M.~L. 1962, \planss, 9, 719, \dodoi{10.1016/0032-0633(62)90129-0}

\bibitem[{{Lucy} \& {Ricco}(1979)}]{Lucy97}
{Lucy}, L.~B., \& {Ricco}, E. 1979, \aj, 84, 401, \dodoi{10.1086/112434}

\bibitem[{{Mardling} \& {Aarseth}(2001)}]{MA2001}
{Mardling}, R.~A., \& {Aarseth}, S.~J. 2001, \mnras, 321, 398, \dodoi{10.1046/j.1365-8711.2001.03974.x}

\bibitem[{{Marks} \& {Kroupa}(2011)}]{Marks11}
{Marks}, M., \& {Kroupa}, P. 2011, \mnras, 417, 1702, \dodoi{10.1111/j.1365-2966.2011.19519.x}

\bibitem[{{Mayo} {et~al.}(2018){Mayo}, {Vanderburg}, {Latham}, {Bieryla}, {Morton}, {Buchhave}, {Dressing}, {Beichman}, {Berlind}, {Calkins}, {Ciardi}, {Crossfield}, {Esquerdo}, {Everett}, {Gonzales}, {Hirsch}, {Horch}, {Howard}, {Howell}, {Livingston}, {Patel}, {Petigura}, {Schlieder}, {Scott}, {Schumer}, {Sinukoff}, {Teske}, \& {Winters}}]{Mayo18}
{Mayo}, A.~W., {Vanderburg}, A., {Latham}, D.~W., {et~al.} 2018, \aj, 155, 136, \dodoi{10.3847/1538-3881/aaadff}

\bibitem[{{Michaely} \& {Perets}(2016)}]{Michaely2016}
{Michaely}, E., \& {Perets}, H.~B. 2016, \mnras, 458, 4188, \dodoi{10.1093/mnras/stw368}

\bibitem[{{Moe} \& {Di Stefano}(2017)}]{Moe17}
{Moe}, M., \& {Di Stefano}, R. 2017, \apjs, 230, 15, \dodoi{10.3847/1538-4365/aa6fb6}

\bibitem[{{Moe} \& {Kratter}(2021)}]{MoeKratter21}
{Moe}, M., \& {Kratter}, K.~M. 2021, \mnras, 507, 3593, \dodoi{10.1093/mnras/stab2328}

\bibitem[{{Montet} {et~al.}(2015){Montet}, {Morton}, {Foreman-Mackey}, {Johnson}, {Hogg}, {Bowler}, {Latham}, {Bieryla}, \& {Mann}}]{Montet15}
{Montet}, B.~T., {Morton}, T.~D., {Foreman-Mackey}, D., {et~al.} 2015, \apj, 809, 25, \dodoi{10.1088/0004-637X/809/1/25}

\bibitem[{{Mowlavi} {et~al.}(2023){Mowlavi}, {Holl}, {Lecoeur-Ta{\"\i}bi}, {Barblan}, {Kochoska}, {Pr{\v{s}}a}, {Mazeh}, {Rimoldini}, {Gavras}, {Audard}, {Jevardat de Fombelle}, {Nienartowicz}, {Garc{\'\i}a-Lario}, \& {Eyer}}]{gaia_ecl}
{Mowlavi}, N., {Holl}, B., {Lecoeur-Ta{\"\i}bi}, I., {et~al.} 2023, \aap, 674, A16, \dodoi{10.1051/0004-6361/202245330}

\bibitem[{{Mu{\~n}oz} \& {Petrovich}(2020)}]{Munoz20}
{Mu{\~n}oz}, D.~J., \& {Petrovich}, C. 2020, \apjl, 904, L3, \dodoi{10.3847/2041-8213/abc564}

\bibitem[{{Mushkin} \& {Katz}(2020)}]{Mushkin20}
{Mushkin}, J., \& {Katz}, B. 2020, \mnras, 498, 665, \dodoi{10.1093/mnras/staa2492}

\bibitem[{{Naoz}(2016)}]{Naoz2016}
{Naoz}, S. 2016, \araa, 54, 441, \dodoi{10.1146/annurev-astro-081915-023315}

\bibitem[{{Naoz} \& {Fabrycky}(2014)}]{Naoz2014}
{Naoz}, S., \& {Fabrycky}, D.~C. 2014, \apj, 793, 137, \dodoi{10.1088/0004-637X/793/2/137}

\bibitem[{{Naoz} {et~al.}(2013){Naoz}, {Farr}, {Lithwick}, {Rasio}, \& {Teyssandier}}]{Naoz2013sec}
{Naoz}, S., {Farr}, W.~M., {Lithwick}, Y., {Rasio}, F.~A., \& {Teyssandier}, J. 2013, \mnras, 431, 2155, \dodoi{10.1093/mnras/stt302}

\bibitem[{{Naoz} {et~al.}(2016){Naoz}, {Fragos}, {Geller}, {Stephan}, \& {Rasio}}]{NaozLMXB}
{Naoz}, S., {Fragos}, T., {Geller}, A., {Stephan}, A.~P., \& {Rasio}, F.~A. 2016, \apjl, 822, L24, \dodoi{10.3847/2041-8205/822/2/L24}

\bibitem[{{Niu} {et~al.}(2022){Niu}, {Yuan}, {Wang}, \& {Liu}}]{Niu22_WBmetals}
{Niu}, Z., {Yuan}, H., {Wang}, Y., \& {Liu}, J. 2022, \apj, 931, 124, \dodoi{10.3847/1538-4357/ac6c84}

\bibitem[{{Ochi} {et~al.}(2005){Ochi}, {Sugimoto}, \& {Hanawa}}]{Ochi05}
{Ochi}, Y., {Sugimoto}, K., \& {Hanawa}, T. 2005, \apj, 623, 922, \dodoi{10.1086/428601}

\bibitem[{{O'Connor} {et~al.}(2021){O'Connor}, {Liu}, \& {Lai}}]{OConnor21}
{O'Connor}, C.~E., {Liu}, B., \& {Lai}, D. 2021, \mnras, 501, 507, \dodoi{10.1093/mnras/staa3723}

\bibitem[{{Offner} {et~al.}(2010){Offner}, {Kratter}, {Matzner}, {Krumholz}, \& {Klein}}]{Offner10}
{Offner}, S. S.~R., {Kratter}, K.~M., {Matzner}, C.~D., {Krumholz}, M.~R., \& {Klein}, R.~I. 2010, \apj, 725, 1485, \dodoi{10.1088/0004-637X/725/2/1485}

\bibitem[{{Offner} {et~al.}(2023){Offner}, {Moe}, {Kratter}, {Sadavoy}, {Jensen}, \& {Tobin}}]{Offner23}
{Offner}, S.~S.~R., {Moe}, M., {Kratter}, K.~M., {et~al.} 2023, in Astronomical Society of the Pacific Conference Series, Vol. 534, Protostars and Planets VII, ed. S.~{Inutsuka}, Y.~{Aikawa}, T.~{Muto}, K.~{Tomida}, \& M.~{Tamura}, 275, \dodoi{10.48550/arXiv.2203.10066}

\bibitem[{{Pecaut} \& {Mamajek}(2013)}]{Pecaut13}
{Pecaut}, M.~J., \& {Mamajek}, E.~E. 2013, \apjs, 208, 9, \dodoi{10.1088/0067-0049/208/1/9}

\bibitem[{{Perets}(2025)}]{Perets25}
{Perets}, H.~B. 2025, arXiv e-prints, arXiv:2504.02939, \dodoi{10.48550/arXiv.2504.02939}

\bibitem[{{Perets} \& {Fabrycky}(2009)}]{Perets09}
{Perets}, H.~B., \& {Fabrycky}, D.~C. 2009, \apj, 697, 1048, \dodoi{10.1088/0004-637X/697/2/1048}

\bibitem[{{Perets} \& {Kratter}(2012)}]{Perets12}
{Perets}, H.~B., \& {Kratter}, K.~M. 2012, \apj, 760, 99, \dodoi{10.1088/0004-637X/760/2/99}

\bibitem[{{Perpiny{\`a}-Vall{\`e}s} {et~al.}(2019){Perpiny{\`a}-Vall{\`e}s}, {Rebassa-Mansergas}, {G{\"a}nsicke}, {Toonen}, {Hermes}, {Gentile Fusillo}, \& {Tremblay}}]{Perpinya19_resolvedtripWD}
{Perpiny{\`a}-Vall{\`e}s}, M., {Rebassa-Mansergas}, A., {G{\"a}nsicke}, B.~T., {et~al.} 2019, \mnras, 483, 901, \dodoi{10.1093/mnras/sty3149}

\bibitem[{{Petigura} {et~al.}(2017){Petigura}, {Sinukoff}, {Lopez}, {Crossfield}, {Howard}, {Brewer}, {Fulton}, {Isaacson}, {Ciardi}, {Howell}, {Everett}, {Horch}, {Hirsch}, {Weiss}, \& {Schlieder}}]{Petigura17}
{Petigura}, E.~A., {Sinukoff}, E., {Lopez}, E.~D., {et~al.} 2017, \aj, 153, 142, \dodoi{10.3847/1538-3881/aa5ea5}

\bibitem[{{Raghavan} {et~al.}(2010){Raghavan}, {McAlister}, {Henry}, {Latham}, {Marcy}, {Mason}, {Gies}, {White}, \& {ten Brummelaar}}]{Raghavan2010}
{Raghavan}, D., {McAlister}, H.~A., {Henry}, T.~J., {et~al.} 2010, \apjs, 190, 1, \dodoi{10.1088/0067-0049/190/1/1}

\bibitem[{{Reipurth} \& {Mikkola}(2012)}]{Reipurth12}
{Reipurth}, B., \& {Mikkola}, S. 2012, \nat, 492, 221, \dodoi{10.1038/nature11662}

\bibitem[{{Shariat} {et~al.}(2025{\natexlab{a}}){Shariat}, {El-Badry}, {Naoz}, {Rodriguez}, \& {van Roestel}}]{Shariat25CV}
{Shariat}, C., {El-Badry}, K., {Naoz}, S., {Rodriguez}, A.~C., \& {van Roestel}, J. 2025{\natexlab{a}}, arXiv e-prints, arXiv:2501.14025, \dodoi{10.48550/arXiv.2501.14025}

\bibitem[{{Shariat} {et~al.}(2025{\natexlab{b}}){Shariat}, {Naoz}, {El-Badry}, {Rocha}, {Kalogera}, {Stephan}, {Burdge}, \& {Angelo}}]{Shariat24LMXB}
{Shariat}, C., {Naoz}, S., {El-Badry}, K., {et~al.} 2025{\natexlab{b}}, \apj, 983, 115, \dodoi{10.3847/1538-4357/adbf01}

\bibitem[{{Shariat} {et~al.}(2025{\natexlab{c}}){Shariat}, {Naoz}, {El-Badry}, {Rodriguez}, {Hansen}, {Angelo}, \& {Stephan}}]{Shariat25Merge}
---. 2025{\natexlab{c}}, \apj, 978, 47, \dodoi{10.3847/1538-4357/ad944a}

\bibitem[{{Shariat} {et~al.}(2023){Shariat}, {Naoz}, {Hansen}, {Angelo}, {Michaely}, \& {Stephan}}]{Shariat23}
{Shariat}, C., {Naoz}, S., {Hansen}, B. M.~S., {et~al.} 2023, \apjl, 955, L14, \dodoi{10.3847/2041-8213/acf76b}

\bibitem[{{Stephan} {et~al.}(2021){Stephan}, {Naoz}, \& {Gaudi}}]{Stephan21}
{Stephan}, A.~P., {Naoz}, S., \& {Gaudi}, B.~S. 2021, \apj, 922, 4, \dodoi{10.3847/1538-4357/ac22a9}

\bibitem[{{Sterzik} \& {Tokovinin}(2002)}]{Sterzik2002}
{Sterzik}, M.~F., \& {Tokovinin}, A.~A. 2002, \aap, 384, 1030, \dodoi{10.1051/0004-6361:20020105}

\bibitem[{{Thomasson} {et~al.}(2024){Thomasson}, {Joncour}, {Moraux}, {Motte}, {Louvet}, {Gonz{\'a}lez}, \& {Nony}}]{Thomasson24}
{Thomasson}, B., {Joncour}, I., {Moraux}, E., {et~al.} 2024, \aap, 689, A133, \dodoi{10.1051/0004-6361/202449649}

\bibitem[{{Tian} {et~al.}(2020){Tian}, {El-Badry}, {Rix}, \& {Gould}}]{Tian20}
{Tian}, H.-J., {El-Badry}, K., {Rix}, H.-W., \& {Gould}, A. 2020, \apjs, 246, 4, \dodoi{10.3847/1538-4365/ab54c4}

\bibitem[{{Tobin} {et~al.}(2016){Tobin}, {Looney}, {Li}, {Chandler}, {Dunham}, {Segura-Cox}, {Sadavoy}, {Melis}, {Harris}, {Kratter}, \& {Perez}}]{Tobin16}
{Tobin}, J.~J., {Looney}, L.~W., {Li}, Z.-Y., {et~al.} 2016, \apj, 818, 73, \dodoi{10.3847/0004-637X/818/1/73}

\bibitem[{{Tokovinin}(2008)}]{Tokovinin08}
{Tokovinin}, A. 2008, \mnras, 389, 925, \dodoi{10.1111/j.1365-2966.2008.13613.x}

\bibitem[{{Tokovinin}(2014{\natexlab{a}})}]{Tokovinin14a}
---. 2014{\natexlab{a}}, \aj, 147, 86, \dodoi{10.1088/0004-6256/147/4/86}

\bibitem[{{Tokovinin}(2014{\natexlab{b}})}]{Tokovinin14b}
---. 2014{\natexlab{b}}, \aj, 147, 87, \dodoi{10.1088/0004-6256/147/4/87}

\bibitem[{{Tokovinin}(2020)}]{Tokovinin20_ecc}
---. 2020, \mnras, 496, 987, \dodoi{10.1093/mnras/staa1639}

\bibitem[{{Tokovinin}(2021)}]{Tokovinin21}
---. 2021, Universe, 7, 352, \dodoi{10.3390/universe7090352}

\bibitem[{{Tokovinin}(2022)}]{Tokovinin22_resolvedtriples}
---. 2022, \apj, 926, 1, \dodoi{10.3847/1538-4357/ac4584}

\bibitem[{{Tokovinin}(2023)}]{Tokovinin23_gaia}
---. 2023, arXiv e-prints, arXiv:2303.17620, \dodoi{10.48550/arXiv.2303.17620}

\bibitem[{{Tokovinin} \& {Kiyaeva}(2016)}]{Tokovinin16_ecc}
{Tokovinin}, A., \& {Kiyaeva}, O. 2016, \mnras, 456, 2070, \dodoi{10.1093/mnras/stv2825}

\bibitem[{{Tokovinin} {et~al.}(2006){Tokovinin}, {Thomas}, {Sterzik}, \& {Udry}}]{Tokovinin06}
{Tokovinin}, A., {Thomas}, S., {Sterzik}, M., \& {Udry}, S. 2006, \aap, 450, 681, \dodoi{10.1051/0004-6361:20054427}

\bibitem[{{Tokovinin}(1997)}]{Tokovinin1997}
{Tokovinin}, A.~A. 1997, Astronomy Letters, 23, 727

\bibitem[{{Tokovinin}(2000)}]{Tokovinin00}
---. 2000, \aap, 360, 997

\bibitem[{{Toonen} {et~al.}(2022){Toonen}, {Boekholt}, \& {Portegies Zwart}}]{Toonen2022}
{Toonen}, S., {Boekholt}, T.~C.~N., \& {Portegies Zwart}, S. 2022, \aap, 661, A61, \dodoi{10.1051/0004-6361/202141991}

\bibitem[{{Toonen} {et~al.}(2018){Toonen}, {Perets}, \& {Hamers}}]{Toonen18}
{Toonen}, S., {Perets}, H.~B., \& {Hamers}, A.~S. 2018, \aap, 610, A22, \dodoi{10.1051/0004-6361/201731874}

\bibitem[{{Toonen} {et~al.}(2020){Toonen}, {Portegies Zwart}, {Hamers}, \& {Bandopadhyay}}]{Toonen20}
{Toonen}, S., {Portegies Zwart}, S., {Hamers}, A.~S., \& {Bandopadhyay}, D. 2020, \aap, 640, A16, \dodoi{10.1051/0004-6361/201936835}

\bibitem[{{Van Eylen} {et~al.}(2016){Van Eylen}, {Nowak}, {Albrecht}, {Palle}, {Ribas}, {Bruntt}, {Perger}, {Gandolfi}, {Hirano}, {Sanchis-Ojeda}, {Kiilerich}, {Prieto-Arranz}, {Badenas}, {Dai}, {Deeg}, {Guenther}, {Monta{\~n}{\'e}s-Rodr{\'\i}guez}, {Narita}, {Rogers}, {B{\'e}jar}, {Shrotriya}, {Winn}, \& {Sebastian}}]{Van-Eylen16}
{Van Eylen}, V., {Nowak}, G., {Albrecht}, S., {et~al.} 2016, \apj, 820, 56, \dodoi{10.3847/0004-637X/820/1/56}

\bibitem[{{Vanderburg} {et~al.}(2020){Vanderburg}, {Rappaport}, {Xu}, {Crossfield}, {Becker}, {Gary}, {Murgas}, {Blouin}, {Kaye}, {Palle}, {Melis}, {Morris}, {Kreidberg}, {Gorjian}, {Morley}, {Mann}, {Parviainen}, {Pearce}, {Newton}, {Carrillo}, {Zuckerman}, {Nelson}, {Zeimann}, {Brown}, {Tronsgaard}, {Klein}, {Ricker}, {Vanderspek}, {Latham}, {Seager}, {Winn}, {Jenkins}, {Adams}, {Benneke}, {Berardo}, {Buchhave}, {Caldwell}, {Christiansen}, {Collins}, {Col{\'o}n}, {Daylan}, {Doty}, {Doyle}, {Dragomir}, {Dressing}, {Dufour}, {Fukui}, {Glidden}, {Guerrero}, {Guo}, {Heng}, {Henriksen}, {Huang}, {Kaltenegger}, {Kane}, {Lewis}, {Lissauer}, {Morales}, {Narita}, {Pepper}, {Rose}, {Smith}, {Stassun}, \& {Yu}}]{Vanderburg20}
{Vanderburg}, A., {Rappaport}, S.~A., {Xu}, S., {et~al.} 2020, \nat, 585, 363, \dodoi{10.1038/s41586-020-2713-y}

\bibitem[{{von Zeipel}(1910)}]{vonZeipel1910}
{von Zeipel}, H. 1910, Astronomische Nachrichten, 183, 345, \dodoi{10.1002/asna.19091832202}

\bibitem[{{Weldon} {et~al.}(2024){Weldon}, {Naoz}, \& {Hansen}}]{Weldon24}
{Weldon}, G.~C., {Naoz}, S., \& {Hansen}, B. M.~S. 2024, \apj, 974, 302, \dodoi{10.3847/1538-4357/ad77a9}

\bibitem[{{White} \& {Ghez}(2001)}]{White01}
{White}, R.~J., \& {Ghez}, A.~M. 2001, \apj, 556, 265, \dodoi{10.1086/321542}

\bibitem[{{Winters} {et~al.}(2019){Winters}, {Henry}, {Jao}, {Subasavage}, {Chatelain}, {Slatten}, {Riedel}, {Silverstein}, \& {Payne}}]{Winters19}
{Winters}, J.~G., {Henry}, T.~J., {Jao}, W.-C., {et~al.} 2019, \aj, 157, 216, \dodoi{10.3847/1538-3881/ab05dc}

\bibitem[{{Worley}(1967)}]{Worley67}
{Worley}, C.~E. 1967, in On the Evolution of Double Stars, ed. J.~{Dommanget}, Vol.~17, 221

\bibitem[{{Yang} {et~al.}(2025){Yang}, {Su}, \& {Winn}}]{Yang2025}
{Yang}, E., {Su}, Y., \& {Winn}, J.~N. 2025, arXiv e-prints, arXiv:2505.07927, \dodoi{10.48550/arXiv.2505.07927}

\bibitem[{{Young} \& {Clarke}(2015)}]{Young15}
{Young}, M.~D., \& {Clarke}, C.~J. 2015, \mnras, 452, 3085, \dodoi{10.1093/mnras/stv1512}

\bibitem[{{Zhang} {et~al.}(2023){Zhang}, {Naoz}, \& {Will}}]{Zhang23}
{Zhang}, E., {Naoz}, S., \& {Will}, C.~M. 2023, arXiv e-prints, arXiv:2301.08271, \dodoi{10.48550/arXiv.2301.08271}

\end{thebibliography}
\end{document}